\documentclass[12pt,authoryear]{elsarticle}
 \usepackage{subfigure}
 \usepackage{hyperref}
\usepackage{amssymb}
\usepackage{amsfonts}
\usepackage{dsfont}
\usepackage{amsmath}
\usepackage{bm}
\usepackage{multirow}
\usepackage{color}
\usepackage{array}

\newcommand{\be}{\begin{equation}}
\newcommand{\ee}{\end{equation}}
\newcommand{\bq}{\begin{eqnarray}}
\newcommand{\eq}{\end{eqnarray}}

\newcommand{\bc}{\begin{center}}
\newcommand{\ec}{\end{center}}
\newcommand{\beq}{\begin{equation}}
\newcommand{\eeq}{\end{equation}}
\newcommand{\bea}{\begin{eqnarray}}
\newcommand{\eea}{\end{eqnarray}}

\newcommand{\ket}[1]{\ensuremath{| #1 \rangle}}

\journal{Advances in Atomic, Molecular and Optical Physics}

\begin{document}

\begin{frontmatter}

\title{Ultracold ion-atom experiments: \\cooling, chemistry, and quantum effects}
\date{}
 \author{Rianne S. Lous}
 \address{Van der Waals-Zeeman Institute, Institute of Physics,
University of Amsterdam, 1098 XH Amsterdam, the Netherlands}
 \author{Ren\'e Gerritsma}
 \address{Van der Waals-Zeeman Institute, Institute of Physics,
University of Amsterdam, 1098 XH Amsterdam, the Netherlands}
\address{QuSoft, Science Park 123, 1098 XG Amsterdam, the Netherlands}

\begin{abstract}
Experimental setups that study laser-cooled ions immersed in baths of ultracold atoms merge the two exciting and well-established fields of quantum gases and trapped ions. These experiments benefit both from the exquisite read-out and control of the few-body ion systems as well as the many-body aspects, tunable interactions, and ultracold temperatures of the atoms. However, combining the two leads to challenges both in the experimental design and the physics that can be studied. Nevertheless, these systems have provided insights into ion-atom collisions, buffer gas cooling of ions and quantum effects in the ion-atom interaction. This makes them promising candidates for ultracold quantum chemistry studies, creation of cold molecular ions for spectroscopy and precision measurements, and as test beds for quantum simulation of charged impurity physics. In this review we aim to provide an experimental account of recent progress and introduce the experimental setup and techniques that enabled the observation of quantum effects.  
\end{abstract}


\end{frontmatter}

%DRAFT: \today

%\maketitle
\newpage
\tableofcontents

\newpage
\section{Introduction}
\label{sec:intro}
 %motivation: why ultracold atoms and ions? 
%include a bit of history of the field 

Over the last decade, a new field of cold atomic physics has emerged in which single cold ions are immersed into ultracold atomic gases. These experiments were motivated by prospects to buffer gas cool trapped ions, study quantum chemistry between atoms and ions and investigate cold collisions between the particles. Moreover they offer prospects for simulating quantum many-body physics and as applications in quantum information processing. In this review, we aim to give an up-to-date experimental account of where the field is currently standing and where it may potentially go. We put a strong emphasis on experimental techniques and results that lead to the recent breakthroughs.  Evidently, the merger of neutral and charged particles is by no means new in atomic and molecular physics and to give a complete review is not our aim here. Instead, we choose to make a clear distinction in terms of energy: we only consider the coldest systems available. This means in practical terms that we omit the beautiful work done with atoms in magneto optical traps and focus instead on atoms in optical dipole traps with temperatures in the (sub)~$\mu$K regime. These ultracold atoms are typically obtained through evaporative cooling, as their temperatures are sub-Doppler, and are in or close to a degenerate quantum gas. Furthermore, we focus on one or a few ions trapped in a many-body bath of atoms.

\begin{figure}[!t]
\center
\includegraphics[height=0.6\textwidth]{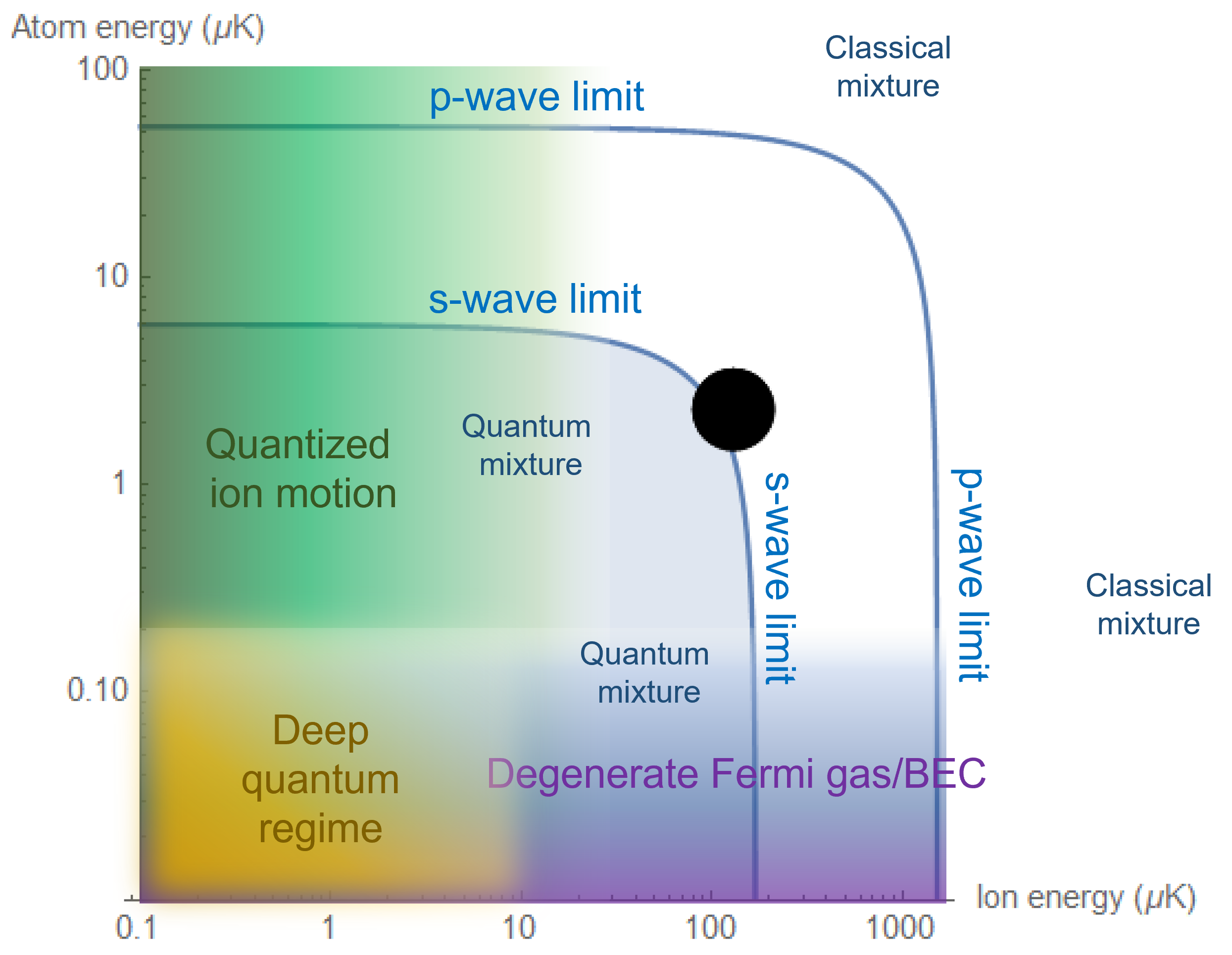}
\caption{Possibilities of ultracold ion-atom mixtures when cooling further towards the quantum regime. The equi-collision-energy lines correspond to the combination $^{171}$Yb$^+$-$^6$Li and the black dot represents the energy reported in Ref.~\citep{Feldker2020bgc}.} \label{Overview}
%\vspace{-5cm}
\end{figure}

The ultracold ion-atom systems we consider can be illustrated by Fig.~\ref{Overview}, which shows the relevant energy scales of the Yb$^+$-Li system studied in Amsterdam\footnote{Note that we work with the combination of Yb$^+$-Li and thus write this review from the perspective of our group in Amsterdam.}. The collision energy between the atom and ion in this system was observed to reach the quantum, or $s$-wave regime in which the collisional angular momentum is quantized and only allows for $s$-wave collisions. In this regime, the description of interacting atoms and ions requires quantum mechanics and quantum effects can be observed. Each of the separate systems of atoms and ions have their own quantum limits. For the ions, the quantization of motion in its trap occurs once its energy drops below that of a single quantum of motion given by $\hbar\omega$, with $\omega$ the ion trap frequency and $\hbar$ the reduced Planck constant. For the atoms, the crossover occurs once the system turns into a quantum gas. Then the atom-atom collisions are described by a single parameter, the $s$-wave scattering length, and the collisions require an explicit quantum mechanical treatment. Cooling the atoms even further, such that the interparticle spacing is smaller than the de Broglie wavelength, quantum degeneracy is reached. In the case of bosons a Bose-Einstein condensate is formed and for fermions a degenerate Fermi gas. The `deep quantum regime' in which both atoms, ions and their interactions are in the quantum regime, has not been reached in a sustained manner and remains an outstanding goal for the future with applications in e.g. quantum impurity physics.

Since the onset, the holy grail of the field has been to observe quantum effects in these hybrid ion-atom systems and most of the mentioned applications benefit from or require reaching the so-called quantum limit (see Fig.~\ref{Overview}). However, efforts to reach this regime have long been hindered and only recently have two groups reported observations of quantum effects in a trapped ion interacting with an ultracold cloud of atoms. In the first~\citep{Feldker2020bgc}, the collision energy of the particles was determined to be on the order of the so-called $s$-wave regime, in which the angular momentum in ion-atom collisions freezes out, and spectral features in ion-atom spin exchange allowed for a first estimate of ion-atom $s$-wave scattering lengths. In the second, Feshbach resonances between atoms and trapped ions were observed for the first time~\citep{Weckesser2021oof}. Feshbach resonances are ubiquitous in ultracold neutral atom systems where they are used to tune the interaction between atoms, for the creation of ultracold molecules and in studies of quantum many-body physics~\citep{Chin2010fri}. The observation of Feshbach resonances in ion-atom systems is important for extending these techniques to neutral-charged mixtures with exciting possibilities for e.g. quantum impurity physics, quantum chemistry and quantum simulation.

However, singling out these two very recent results does not do full justice to the rapid developments in the whole field. Other important examples of breakthroughs include the observation of signatures of $s$-wave spin dynamics far above the $s$-wave energy~\citep{Sikorsky2018plb,Cote2018sot,Fuerst2018doa}, rapid improvements in buffer gas cooling~\citep{Haze2018cdo,Schmidt2020otf,Hirzler2020esf,Dieterle2021toa,Trimby2022bgc}, observations of interactions between Rydberg atoms and ions~\citep{Engel2018oor,Ewald:2019,Haze:2019,Deiss2021lra,Zuber2021sio}, the accurate experimental breakdown of cold chemistry reactions paths in ion-atom mixtures~\citep{Silva:2017,Sikorsky2018qca,Benshlomi2020doo,Mohammadi:2021,BenShlomi2021her}, quantum logic spectroscopy of ion-atom chemistry~\citep{Katz2022qld} and recently the observation of interactions between atomic Feshbach dimers and trapped ions~\citep{Hirzler2020ctn, Hirzler2022ooc}, just to mention a few. 

These experiments build on the earlier experimental work on hybrid systems~\citep{Smith2005cin, Grier2009ooc,Zipkes2010ats,Schmid2010doa, Ravi2012cas,Haze:2015, Meir2016doa, Schowalter2016bsb, Kleinbach2018iii}, which looked the reactions that take place~\citep{Harter2012sia,Ratschbacher:2012,Ratschbacher2013doa,Haze:2015,Kruekow2016,Saito:2017} and studied the cold ion-atom collisions~\citep{Zipkes2011koa,Meir2016doa,Fuerst2018por}. The ongoing work is further stimulated by the recent theory insights into, e.g., ion-atom interactions~\citep{Tscherbul2016soi,Schmid2018rmf, Secker:2017, Cote2018sot, Smialkowski2020iac, Tomza2020iac, Wang2020oco, Doerfler2020rsc, Bosworth2021spo, PerezRios2021efd}, quantum simulation of impurity and many-body physics~\citep{Bissbort:2013,Negretti:2014,Schurer2014,Midya2016roc,PerezRios2021cca, Astrakharchik2021ipi,Christensen2021cpa,Christensen2022mii,Oghittu2021doa,ding2022mib} and proposals for applications in quantum information processing~\citep{Doerk:2010,Secker:2016,Ebgha2019cai}.  For an in-depth theory overview or a more historical account, we refer to the previous reviews which can be found in Refs.~\citep{Harter2014cai,Sias2014hqs, Willitsch:2015,Cote2016uha,Zhang:2017,Tomza2019chi}. Here, we review the latest experimental progress.

Here, we start by providing a detailed review of the most important experimental techniques used for preparation of and measurements on ion-atom systems, followed by followed by a discussion of the most recent experimental results. In Sec.~\ref{Langevin}, we introduce the ion-atom interaction and collisions that play a role in ion-atom mixtures. Then in Sec.~\ref{setup}, we focus on the experimental techniques for preparation and detection of the ion and ultracold quantum gas, before discussing the ion-atom mixtures available today and the challenges that come with combining the two. This sets the stage to discuss the latest experiments with hybrid ion-atom systems that have focus especially on buffer gas cooling, chemistry, quantum effects and ion transport in Sec.~\ref{results}. We end the review with a short outlook on what is next.   %in 
 
\section{Ion-Atom Interactions and Collisions}
\label{Langevin}
The opportunities offered  by hybrid ion-atom systems depend greatly on the understanding and control of the interactions at play between the ion and the gas of atoms. The charge of the ion induces a dipole moment on the atom as the electric field of the ion distorts the distribution of charge in the atom. The amount of distortion depends on the
polarizibility $\alpha$ of the atom. This is the leading order contribution to the ion-atom interaction potential. Next-order contributions come from the interplay between the charge of the ion and the induced electric quadrupole moment of the atom as well as the dispersion interaction. Furthermore, if the ion and atom have nonzero orbital angular momentum, the interaction potential becomes more complex\footnote{See ~\cite{Cote2016uha} or ~\cite{Tomza2019chi} for an in-depth theory review}. 

%ion-atom interaction potential
The long-range behavior of the ion-atom interaction potential is typically described by 
\begin{equation}
V_\text{ia} (r)=-\frac{C_4}{r^4},
\end{equation}
with $r$ the distance between the ion and atom and the coefficient $C_4=\alpha/2$ in atomic units is related to the polarizibility of the atom~\footnote{Classically, if you consider the ion to be a point charge, one can derive this behavior from $V_\mathrm{ia}(r)= -\int_0^{\mathcal{E}}{p}\,d{\mathcal{E}}= -\frac{1}{2} \alpha{\mathcal{E}^2}$. Here $p$ is the induced dipole moment and $\mathcal{E}$ the electric field.}. A repulsive potential is often added to capture the hard core potential for short-range interactions\footnote{For example, a $1/r^{6}$ term can be added as in~\cite{Fuerst2018por}. However, this is only one model an in accordance with the Lennard-Jones version of molecular potentials a $1/r^{8}$ repulsive term could also be used.}. This ion-atom interaction potential is of intermediate range as compared to the van-der-Waals potential ($\propto 1/r^6$), the dipole-dipole ($\propto 1/r^3$) or the Coulomb potential ($\propto 1/r$). The real interaction potential is more complex and contains a multitude of effects. It requires both theory and experiment to be fully characterized. In particular, when the atom and ion are so close together that their electronic clouds overlap, the interactions should be obtained through quantum chemistry calculations which take into account all contributions from electronic correlation and exchange. These calculations provide the potential energy curves of the complex structure of ion-atom molecular states, which determine the reaction pathways and collision rates of the processes found in ion-atom systems. Experimental input is needed to benchmark these calculations and improve the accuracy of interaction potentials, cross sections and rate constants of all the collisions and reactions that take place.

%\subsection{Types of collisions}
%Collisions, which types are there, what they do and collision energy.
The ion-atom collisions in the system can be elastic, inelastic or reactive. This leads to changes of the initial colliding partners in momentum, quantum state or chemical species. In elastic collisions the kinetic energy is conserved and the momentum of the particles changes, while the internal state of the colliding particles does not change. In other words, the translational degrees of freedom are decoupled from the internal degrees of freedom of the colliding particles and the latter is kept unchanged. However, because of the momenta changes, elastic collisions can lead to heating of the atom cloud and atom loss from the trap as well as cooling or heating of the ion. Away from the quantum regime, the elastic cross section was found to follow
%\begin{equation}
$\sigma_\mathrm{el}(E_\mathrm{col})=\pi \left(\frac{4\mu C_4^2}{\hbar^2}\right)^{1/3}\left(1+\frac{\pi^2}{16}\right){\left(E_\mathrm{col}\right)^{-1/3}}, $
%\end{equation}
using a semi-classical approach describing many partial waves~\citep{Cote2000uai}.  Here, $\mu=\frac{m_\text{i} m_\text{a}}{m_\text{i}+m_\text{a}}$ is the reduced mass and $E_\mathrm{col}$ the collision energy. In the relative frame of two particles colliding, the collision energy corresponds to the relative kinetic energy between an ion and an atom~\citep{Gerlich2008tso}. Thus, the mean collision energy is given by 
 \begin{equation}
 E_\text{col}=\frac{\mu}{m_\text{i}}E_\text{i}+\frac{\mu}{m_\text{a}}E_\text{a},
 \label{Ecol}
 \end{equation}
  with $E_\text{i}$ the ion energy and $E_\text{a}$ the atom energy\footnote{For a single collision $E_\text{col}=\frac{1}{2} \mu \left(v_\text{i}^2+v_\text{a}^2-2v_\text{i}v_\text{a} \text{cos}\theta \right)$, with $\theta$ the angle between the colliding ion of velocity $v_\text{i}$ and atom of velocity $v_\text{a}$. By averaging this leads to $\langle E_\text{col}\rangle =\frac{1}{2} \mu\left( \langle v_\text{i}\rangle^2 + \langle v_\text{a}\rangle^2\right)$ and thus Equation~\ref{Ecol}}. The latter is related to the temperature $T_\text{a}$ of the atom cloud $E_\text{a}=\frac{3}{2}k_\text{B} T_\text{a}$, with the Boltzmann constant $k_\text{B}$.

Inelastic and reactive collisions come with a change in kinetic and/or internal energy of the particles. The internal and translational degrees of freedom are coupled and the internal states of the colliding particles can change.  An exothermic reaction results in a release of energy when the reaction takes place and an endothermic reaction requires energy to happen. The total energy during a reaction is conserved, so the difference in internal energy between the reactants and product states will be translated into kinetic energy. Collisions can thus take away or add kinetic energy or internal energy to the systems and lead to cooling or heating. For instance an ion relaxes from an excited hyperfine state back to the lower manifold, thereby releasing the hyperfine energy difference. Inelastic collisions lead to a change of the quantum state. For instance, spin changing collisions can lead to both spin conserving and spin non-conserving changes in the spin states of the initial colliding partners. The first is commonly called spin exchange and the second spin relaxation, respectively. Reactive collisions are chemical reactions, where the initial reactants are transformed into different final products. A prime example is resonant charge exchange where the electron is transferred and the ion becomes an atom and vice versa via A+B$^+ \rightarrow$ A$^+$+B. Another example is molecular ion formation whereby A+B$^+ \rightarrow$ AB$^+$, which can happen radiatively or via three-body recombination where the second atom carries away the released energy, i.e. A+ A+ B$^+ \rightarrow$ AB$^+$+ A. Both inelastic and reactive collisions can be radiative or nonradiative. They can happen by absorbing or emitting a photon or through adiabatic crossings and resonances. Especially radiative association and dissociation ($A+B^+\leftrightarrow AB^+ +\gamma$) whereby a molecular ion is formed or the molecular ion breaks into an atom and ion, are aided by light ($\gamma$) present for cooling and trapping of the atomic and ionic species. 

Another common type of ion-atom collisions are quenching collisions, where the internal states of the ion relax to energetically lower lying states. For collisions that involve spin states, this so-called spin relaxation depends on the ion-atom system and the coupling mechanisms between the atom and ion states. They were measured to be higher for Yb$^+$-Rb~\citep{Ratschbacher2013doa} than for Yb$^+$-Li~\citep{Fuerst2018doa} and were found to be even smaller for Sr$^+$-Rb~\citep{Sikorsky2018qca, Sikorsky2018plb} . This could be explained by the difference in second-order spin-orbit coupling~\citep{Tscherbul2016soi}, which is higher for Yb$^+$-Rb than Yb$^+$-Li and Sr$^+$-Rb. These measurements rely on the initial state preparation of the ion in a particular spin state as well as on the creation and preparation of spin-polarized atom clouds in various atomic spin states.

%Langevin vs glancing. 
Ion-atom collisions can be sorted into Langevin or glancing collisions based on how close the ion and atom encounter each other. A Langevin collision happens for a close encounter and comes with a large transfer of energy and/or momentum. Typically, charge exchange and molecular ion formation processes occur via a Langevin collision. A glancing collision is a far away encounter and only leads to a slight deflection of the particles' trajectories. However, it can still facilitate processes like resonant charge exchange that leads to cooling in homonuclear mixtures~\citep{Ravi2012cas, Mahdian2021doo}. 

Most inelastic and reactive collisions in ion-atom mixtures, happen through a close encounter and can be described by the Langevin model. This model assumes that given the impact parameter $b$ between the ion and the atom, a Langevin collision inevitably occurs when $b<b_c$ and a glancing collision happens when $b>b_\text{c}$. The critical impact parameter $b_\text{c}$ depends on the collision energy and the atom polarizibility as $b_\text{c}=\left(4C_4/E_\text{col}\right)^{1/4}$.

% Langeving Model 
 The Langevin capture model predicts a Langevin rate which is independent of the collision energy. The Langevin rate in units of m$^{-3}$s$^{-1}$ is given by
\begin{equation}
    \Gamma_{L}=\sigma_{L} v =2\pi \sqrt{\frac{2C_4}{\mu}}
\end{equation}
whereby the cross section $ \sigma_{L}=2\pi \sqrt{\frac{C_4}{E_\text{col}}} $.~\footnote{The Langevin rate can also be given in units of s$^{-1}$, which means that it has been multiplied by the atom density.} Here, $v$ represents the relative velocity of the ion and atom. At large distances and for the kinetic energy dominating the interaction energy of the system, $E_\text{col}=\frac{1}{2}\mu v^2$. Filling this in, leads to the right side of the equation and the collision energy drops out. The total number of Langevin collisions that can take place between the ion and atom cloud depends on the density of the atoms $n_a$, the Langevin rate and the interaction time $\tau$ of how long the ion is immersed in the bath. Therefore the probability of a collision to happen is given by $ P=1-e^{-\Gamma_{L} n_\text{a} \tau}$ with $n_\text{a}$ the atomic density.

%S-wave regime and partial waves
So far the collisions were described classically, however in the quantum regime, the classical capture picture dissolves and the reaction pathways occur via tunneling, reflection, and resonances. The quantum regime can be reached by cooling down both the ion and atoms degrees of freedom (see Fig.~\ref{Overview}). 
Classically, the ion-atom interaction is described by many partial waves as the total scattering wave function can be decomposed into contributions of different partial waves, i.e. different angular momenta of the rotational motion. However, in the quantum regime all partial waves except the $s$-wave are frozen out~\footnote{This is in analogy with ultracold gases, which are routinely cooled to the $s$-wave regime, see Sec.~\ref{setupatom}.}. The $s$-wave regime is reached when the ion-atom collision energy of the system is smaller than the lowest centrifugal barrier. All partial waves, except for the lowest $s$-wave, have a centrifugal barrier in their potential. The range and energy of the centrifugal barrier for a partial wave with angular momentum $l$ are given by
$
R_l=\sqrt{\frac{2}{l(l+1)}}\sqrt{\frac{2\mu C_4}{\hbar^2}}
$
and 
$
E_l=\frac{l^2(l+1)^2}{4}\frac{\hbar^4}{4\mu^2C_4}. 
$
Here, $R_\text{l=1}$ defines the characteristic range $R^*$ which e.g. for the Yb$^+$-Li and Ba$^+$-Li is about 70\, and 69\, nm, respectively. This is an order of magnitude larger than the typical range of the van der Waals interactions for neutral atoms, e.g. for Rb and Li $R_\text{vdW}= 4.4$ and 1.7\,nm ~\citep{Chin2010fri}. Similary, $E_\text{l=1}$ represents the  $p$-wave centrifugal barrier and is called the characteristic energy $E^*$. For collision energies below this energy all partial waves freeze out and only the isotropic $s$-wave contribution to the scattering survives. It is therefore also called the $s$-wave limit, i.e. $E_{s}=\frac{\hbar^4}{4\mu^2C_4}$. Below this limit, the ultracold collisions are entirely determined by single partial-wave scattering in the incoming collision channel and a quantum treatment of the interactions is necessary. Here, one expects to see quantum effects, such as deviation from the Langevin rate and Feshbach resonances (Sec.~\ref{QE}). The $s$-wave energy depends on both the reduced mass as well as the polarizibility of the atoms and depends greatly on the choice of ion-atom system as we will discuss in Sec.~\ref{sec:AImixtures}

\section{Setup and Techniques}
\label{setup}
Creating hybrid ion-atom experiments requires the combination of a trapped ion setup with that of a cold quantum gas machine. Both are ultra-high vacuum setups, however the ion requires electric feedthroughs for the operation of the electrodes, whereas the atoms require a lot of optical access to fit in all the beams required for magneto-optical trapping, optical trapping and absorption imaging. Both atoms and ions can be cooled by laser light, however their frequencies differ as do their detection requirements. Ions are detected via fluorescence imaging and require an efficient collection of the light to measure a single ion. In the ion-atom mixture, the ion represents a single well-controlled reaction center and most of the measurements rely on the the exquisite control and read-out for which trapped ions are known. From the atom cloud, the relevant parameters are typically the temperature, atom number and density probed by the ion. The number of atoms in the cloud is determined by absorption imaging where a shadow image of the cloud is made. This requires a good imaging system, as also the width of the cloud is needed to obtain the density and temperature of the atoms. Moreover, this is a destructive imaging method and afterwards a new atom cloud needs to be loaded. Thus, ultracold atom setups rely on the reproducibility of the atom cloud, while the same trapped ion can be probed numerous times. When designing these hybrid setups an intricate balance between the various wishes needs to be found. 

\begin{figure}[!t]
\center
\includegraphics[width=\textwidth]{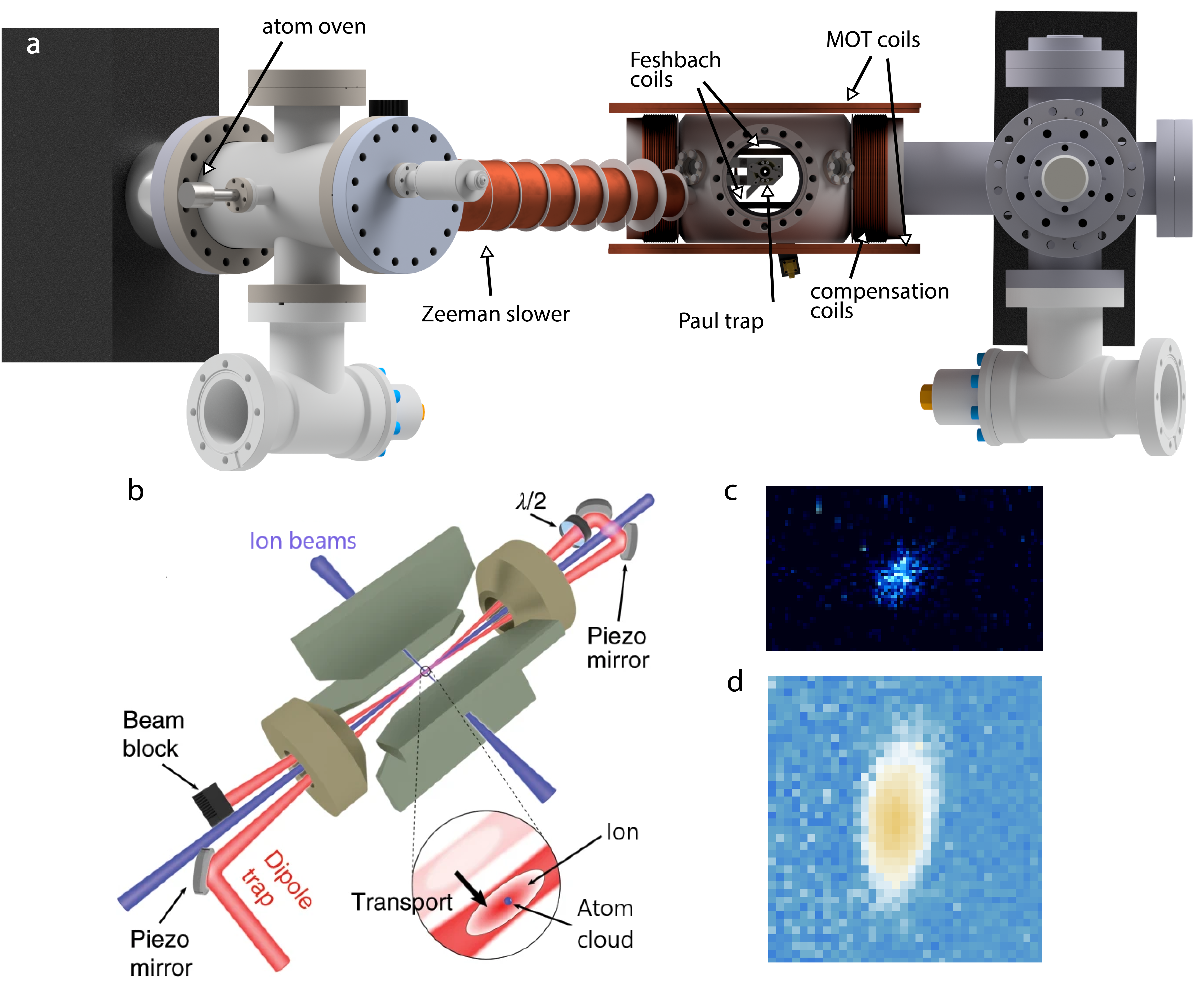}
\caption{Schematics of the setup of an ion-atom experiment. a) Overview of the entire ultra-high vacuum setup. Shown are the atom-part consisting of oven and Zeeman slower as well as the central part where the Paul trap is located. The main chamber is surrounded by various magnetic field coils, which can be used to tune the atom-atom interactions. b) The heart of the setup depicting the key elements of a hybrid ion-atom experiment: laser beams for atom trapping, laser beams for ion cooling and detection and a radiofrequency Paul trap for ion trapping. Not shown are the camera to take fluorescence images of the ion and the camera and laser beam for taking absorption images of the atoms, which are shown in d and c, respectively. Adapted from~\cite{Hirzler2020esf} and ~\cite{Feldker2020bgc}} %\rsl{pic needs to be made nicely}
\label{Setup}
%\vspace{-5cm}
\end{figure}

In this section we discuss the ion and atom preparation and detection as well as the experimental mixtures available and the challenges of combining ions and atoms. As an example of the ion-atom setup to keep in mind, Fig.~\ref{Setup}a gives an overview of the design of the Yb$^+$-Li setup. Here the atom-part on the left consists of an oven followed by a Zeeman slower, before the atomic beam enters the core of the setup. The ion-oven is located close to the Paul trap. The heart of the setup is the central chamber with various optical viewports for good optical access. Fig.~\ref{Setup}b illustrates two key elements of a hybrid ion-atom experiment. Lasers are used for both ion and atom cooling as well as trapping, while electric fields aid in controlling the ion. Here, a radio frequency Paul trap for ion trapping is shown, which is the most commonly used trap. Other elements are magnetic coils, which can be used for atom trapping and to tune the atom-atom interactions via a Feshbach resonance (FR) and, finally, cameras to take fluorescence images of the ion (Fig~\ref{Setup}c) and absorption images of the atoms (Fig~\ref{Setup}d).

\subsection{Ion preparation and detection}
\label{setupion}

Ions used in hybrid ion-atom experiments usually are singly charged earth-alkaline or alkali ions. The former has the electron structure of neutral alkali atoms, although the heavier ions distinguish themselves by featuring low lying metastable $^2D$ levels. The low energy levels of Yb$^+$ also are alkali-like but feature in addition two metastable $^2D$ levels and the extremely long lived $^2F_{7/2}$ state~\citep{Lange:2021}. On the one hand, the metastable states complicate laser cooling since more repump lasers are required than for the neutral alkalis, but on the other hand the availability of narrow line width quadrupole transitions, $^2S\rightarrow {^2D}$, facilitate sensitive detection methods as described below. 

The ions are typically created by photoionizing neutral atoms emerging from an oven in an ultrahigh vacuum environment. Usually, two-photon ionization is employed. In this method, the isotope shifts in the ion species can be easily resolved such that the process is isotope selective. Since ion traps are usually extremely deep, precooling of the atoms is not needed. Doppler laser cooling after ionization localizes the ions in the center of the trap at temperatures of $\sim$~mK. Fluorescence during laser cooling can be collected on a camera or photon detector to infer the presence of the ions. In typical traps, the ions form crystaline structures due to their Coulomb interactions with ion separations in the 2-20~$\mu$m range, which are easily resolved with a modest microscope. The number of ions can be readily controlled by inspection and turning off the photo-ionization laser. When too many ions were loaded by accident, temporarily lowering the trap allows for a surprisingly easy method to remove ions one by one with relatively large success rate. 

Further laser manipulation allows to cool the ions to near their ground state of motion using e.g. resolved sideband cooling. Moreover, the electronic, hyperfine and Zeeman state of the ions can be prepared and read out with laser pulses as well. The closed-shell alkali ions do not feature practical optical transitions and are usually created by ionizing an ultracold alkali atom and are detected on multi channel plates. 
For further reading, we refer the reader to the various detailed reviews on trapped ions showcasing the toolbox at hand especially for their use as quantum simulators and qubits for quantum computation, e.g.~\citep{Leibfried:2003, Monroe2021pqs, Bruzewicz2019tiq}. 

\subsubsection{Paul trap}
\label{paultrap}

The most common method of trapping ions is by using a Paul trap~\citep{Paul:1990}. In it, the ions are confined by a combination of static and radio frequency electric fields, $\bm{\mathcal{E}}_\text{s}$ and $\bm{\mathcal{E}}_\text{rf}(t)$. The total electric field is given by $\bm{\mathcal{E}}(t)=\bm{\mathcal{E}}_\text{s}+\bm{\mathcal{E}}_\text{rf}(t)$. In an ideal Paul trap, both electric fields disappear in the trap center and we can expand the fields to first order in the ion coordinate $\mathbf{r}$ around the origin: $\bm{\mathcal{E}}(t)\approx (\mathbf{G}_\text{s} +\mathbf{G}_\text{rf}(t))\cdot \mathbf{r}$. Here, $\mathbf{G}_\text{s}$ and $\mathbf{G}_\text{rf}(t)$ are the 3$\times$3 matrices representing the gradient of the vector fields $\bm{\mathcal{E}}_\text{s}$ and $\bm{\mathcal{E}}_\text{rf}$ with elements
\begin{align}
G_{\text{s},i,j}=\frac{d\mathcal{E}_{\text{s},i}}{dr_j}\\
G_{\text{rf},i,j}(t)=\frac{d\mathcal{E}_{\text{rf},i}(t)}{dr_j}
\end{align}
and $i,j\in \{x,y,z\}$. The Maxwell equation $\nabla \cdot \bm{\mathcal{E}}_\text{s}=0 $ imposes the constraint $\text{Tr}\left(\mathbf{G}_\text{s}\right)=0$, suggesting that charged particles cannot be trapped by static fields alone since the retaining force $\mathbf{F}=e\bm{\mathcal{E}}_\text{s}$ will have to be of the wrong sign in at least one direction. Here $e$ is the elementary charge. The result is know as Earnshaw's theorem~\citep{Earnshaw:1842}, for a discussion and a full proof see~\cite{Weinstock:1976}. As we will see later in Section~\ref{sec:AImixtures}, the time-dependent field $\bm{\mathcal{E}}_\text{rf}(t)$ can have serious consequences when combining ions with ultracold atoms. However, Earnshaw's theorem makes it impossible to simply switch it off, unless we consider other means of trapping ions such as optical trapping or Penning traps.

The radio frequency electric field generally has a simple harmonic time dependence $G_{\text{rf},i,j}(t)=G_{\text{rf},i,j}\cos(\Omega_\text{rf} t)$ and we call $\Omega_\text{rf}$ the trap drive frequency, with the subscript rf denoting radio frequency. In experiments we typically have $\Omega_\text{rf}/(2\pi)$ between 100~kHz and 50~MHz. It is convenient to introduce unitless parameters $a_{ij}$ and $q_{ij}$ that relate the electric field gradients to $\Omega_\text{rf}$ such that $G_{\text{s},i,j}=-(m_\text{i}\Omega_\text{rf}^2/e)a_{ij}$ and $G_{\text{rf},i,j}=-(m\Omega^2/(2e))q_{ij}$, with $m_\text{i}$ the mass of the ion.
%\footnote{For simplicity in this section we remove subscript of the mass of the ion, i.e. $m=m_\text{i}$}. 
The parameters $a_{ij}$ and $q_{ij}$ are known as the static and dynamic stability parameters, respectively, because they determine whether an ion can be trapped stably. In case all $a_{ij}=0$, we have that the eigenvalues of $\mathbf{q}$ should all be $<0.908$~\citep{Ghosh:1995} for the ion to be trapped. The more general case requires solving the coupled Mathieu equations $m_\text{i}\mathbf{\ddot{r}(t)}=e(\mathbf{G}_\text{s} +\mathbf{G}_\text{rf}(t))\cdot \mathbf{r}$ with $e$ the (unit) charge of the ion.

The experimentally most relevant case is the so-called linear Paul trap. Here, the radio frequency field supplies confinement in the two radial directions, which we set as the $x$- and $y$-direction here, while the static field supplies confinement in the $z$-direction. We require $q_x=-q_y=q$ and $a_z=-a_x-a_y>0$ due to the Maxwell equations and to assure trapping in the $z$-direction. We only consider positively charged ions here and have assumed for simplicity that $\mathbf{G}_\text{s}$ and $\mathbf{G}_\text{rf}$ can be simultaneously diagonalized such that the equations of motion decouple in each of the independent directions of motion. For the linear Paul trap, the equations of motion take the form of Mathieu equations and can be solved analytically using a Floquet Ansatz~\citep{Ghosh:1995,Leibfried:2003}. To lowest order in the stability parameters, the solutions can be approximated as:
\begin{equation}\label{rt_lowest_order}
r_j(t)\approx A_j\cos(\omega_j t+\phi_j)\left(1+\frac{q_j}{2}\cos(\Omega_\text{rf} t)\right).
\end{equation}
\noindent This equation describes harmonic motion with amplitude $A_j$ and phase $\phi_j$ on top of a fast \textit{intrinsic micromotion} (IMM) of amplitude $A_jq_j/2\ll A_j$ and frequency $\Omega_\text{rf}$. The secular trap frequencies are given, to lowest order, by ${\omega_j\approx (\Omega_\text{rf}/2) \sqrt{a_j+q_j^2/2}\ll \Omega_\text{rf}}$. 

Since the micromotion has a small amplitude as compared to the secular motion, it is often neglected in a process called the secular approximation. However, whether this approximation is justified depends a lot on the situation at hand. It is known that it can fail dramatically when the ion is interacting with neutral particles~\citep{Major:1968,Itano:1995,Gerlich:1995,DeVoe2009pld,Zipkes2011koa,Cetina2012flt,Chen2014ngs,Holtkemeier2016bgc,Meir2016doa,Rouse2017sed,Fuerst2018por,Rouse2019ted}. 

For the ideal linear Paul trap, $q_z=0$ such that the ion motion in this direction does not have micromotion. In practical Paul traps however, $|q_z|>0$. This has an effect that needs consideration when minimizing so-called \textit{excess micromotion}. Excess micromotion (EMM) is the additional micromotion in the trap that stems from imperfections in the electric field as real setups deviate from the ideal linear Paul trap scenario. It can be well-characterized and compensated to a certain degree as described in the next section. 

For non-negligible $q$, it is worthwhile to consider the next higher order in $q$ to accurately describe the ion trajectories~\citep{Meir:2016PhD}:
\begin{align}
\begin{split}
r_j(t)\approx & A_j\left(\cos(\omega_j t+\phi_j)\left(1+\frac{q_j}{2}\cos(\Omega_\text{rf} t)+\frac{q^2_j}{32}\cos(2\Omega_\text{rf} t)\right)\right)\\
&+A_j\left(\frac{q_j\omega_j}{\Omega_\text{rf}}\sin(\omega_jt+\phi_j)\sin(\Omega_\text{rf} t)\right)
\end{split}
\end{align}
and
\begin{equation}
\omega_j\approx \frac{\Omega_\text{rf}}{2}\sqrt{\frac{a_j+q_j^2\left(\frac{1}{2}+\frac{a_j}{8} \right)+\frac{q_j^4}{128}}{1-q_j^2\left(\frac{3}{8}+\frac{5a_j}{16}\right)}}
\end{equation}
In practical experiments, usually $q<0.5$ because higher values may result in parameteric excitation and heating of the ions. This effect results from anharmonic terms in the trapping fields that we neglected in the first order expansion of the trapping field~\citep{Alheit1995OIP,Pedregosa2010ACR,DengDBS2015}.

Finally, let us consider the amount of energy that is stored in the micromotion. Using Eq.~\ref{rt_lowest_order}, we can obtain the average kinetic energy $m_\text{i}\langle v_j^2 \rangle/2\approx m_\text{i}\omega_j^2A_j^2/2$ for $j=x,y$ while $m_\text{i}\langle v_z^2 \rangle/2\approx m\omega_z^2A_z^2/4$ in accordance with a normal harmonic oscillator. Here, we assumed $a\ll q^2$ such that $\omega_{x,y}\rightarrow \Omega q_{x,y}/(2\sqrt{2})$ and $v_j$ is the velocity in direction $j$. Therefore, the two directions with micromotion add an additional $\sim k_\text{B}T/2$ to the total kinetic energy each, a result that was obtained in~\cite{Berkeland:1998}. 
In summary, the total kinetic energy of the ion in a linear Paul trap is given by $E_\text{kin}^j=E^j_\text{sec}+E^j_\text{IMM}+E_\text{EMM}^j$, for each direction. It consists of the secular energy and the energy due to the intrinsic and excess micromotion. Thus, micromotion is of importance when dealing with collisions between ions and neutral atoms. The relative contribution of micromotion along the $z$-axis is of order $\sim q_z^2/(8a_z)\ll 1$ as long as $a_z\gg q_z^2$. Lastly, note that intrinsic micromotion occurs mostly at frequencies $\Omega_\text{rf} \pm \omega_j$ as is clear from Eq.~\ref{rt_lowest_order}. This distinguishes intrinsic micromotion from excess micromotion that occurs at (multiples of) $\Omega_\text{rf}$ as described below.

\subsubsection{Excess micromotion}
\label{micromotion}
Up until now, we have considered ideal Paul traps, but there are several experimental imperfections that need to be taken into account before we can look at realistic ion trapping. Here, the most important of these is without doubt excess micromotion. Intrinsic micromotion in an ideal Paul trap adds kinetic energy on the order of $k_\text{B}T$ to the ion that needs to be taken into account when it interacts with another particle~\citep{Berkeland:1998}. Excess micromotion can further increase this energy, independently of $T$ and poses a serious restriction in studying collisions and chemistry between ultracold atoms and ions in Paul traps. There are three types of excess micromotion commonly named radial, axial and quadrature (or phase) excess micromotion. All originate from deviations in the ideal electric field of the Paul trap,  are independent of the secular motion amplitudes $A_j$ and occur at (multiples of) the trap drive frequency $\Omega$. 

\paragraph{Radial micromotion}
Here, the position at which each field $\bm{\mathcal{E}}_\text{s}$ and $\bm{\mathcal{E}}_\text{rf}$ vanish do not overlap. Since the latter is the more important for radial confinement, we quantify the electric offset field as the static field $\bm{\mathcal{E}}_\text{offs}=\bm{\mathcal{E}}_\text{s}(\mathbf{r}=0)$ and $\bm{\mathcal{E}}_\text{rf}(\mathbf{r}=0)=0$. The offset field has the effect of pushing the ion slightly away from the radio frequency trap center, by an amount $d_j\sim e \mathcal{E}_{\text{offs},j}/(m\omega_j^2)$. Here, the ion undergoes excess micromotion of amplitude $\sim q_j d_j/2$, corresponding to an average kinetic energy of
\begin{equation}
\label{Eoffset}
E_\text{kin}^{\text{rad}}\sim m_\text{i}\Omega_\text{rf}^2 q_j^2 d_j^2/16\sim 4 e^2\mathcal{E}_{\text{offs},j}^2/(m_\text{i}\Omega_\text{rf}^2q_j^2)
\end{equation} 
to lowest order, where we used $q_{x,y}^2\gg |a_{x,y}|$. Taking e.g. a $^{40}$Ca$^+$ with $\Omega_\text{rf}=2\pi\times$~20\,MHz, $q=0.2$ and $\mathrm{E}_{\text{offs},j}=1$\,V/m, we obtain $E_\text{kin}^{\text{rad}}\sim k_\text{B}\times$~4\,mK, an order of magnitude higher than the Doppler limit for laser cooling\footnote{The Doppler limit gives the lower limit to the temperature of the ion after laser cooling on the optical transition with linewidth $\gamma$ and is given by $T_D=\hbar\gamma/\left(2k_B\right)$. See Sec.~\ref{lasercooling}.}. By getting a better control on the electric field or compensating for the offset field down to $<$~0.1\,V/m, excess micromotion energies in the $\mu$K regime are allowed.

Reducing radial excess micromotion can be done by tighter ion confinement since $E_\text{kin}^{\text{rad}}\propto 1/\omega_j^2$. However,the other types of micromotion do not share this property and care needs to be taken when minimizing the total excess micromotion. Moreover, the effect of micromotion on ion-atom collisions does also not prefer a tight confinement (Sec.~\ref{sec:AImixtures}). 

\paragraph{Axial micromotion}
Another type of excess micromotion occurs when there is no point where the axial rf field disappears, but rather there is a homogeneous rf field in the axial direction. Even with $q=q_x=-q_y$, this field is permitted by the Maxwell equations. This type of micromotion is best characterized by the strength of the oscillating field $\mathcal{E}_\text{EMM}^{ax}\propto q$. Axial micromotion has an amplitude of $\sim e \mathcal{E}_\text{EMM}^{ax}/(m_\text{i}(\Omega_\text{rf}^2-\omega_z^2))$ and an average kinetic energy of $E_\text{kin}^{\text{ax}}\sim e^2 \left(\mathcal{E}_\text{EMM}^{ax}\right)^2\Omega_\text{rf}^2/(4m_\text{i}(\Omega_\text{rf}^2-\omega_z^2)^2)$. Considering again $^{40}$Ca$^+$ with $\Omega_\text{rf}=2\pi\times$ 20\,MHz, $\omega_z=2\pi\times$ 1\,MHz and $\mathcal{E}_\text{EMM}^{ax}=1$\,V/m, we get $E_\text{kin}^{\text{ax}}\sim k_\text{B}\times$~200\,nK. Axial micromotion can occur for instance due to a small angle between the rf electrodes.The electric field that causes the axial excess micromotion is proportional to the trapping electric fields. Thus, reducing the effect can be done by using a low $\Omega_\text{rf}$ and $q$, which unfortunately leads to a larger effect of radial excess micromotion.

\paragraph{Quadrature micromotion}
Phase- or quadrature micromotion~\citep{Berkeland:1998,Meir2017eaf} originates from a phase
difference $\delta\phi_{\rm{rf}}$ between the radiofrequency voltages on different rf-electrodes. This can be seen as a time dependence in the location of the rf null. It causes an out-of-phase micromotion of amplitude $\sim q_jR_{\rm{trap}}\delta\phi_{\rm{rf}}$, where $R_{\rm{trap}}$ is half the distance between the two rf-electrodes.  This results in an average kinetic energy of 
$
  E_{\rm{phmm}} = \frac{1}{64}m_\text{i}\left(q_j R_{\rm{trap}}\delta\phi_{\rm{rf}}\Omega_\text{rf}\right)^2\,.
$
Quadrature micromotion originates either from impedance or mechanical mismatches between the electrodes. For instance a simple difference in the length of the wires supplying voltage to the electrodes results in a phase difference. Like with axial micromotion, the out-of-phase field is proportional to the trapping field and reducing its effect can thus be done by using a low $\Omega_\text{rf}$ and $q$. Setting a low $\Omega_\text{rf}$ makes it easier to reduce $\delta\phi_{\rm{rf}}$ because the wavelength of the rf increases. The scaling with $\propto q_jR_{\rm{trap}}\delta\phi_{\rm{rf}}$ is very unfavorable. To have a phase micromotion energy below the Doppler limit, $\delta\phi_{\rm{rf}}\ll 0.01^{\circ}$ for the typical $R_{\rm{trap}}$ in the mm range and taking again $^{40}$Ca$^+$ with $\Omega_\text{rf}=2\pi\times$~20\,MHz, $q=0.2$, this corresponds to path length differences much smaller than ~0.5\,mm. At the same time, phase micromotion cannot be inferred from indirect measurements, as is often done with radial micromotion and will be discussed next. Because of all of this, phase micromotion is often the dominant form of excess micromotion.

\paragraph{Prevention}
Excess micromotion poses challenges to a wide variety of applications of trapped ions including for atomic clocks and in quantum computing~\citep{Berkeland:1998,Keller:2015} and a large amount of work has been done in either preventing or undoing its effects. In this section we describe a few known measures to prevent excess micromotion.

Radial excess micromotion occurs because of uncontrolled static electric fields in the experiments. The laser used for creating, cooling and detecting ions typically operate in the UV and can easily free electrons from any metal that the beams hit~\citep{Harlander:2010}. These electrons are accelerated in the electric field of the Paul trap and may accumulate on areas that are non-conducting, causing electric offset fields. Therefore, non-conducting elements must be `hidden' from the ion's line of sight as much as possible. In the case of hybrid ion-atom experiments however, there is an additional danger. The atoms used in the experiment may land on the ion trap electrodes and can, after some time, oxidize to form a non-conductive layer on which electrons may stick and cause electric offset fields. In the group in Amsterdam, we periodically clean the trapping electrodes using a high power IR laser, to evaporate any material sticking on it, eliminating the electric offset fields~\citep{Hirzler2020esf} for a few weeks. Large Paul traps help mitigate these problems, such that the atoms are kept away from the electrodes as much as possible.

Phase and axial micromotion can be prevented by accurate mechanical and electrical construction of the Paul trap. This means that asymmetries, angles and differences in wire lengths need to be avoided. Adequate filtering can prevent radio frequency pickup in the electrodes. Numerical simulations of the trapping fields are a valuable tool for checking the resilience of the trap to realistic asymmetries and angles between the electrodes. Furthermore, optimization of the trap parameters ($q$ and $\Omega$) can be helpful as the various types of excess micromotion depend differently on them as we previously showed. However, other demands also play a role when designing ion traps, such as optical access or the need to reach the so-called Lamb-Dicke regime for ion-laser interactions (See Sec.~\ref{Iondetection}).

\paragraph{Detection and compensation}
Radial offset fields that can cause excess micromotion are easily detected and compensated by applying additional electric fields. These fields typically originate from strategically placed dedicated compensation electrodes. While radial excess micromotion can be compensated with static electric fields, compensating axial and phase micromotion requires the application of time-dependent compensation fields with full phase control. This is typically quite challenging and so prevention is the preferred approach. Here, we describe the various techniques available to detect radial offset fields and micromotion in general.

For detecting radial offset fields, the fact that the ion position in the trap under the influence of an offset field depends on the trapfrequency $\omega_j$ according to  $d_j\sim e \mathcal{E}_{\text{offs},j}/(m_\text{i}\omega_j^2)$, is often used. The trap frequency may be controlled by changing the rf voltage while the distance that a fluorescing ion moves as a function of it may be tracked with a camera. Crucially, both $\omega_j$ and the magnification of the camera (and thereby $d_j$) can be accurately determined by independent measurement. The former is measured by applying a small oscillating electric field and observing the fluorescence of the ion, upon hitting the resonance at $\omega_j$ the fluorescence abruptly changes due to the Doppler effect as the ion motion gets excited. The magnification can be determined by loading two ions in the trap at a known axial trapfrequency $\omega_z$ and noticing that the distance between them is given by~\citep{James:1998}:
\begin{equation}
d_\text{2-ions}=\left( \frac{e^2}{2\pi\epsilon_0m_\text{i}\omega_z^2} \right)^{1/3}.
\end{equation}
The radial excess micromotion may be compensated by applying additional electric fields until the ion does not move upon changing $\omega_j$. 

In these camera-position measurements, a potential systematic error is the radiation pressure of the laser that makes the ion fluoresce. This causes a force  of $\sim \hbar k \gamma_\text{ph}$ with $\gamma_\text{ph}$ the photon scattering rate and $k$ the wavenumber. For $\gamma_\text{ph} \sim \Gamma/2$ with $\Gamma$ the transition linewidth of typically $\Gamma \sim 2\pi\times$~20\,MHz for alkali earth ions and typical wavelengths $\lambda\sim 400$\.nm, the radiation pressure is equivalent to the effect of an offset field of $\sim$~1\,V/m. Its effect can be reduced by aligning the laser along an axis in which the ion does not move or by reducing its intensity to far below saturation.

The technique does not work for ion movement in the direction of the camera. However, tricks are available to map the information of the ion position onto the internal state of the ion. This may for instance be done by applying a magnetic field gradient and detecting the energy shift between states with a magnetic moment as a function of $\omega_j$~\citep{Joger:2017,Hirzler2020esf}. One can also detct the position dependent phase of a laser. These phase differences may in turn be mapped onto the internal states of ions~\citep{Higgins:2021}. These techniques have the benefit of not suffering from radiation pressure.

Another method to detect an offset between the static and rf fields is to mix in a voltage at frequency $\omega_j$ in the rf signal. If the ion is not situated in a position where $\bm{\mathcal{E}}_\text{rf}=0$, this field can parametrically excite the ion, which will be visible in the fluorescence~\citep{Ibaraki:2011,Narayanan:2011,Tanaka:2012,Keller:2015,Nadlinger:2021}. 

The micromotion in the trapped ion can be directly probed by making use of the Doppler effect. The `photon correlation method' works by detecting photons on e.g. a photomultiplier tube. Due to micromotion and the Doppler effect, the ion fluoresence shows a modulation at the trapdrive frequency $\Omega_\text{rf}$ that can be made visible by averaging over many periods and making sure the detection and the trapdrive are exactly in sync. The success of compensation methods can be checked by flattening this fluoresence curve. The method also gives access to the relative phase between the micromotion and the trapdrive and can thus distinguish between phase and axial/radial micromotion~\citep{Berkeland:1998,Keller:2015}.

Another method based on the Doppler effect is to look at the fluorescence in the spectral domain. In the rest frame of the ion, the laser field is frequency modulated with frequency $\Omega_\text{rf}$ and modulation index $\beta = k x_\text{EMM}$ with $x_\text{EMM}$ the amplitude of the micromotion~\citep{Berkeland:1998} and $k$ the wavevector projected on the direction of the micromotion. This causes sidebands at multiples of $\Omega_\text{rf}$ in the spectral response of the ion when probed with near resonant light. The success of compensation methods can be checked by minimizing the strength of these sidebands. However in ion-atom mixtures, the sidebands are usually not resolved as the ions typically have linewidths $\Gamma\sim 2\pi\times$~20~MHz on their main cooling transitions and these mixtures favor low $\Omega_\text{rf}$. One solution is to rather perform the measurement with a weaker repump transition~\citep{Joger:2017,Goham:2021}.

A widely applied variation is to make use of narrow linewidth transitions and directly probe the coupling strength of the laser on a number of the sidebands~\citep{Berkeland:1998}. This method is sometimes called the Rabi method since the coupling strengths are usually derived from measured Rabi flops. The Rabi frequencies on the carrier $\Omega_\mathrm{car}$ and first micromotion sideband $\Omega_\mathrm{sb}$ are related by

\begin{equation}\label{Eq_MMJ}
\frac{\Omega_\text{sb}}{\Omega_\text{{car}}}=\frac{J_\text{1}(\beta_\text{mm})}{J_\text{0}(\beta_\text{mm})},
%\frac{\Omega_\text{sb}}{\Omega_\text{{car}}}\approx \frac{\vec{k}_{411}\cdot\vec{r}_\text{mm}}{2},
\end{equation}

with $J_\text{1}(\beta_\text{mm})$ and $J_\text{0}(\beta_\text{mm})$ Bessel functions of the first kind. The heavy earth-alkaline ions and Yb$^+$ used in ultracold ion-atom experiments have narrow linewidth quadrupole transitions available to the low lying $^2D$ states that can be used for this type of micromotion detection. Potential systematic shift can originate from Stark shifts for weak sideband transitions (i.e. close to perfect compensation) and modulation of magnetic substates due to magnetic fields generated by the rf trapping fields~\citep{Meir2017eaf}.

For ion-atom experiments, the particulars of the experimental setup allow for very precise determination of excess micromotion. This is because the outcome of ion-atom collisions depends very strongly on the amount of excess micromotion as we describe in section~\ref{sec:AImixtures}. Excess micromotion can then be detected either by measuring the loss of atoms in energetic collisions~\citep{Mohammadi:2019} or by measuring the energy of the ion after interacting with atoms~\citep{Hirzler2020esf}. Both methods are among the most accurate available and radial excess micromotion compensation $\bm{\mathcal{E}}_\text{offs}\lesssim 0.02$~V/m have been reported. The work reported in~\cite{Mohammadi:2019} also demonstrates the measurement and compensation of the other forms of micromotion. In \cite{Huber:2014}  $\bm{\mathcal{E}}_\text{offs}\lesssim 0.0087$~V/m was reported by optimizing the transfer of an ion in a Paul trap to an optical dipole trap. Since the dipole trap is very shallow, any stray electric field will lead to ions loss, making the system an extremely sensitive probe for radial excess micromotion.

\subsubsection{Background heating}

Electric field noise originating either from technical noise or room temperature electrodes in the vicinity of the ions may cause background heating rates that are independent of the presence of atoms. Unlike buffer gas cooling (see Sec.~\ref{bgcool}), this heating is nearly independent of the temperature of the ions in the cold regime. Simply equating the cooling and heating rates shows that background heating limits buffergas cooling by  $\Delta T_\text{i}=\Gamma_\text{heat}/\gamma_\text{cool}$ with $\Gamma_\text{heat}$ the background heating rate in K/s and $\gamma_\text{cool}$ the buffer gas cooling rate such that $dT_\text{i}/dt = -\gamma_\text{cool}(T_\text{i}-T_\infty)+\Gamma_\text{heat}$ with $T_\infty$ the equilibrium temperature of the ion in the buffer gas without background heating~\citep{Hirzler2020esf}. Note that this simple analysis can only be expected to hold in the large mass ratio limit since otherwise the ion energy distribution will deviate from a thermal distribution~\citep{DeVoe2009pld,Zipkes2011koa,Chen2014ngs,Meir2016doa,Holtkemeier2016bgc,Rouse2019ted,Feldker2020bgc, Pinkas2020eoi}. For low densities, the cooling force of the ultracold buffer gas is small and the background heating can be a limiting factor with e.g.~\citep{Feldker2020bgc} citing $\Delta T_\text{i}=20-40~\mu$K. In general, background heating can be mitigated by employing large Paul traps, in which the ions are far separated from any surfaces~\citep{Brownnutt:2015}.

\subsubsection{State detection and thermometry}
\label{Iondetection}

Detection of one electron ions such as the earth alkaline ions and Yb$^+$ can be easily done by collecting laser induced fluorescence on a camera or photon detector. With photon scattering rates of several MHz and collection efficiencies of up to $\sim$~1\% for nearby lenses and photon detectors, the presence of an ion can be determined within milliseconds~\citep{Leibfried:2003}. 

Typically, the ions used feature multiple states that can be driven with laser fields to make it possible to read out the internal state of the ion~\citep{Dehmelt:1975}. For instance, one of the states may lead to fluorescence while the other(s) do not~\citep{Leibfried:2003}. The dark and bright states may correspond to hyperfine ground states of an ion and the optical transition used for detection may be closed via selection rules. The ions Ca$^+$, Sr$^+$, Ba$^+$ and Yb$^+$ also feature low lying $^2D_{5/2}$ states that may remain dark during fluorescence detection. This feature can be used for state detection with a bright ion corresponding to an ion measured to be in the $^2S_{1/2}$ ground state and a dark ion corresponding to $^2D_{5/2}$. Additional laser pulses operating on the narrow $^2S_{1/2}\rightarrow ^2D_{5/2}$ quadrupole transition may be used to map other states or information about e.g. ion motion to the dark state. Using this electron shelving method electronic, hyperfine and Zeeman substates may be distinguished with very high accuracy~\citep{Myerson:2008}. 

For closed shell ions such as Rb$^+$, that are typically created by ionizing a buffer gas atom or via charge transfer from a one electron ion, there are no practical optical transitions available. In this case, the ions may be detected via a multi channel plate. Here, the application of extraction fields and time-of-flight analysis gives access to the energy and position of the ions~\citep{Dieterle2021toa}. Another method to detect closed shell ions is by detecting the back-action that they have on an atomic gas, for instance in the form of atom loss due to energetic collisions~\citep{Schmid2010doa}.

\paragraph{Co-trapping of an ion}
A very common method in ion trapping is to co-trap ions, or even molecular ions with ions that allow for straightforward fluorescence detection. Sympathetic laser cooling allows the formation of an ion crystal and the impurity ions show up as dark vacancies in a crystal. Moreover, the trap frequencies of the ion crystal depend on the masses and composition of the crystal. In a linear Paul trap, $\omega_z\propto 1/\sqrt{m_\text{i}}$, while in the radial directions $\omega_\perp\propto 1/m_\text{i}$. This effect can be used to accurately determine the atomic mass of the impurity ions with resolution on the order of or even below an atomic mass unit~\citep{Morigi:2001,Schmid2010doa,Home:2013}. 

The motional modes of ion crystals containing ions with different masses have been described in Refs.~\citep{Morigi:2001,Home:2013}. In a crystal of two ions with masses $m_1$ and $m_2$ aligned along the radio frequency null of a linear Paul trap, the distance between them is independent of their masses. In contrast, the axial eigenmode frequencies are related via~\citep{Sinhal:2020}
\begin{equation}
   \left( \omega_z^\pm\right)^2 =\omega_{z,m_1}^2 (1+\nu \pm \sqrt{1+\nu^2-\nu}),  
\end{equation}
\noindent with $\omega_{z,m_1}$ the trapfrequency of a single ion of mass $m_1$ in the trap and $\nu=m_1/m_2$. In the case that $m_1=m_2$, this leads to the well known result that the eigenmode frequencies are given by $\omega_z$ and $\sqrt{3}\omega_z$, for the center-of-mass and relative motion respectively~\citep{James:1998}.

Information about the motional state of the ions can be obtained by making use of the Doppler effect which relates the photon scattering rate to the velocity of the ions. The first of these methods is known colloquially as the Doppler re-cooling method and was first proposed and used in~\citep{Epstein:2007,Wesenberg:2007}. This method works over a wide range of energies and can even be used to probe large ($\gg$~K) amounts of energy released e.g. in a chemical reaction in a single run by timing the onset of ion fluorescence as was shown in~\citep{Meir:2017}. However, methods based on the Doppler effect have as a lower limit the Doppler limit, which typically lies in the the $0.5$~mK range for the strong cooling transitions in alkaline earth ions and Yb$^+$. Reaching the quantum regime for interacting ion-atom mixtures requires lower ion temperatures for all experimentally studied systems. 

\paragraph{Measuring the secular energy}

To analyse the kinetic energy of an ion in the ultracold regime, we can couple information about it to its internal state followed by fluorescence detection. This not only gives access to the average ion energy, but can also be used to measure the energy distribution, the amplitude and phase of its motion and the probability distribution in phase space, see e.g.~\citep{Wallentowitz:1995,Meekhof:1996,Leibfried:1996,Leibfried:1998,Leibfried:2003,Lougovski:2006,Santos:2007,Lamata:2007,Schmitz:2009,Gerritsma:2010,Zaehringer:2010,Fluehmann:2020}. These techniques were developed in the context of trapped ion quantum computing and rely on laser-induced qubit-motion coupling. It is important to realise that since readout relies on the measured value of a single qubit, little information is gained per measurement and typically a lot of averaging is required. Moreover, the creation of an ultracold gas of atoms relies on time-consuming evaporation cooling such that typically a lot of measurement time needs to be reserved for measuring the motion of a single ion. These caveats not withstanding, the tools are very powerful and allow for example for an accurate determination of the total ion energy~\citep{Feldker2020bgc}.

The secular energy can be determined via the `Rabi method' of measuring the Rabi flops on a transition. Consider, the Hamiltonian of a two level system with levels $|0\rangle$ and $|1\rangle$ in a harmonic trap with a laser field of frequency $\omega_L$ and wavenumber $\mathbf{k}$. This can be described as
\begin{equation}
    H(t)=2\hbar\Omega\left(e^{i(\mathbf{k}\cdot \mathbf{r}-\omega_L t)}+e^{-i(\mathbf{k}\cdot \mathbf{r}-\omega_L t)}\right)\left(\hat{\sigma}^++\hat{\sigma}^-\right),
\end{equation}
where we introduced the Rabi frequency $\Omega$ with a factor 2 in anticipation of the rotating wave approximation and $\hat{\sigma}^+$ and $\hat{\sigma}^-$ are the usual raising and lowering operators for the state. 
We can go into the interaction picture with the unitary transformation $\hat{U}=e^{iH_0t/\hbar}$ with $H_0=\hbar\sum_{j=1}^3\omega_j\hat{a}_j^{\dag}\hat{a}_j+\hbar\omega_0|1\rangle\langle 1|$ the Hamiltonian in the absence of the laser and $\hat{a}_j^{\dag}$ and $\hat{a}_j$ the creation and annihilation operator belonging to the motional mode $j$ with frequency $\omega_j$. Furthermore, we quantize $\mathbf{r}$ with $\mathbf{k}\cdot \mathbf{\hat{r}}=\sum_{j=1}^3\eta_j\left(\hat{a}_j^{\dag}+\hat{a}_j \right)$ with the Lamb-Dicke parameters $\eta_j =l_j k_j$ with $l_j=\sqrt{\hbar/(2m\omega_j)}$ and $k_j$ the laser wavevector projected on the direction of motion $j$. In cases of interest, $\eta_j\ll 1$ and we make the Lamb-Dicke approximation $e^{i\,\sum_{j=1}^3\eta_j\left(\hat{a}_j^{\dag}+\hat{a}_j \right)}\approx 1+\sum_{j=1}^3i\eta_j\left(\hat{a}_j^{\dag}+\hat{a}_j \right)+...$. 

All this results in the interaction Hamiltonian
\begin{equation}\label{eq_HIt}
    H_I(t)=\hbar\Omega e^{-i\Delta_L t}\left(1+i\sum_{j=1}^3\eta_j\left(\hat{a}_j^{\dag}e^{i\omega_jt}+\hat{a}_je^{-i\omega_jt} \right) \right)\hat{\sigma}^++\rm{c.c.}.
\end{equation}
Here, we introduced the detuning of the laser, $\Delta_L=\omega_L-\omega_0$. Three options are of interest, commonly called carrier ($\Delta_L=0$), red ($\Delta_L=-\omega_j$) and blue sideband ($\Delta_L=+\omega_j$) transitions. 

When we set $\Delta_L=0$, we simply drive the transition $|0\rangle\leftrightarrow |1\rangle$, while the motional state changing terms $\propto \eta_n$ are not resonant and can be discarded as fast rotating terms. It turns out that when we omit the Lamb-Dicke approximation, there are resonant terms coupling to the ion motion also in the carrier. These originate from terms such as $\propto \eta_j^2\hat{a}_j^{\dag}\hat{a}_j$ and $\propto \eta_j^4\hat{a}_j^{\dag}\hat{a}_j\hat{a}_j^{\dag}\hat{a}_j$ in the Lamb-Dicke expansion. The Rabi frequency can be generalised to $\Omega_{n_j}=\Omega\prod_j e^{-\eta_j/2}L_{n_j}(\eta_j^2)$ with $n_j$ the motional quantum number of harmonic oscillator $j$. Measuring the coupling strength on the transition $|0\rangle\leftrightarrow |1\rangle$ thus allows to obtain information about the states of (thermal) motion. A much employed technique is to vary the duration of the laser pulse such that Rabi flops are recorded. Since the Rabi frequency depends on the populations $n_j$ in each run, the recorded flops will dephase at a rate that gives information about the statistical spread $p(n_j)$. Note that the laser direction may be aligned such that it only couples to a single direction of motion and each distribution of $n_j$ may be interrogated separately. In this case, the Rabi frequency may be approximated as $\Omega_{n_j}\approx \Omega(1-\eta_j^2n_j)$. Obtaining the full energy distribution (and determining whether it is e.g. thermal or some other distribution such as a Tsalis distribution) is possible but relatively time consuming~\citep{Meir2016doa}. This is especially true for large average populations $\bar{n}_j$ since a sum of many frequencies needs to be sampled. When some thermodynamic assumptions are justified, such as equipartition of energy and a thermal energy distribution, $p_\text{therm}(n_j)=\bar{n}_j^{n_j}/(1+\bar{n}_j)^{n_j+1}$, average ion energies (`ion temperatures') can be obtained very efficiently~\citep{Feldker2020bgc,Hirzler2020esf}. 

The `Rabi method' discussed above requires $\eta_j^2 \bar{n}_j\sim 1$ to obtain a reasonable signal. In the quantum regime of ion-atom interactions, typically on the order of 1 quanta of motion remain on average. In this case, it can be more efficient to use a blue or red sideband pulse in which $\Delta_L=\pm \omega_j$~\citep{Meekhof:1996}. In this setting the measured direction of motion can be set with the frequency of the laser. The Rabi frequency on the blue and red sideband can be approximates as $\Omega^\text{bsb}_{n_j}=\Omega\eta_j\sqrt{n_j+1}$ and $\Omega^\text{rsb}_{n_j}=\Omega\eta_j\sqrt{n_j}$. Note that the latter equals zero for the ground state of motion (i.e. $n_j=0$). Once more, the energy distribution may be obtained by varying the pulse width and measuring the probability of each state. Here, the recording pulse length scan is often Fourier transformed to obtain the frequency components in it. The sideband method has the benefit that none of the frequencies are commensurate because of the squareroot in the Rabi frequency, making it easier to distinguish them. 

If only the values $\bar{n}_j$ are to be measured, a trick can be used that saves a lot of measurement time. Here, one only records the curvature of the probability of finding the ion in state in $|1\rangle$, starting from $|0\rangle$. This curvature scales as: $-2\bar{n}_j\eta^2\Omega^2$ and can be measured quickly by sampling a handful of small pulse durations $\ll 1/(\eta_j\Omega)$. Moreover, for small $\bar{n}_j$, the strength of the red and blue sideband may be compared in the frequency domain~\citep{Turchette:2000}, giving direct access to $\bar{n}_j$.

Another elegant method of obtaining the average kinetic energy of the ion is by loading it into a shallow optical dipole trap. Here the survival probability is a direct probe of the ion energy~\citep{Schmidt2020otf,Weckesser2021oof}.

\subsection{Atom bath preparation and detection}
\label{setupatom}
%summary of it all
In essence, the atom part of the ion-atom experiments addresses a cold cloud of one atomic species trapped and cooled by laser light and observed by absorption imaging  within an ultra-high vacuum setup. Therefore, the main elements of the atom-side of the setup (see Fig.~\ref{setup}) are optical beams, magnetic field coils and cameras for absorption imaging. In hybrid ion-atom systems, alkali atoms (e.g. Li, Rb, Na or Ca) are commonly used. They are cooled to the $s$-wave regime where the quantum mechanical underpinnig of atom-atom interactions comes into play. The atoms interact with each other through the van der Waals interaction, which stems from an induced electric dipole moment. This second order electric dipole-dipole interaction is short-ranged and isotropic. Beyond alkali atoms, for elements like erbium, dysprosium and chromium which have a magnetic dipole moment, the dipole-dipole interaction also plays a role. Furthermore, ultracold gases offer an unprecedented control in intra-species interaction strength by magnetically-tunable Feshbach resonances (FRs). 

Each measurement begins with the creation of the gas by cooling and trapping the atoms from an atomic beam and ends with the destructive imaging of the cloud of atoms. The techniques behind this make-probe-discard cycle are well-established as are the requirements for the experimental setup. The atomic isotopes selected determine the details of the optical and magnetic fields needed as well as the cooling and trapping cycle required to prepare the mixture at a specific temperature and density. Additionally a radio-frequency (rf) antenna can be implemented to change the spin states of the atoms, create spin polarized mixtures and do both frequency as well as time-domain spectroscopy. 

%Why ultra-high vacuum
A quantum gas is a metastable state of matter and requires an ultra-high vacuum setup. It requires a a wall-free trap for confinement, most commonly a magneto-optical, magnetic or optical trap. These traps prevent the nucleation on surfaces, which triggers the phase transition of gas into a solid. The lifetime of the gas in the trap is determined by two- and three-body loss processes. Through collisions the particles can gain enough kinetic energy to leave the trap and this gives an upper bound on the densities that quantum gases can have. The rate of three-body recombination, where three atoms collide and form a bound molecule and a free atom that carries away the binding energy, scales as the density cubed. It is this inelastic loss process that drives the transition towards chemical equilibrium. A second constraint on the density comes from the elastic scattering between the particles, which enables the gas to rethermalize and reach a kinetic equilibrium. If the density is too low, collisions between particles take a long time to occur and thermalization might not happen within the lifetime of the gas. These constraints lead to the typical densities of about $10^{16}-10^{19}$\,m$^{-3}$ most commonly seen in experiments with ultracold gases.

%for further reading.
Here, we will discuss briefly the core building blocks of many ultracold atom experiments. For further reading, the reader is referred to the excellent in-depth review articles existing in the field about e.g., introduction into experimental methods~\citep{StamperKurn2012emo, Torma2014qge}, preparation of quantum gases~\citep{Jervis2014mau}, laser cooling and trapping~\citep{Metcalf1999book, Schreck2021lcf}, evaporative cooling~\citep{Ketterle1996eco}, optical dipole traps~\citep{Grimm2000odt}, Feshbach resonances ~\citep{Chin2010fri}, collective modes~\citep{Grimm2008ufg}, spectroscopic probes~\citep{Torma2016pou, Vale2021spo} and the properties of Fermi gases~\citep{Inguscio2008ufg, Zwerger2012tbb, Turlapov2012fgo} and Bose gases~\citep{Ketterle1999mpa, StamperKurn2013sbg} in harmonic traps.

\subsubsection{Degeneracy and interactions}
 %\rsl{Define the s-wave regime for atoms}
When discussing ultracold interacting gases, three length scales matter. These are the interparticle distance $d$ between the atoms, the thermal de Broglie wavelength $\lambda_\text{dB}$ and the range $r_0$ of the atom-atom interaction. For alkali atoms at low temperatures, $r_0$ is given by the $s$-wave scattering length parameter $a$. The de Broglie wavelength changes with the atom gas temperature $T_\text{a}$ according to its definition $\lambda_\text{dB}=\sqrt{2\pi\hbar^2/\left(mk_\mathrm{B}T_\text{a}\right)}$, where $\hbar$ is the reduced Planck constant, $m$ the mass of the particle and $k_\mathrm{B}$ the Boltzmann constant. It characterizes the wave nature of particles in the context of the wave-particle duality. For high temperatures, in the regime of $\lambda_\text{dB}\ll r_0\ll d$, the system is described as a classical gas with only pairwise interactions. For colder temperatures, when the $s$-wave scattering picture applies, the collisions between particles are affected by quantum mechanics once $a\ll\lambda_\text{dB}\ll d$ and collisions require explicit quantum mechanical treatment. Therefore the $s$-wave regime can be reached once $\lambda_\text{dB} > a$ and the $s$-wave temperature for atom-atom interactions is given by $2\pi\hbar^2/\left(mk_\mathrm{B}a^2\right)$. For lithium this is about 50\,mK.

%BEC and DFG
Quantum degeneracy is obtained for $a \ll d< \lambda_\text{dB}$ and the system can then be described as a weakly interacting Bose-Einstein condensate (BEC) or a degenerate Fermi gas (DFG) within the mean-field approximation. Here, the waves that describe each particle overlap and interfere, and the distinction between individual waves is lost. The gas becomes degenerate. In terms of phase-space density this happens for $d^{-3}\lambda_\text{dB}^3>1$. Depending on the density, for Lithium this regime is entered for temperatures of a few hundreds of nanokelvin. 

When the particles are indistinguishable, the distinction between bosons or fermions is important. The wavefunction of a many-body system of bosons is symmetric under the exchange of particles, while that of fermions is antisymmetric. For fermions, this results in Pauli's exclusion principle, where no two identical fermions can occupy the same quantum state. Thus a system of $N$ identical fermions will occupy $N$ different quantum states. For temperatures close to zero, these fermions fill the energy levels up from the lowest level to the Fermi energy. The Fermi energy is the energy of the highest filled quantum state, which depends on the number of fermions in the system and benchmarks the DFG. On the contrary, all atoms in a system of $N$ identical bosons can occupy the same quantum state. Moreover, the occupation of the same state is actually favored. When $N$ bosons occupy the same state, the probability to get an additional boson in that state is enhanced by a factor of $(N+1)$. Thus, for temperatures close to zero a macroscopic occupation of a single quantum state occurs and a BEC is formed. 

%strong interactions
By tuning the interaction in an ultracold quantum gas, one can also reach the strongly interacting regime where $d<a<\lambda_T$. Here, the description of the degenerate many-body system as a single macroscopic wavefunction fails and rich physics and complex quantum phases can be expected. At the typical densities ($n=d^{-3}$) of an ultracold gas, the temperatures required to obtain degeneracy are around a few hundred nanokelvin. Of course increasing the density of the gas would give fewer constraints on the temperature, but this shortens the lifetime of the gas and thus the measurement time of an experiment.

Feshbach resonances provide the tunability of interactions for which ultracold gases are famous~\citep{Chin2010fri}. A FR occurs when a molecular bound state of almost no energy couples resonantly to the free state of two colliding atoms~\footnote{Depending on the channels involved, these FRs can be attributed to different partial waves and are called $s$, $p$, etc..-wave FRs. For ultracold temperatures, the $s-$wave FRs are the most important. $p$-wave resonances are especially interesting for spin-polarized DFG, as the Pauli principle prevents them from interacting via s-wave scattering.}. The difference in the magnetic moment between those two states, can be used to tune the states in and out of resonance by changing the magnetic field. Close to the FR, the scattering length as a function of magnetic field is given by $a (B) =a_\text{bg} \left(1-\Delta/(B-B_0)\right)$, with $a_\text{bg}$ the background scattering length and $\delta$ the FR width. At the Feshbach resonance center $B_0$ the scattering length diverges.  Due to a FR, the interaction between atoms can be tuned from weak to strong and from attractive to repulsive. The lifetime of a gas with strong interactions is limited by three-body recombination, and often atom loss measurements are used to characterize FRs. Feshbach resonances can occur between atoms in the same spin state, in different spin states and between spin states of different elements. On the repulsive side of a FR, a weakly bound state exists and this allows the creation of weakly bound pairs~\citep{Koehler2006poc, Ferlaino2009ufm}. Close to $B_0$, the binding energy of the dimer follows $E_b=\hbar^2/\left(2\mu a\right)$, with $\mu=\frac{1}{2}m_a$ for a single species atom bath.  

\subsubsection{Laser cooling}
\label{lasercooling}
%Oven, Zeemanslower, 2D MoT
To create an ultracold gas, a hot gas is formed and cooled down by laser cooling~\citep{Metcalf1999book, Schreck2021lcf}. The atomic beam exiting the oven is slowed down either via a Zeemanslower or via a 2D-MOT stage, both relying on light scattering. The Zeemanslower uses a counterpropagating laser beam and the magnetic fields changes along the atomic pathway to compensate for the Doppler shift and keep the atoms resonant using the Zeeman effect. A 2D-MOT is similar to the typical magneto optical trap (MOT), however in one direction there is the possibility for the atoms to escape, which can be facilitated by using a pushing beam. It depends on the species used which of the two methods are favorable.  This first stage typically slows the atoms down to a few tens of metres per second. 

%Magento Optical trap
The atoms are subsequently caught in a magneto optical trap and further Doppler cooled. The MOT is a combination of counterpropagating red-detuned optical beams from all three directions and a magnetic field gradient. Together they provide a trapping environment for the atoms in which those travelling away from the trap center are brought back through light scattering. As the lasers are tuned to a frequency below the transition frequency, high velocity atoms are more likely to absorb photons from the optical beams counter to the atoms direction of movement, which leads to both cooling and trapping. The temperatures of a MOT are typically limited by the Doppler temperature $T_D=\hbar\gamma/\left(2k_B\right)$ which depends on the linewidth $\gamma$ of the optical transition used. For laser cooling of alkali atoms, this is most commonly the D2 transition ($S_{1/2}\rightarrow P_{3/2}$).  The number of atoms in the MOT depends on the loading time, during which the atoms are captured from the atomic beam, and is limited by the loss of atoms due to photoassociation. 

Although several sub-doppler techniques exist, the most common next step is to load an off-resonant optical dipole trap~\citep{Grimm2000odt} or magnetic trap\citep{perezrios2013hda} and do evaporative cooling to further decrease the temperature. The ultimate limit for laser cooling is set by the recoil temperature $T_r=\frac{h^2}{2 k_\mathrm{B} m \lambda^2}$, which is 3.5\,$\mu$K for lithium.

\subsubsection{Optical trapping}
% optical trapping
In an far-off resonant single beam optical trap~\citep{Grimm2000odt}, the atoms experience a trapping potential which depends on their polarizibility $\alpha$ and the waist $w_0$ and power $P$ of the trapping laser of wavelength $\lambda$. The polarizibility depends on the wavelength used for optical trapping and is typically calculated from theory and verified experimentally\footnote{A useful database thereby is the recently set-up portal for atomic data and computation~\citep{UDportal}.}. Assuming a gaussian beam the potential has a trap depth of $U^{opt}=-\frac{1}{2\epsilon_0 c} \,\text{Re}(\alpha)\frac{2}{\pi \, w_0^2}\,P$, with $\epsilon_0$ the vacuum permittivity and $c$ the speed of light. This leads to a trapping frequency in radial direction of $\omega_{r} =\frac{2}{w_{0}}\sqrt{\frac{ U^{opt}}{m_a}}$ and axially $\omega_{ax}=\sqrt{ 2 U^{opt}\lambda^2/\left( m\,\pi^2\, w_0^4\right)}$. Most commonly a crossed beam setup is used such that the confinement in the axial direction can be increased. The optical trap can be fully characterized by measuring its trap frequencies. Besides an optical potential, the scattering rate of the dipole trap is also important. This highlights the rate with which the trapping light can  be absorb by the atoms and re-emitted. The scattering rate for a gaussian beam is given by $\Gamma_\text{sc}=\frac{1}{\hbar\epsilon_0 c} \,\text{Im}(\alpha)\frac{2}{\pi \, w_0^2}\,P$. To keep the scattering rate as low as possible, traps that are far-off resonant from optical transitions within the atom are typically used. 

%evaporative cooling
Evaporative cooling~\citep{Ketterle1996eco} takes place by lowering the power of the optical dipole trap, which reduces the trap depth $U^{opt}(0)$. It relies on removing energetic particles from the trap and letting the remaining particles rethermalize. The gas then acquires a lower temperature and a higher phase-space density. The limit of evaporative cooling is given by the ratio of inelastic to elastic collisions. Elastic collisions are needed for thermalisation, while inelastic collisions typically shorten the lifetime of the sample.  The trap depth you set in order to truncate an atom cloud of a certain temperature is given by the  truncation parameter $\eta$, i.e. $\eta=U^{opt}(0)/(k_B T_a)$, and is typically kept constant and around 10. Typically a decrease of 3 orders of magnitude in atom number leads to an increase in phase-space density of six orders of magnitude. Therefore, evaporative cooling is a powerful tool to reach quantum degeneracy and get ultracold samples.  

% Densities
For a deep enough trap, the ODT trap seen by the atoms can be treated as an harmonic trap with trap frequencies $\omega^a_i$ for all three directions $i=x,y,z$. The density distribution in the trap is non-uniform as the trapping potential is not a box-potential~\footnote{Box potentials, leading to a homogeneus density, are being explored though ~\citep{Navon2021qgi}.}. The density of a thermal cloud is then given by 
\begin{equation}
n_\mathrm{t}=\frac{N}{L_x L_y L_z} \text{exp} \left (\frac {-m_a}{2kT_a}((\omega^a_x)^2 x^2 + (\omega^a_y)^2 y^2 + (\omega^a_z)^2 z^2) \right )
\end{equation}
with $L_i=\left( \frac{2 \pi k T}{m (\omega^a_i)^2}\right)^{1/2}$, which is related to the width of the cloud by $\sigma_i=\left( \frac{k T}{m (\omega^a_i)^2}\right)^{1/2}$. The density depends on the mass of the atom, the total atomnumber $N$ and the temperature. The thermal peak density $\hat{n}_\mathrm{t}$ in an harmonic trap is given by
\begin{equation}
\label{peakdensity}
\hat{n}_t=\left(\frac{(\bar{\omega}^a)^2\,m_a}{2\pi\, k_\mathrm{B}\, T}\right)^{3/2}N.
\end{equation}
Here,$(\bar{\omega}^a)^2$ is the geometrical average of the trap frequency and can be calculated as $\bar{\omega}^a=(\omega^a_x\omega^a_y \omega^a_z)^{1/3}$ for a cigar-shaped cylindrical trap. The mean density is given by,$\bar{n}_\mathrm{t}= \frac{1}{\sqrt{8}} \hat{n}_\mathrm{t}$. 

For a \textbf{single}-component Fermi Gas in an harmonic trap
\begin{equation}
\hat{n}_\mathrm{f} = \frac{2 \sqrt{N}}{\sqrt{3}\, \pi^2}\left( \frac{\bar{\omega}^a \,m_a} {\hbar}  \right)^{3/2}
\end{equation}
and for a BEC
\begin{equation}
\hat{n}_b=\frac{15^{2/5}}{8\pi}\left( \frac{\bar{\omega}^a m_a}{\hbar \sqrt{a_\mathrm{bb}}}\right)^{6/5} N^{2/5}
\end{equation}
with $a_\mathrm{bb}$ the bose-bose intraspecies scattering length.

%Transport 
Usually the cold atoms are created far away from the ions in order not to contaminate the ion trap electrodes and a form of transport is needed to overlap the atoms with the ions. Typically this is done by optical transport, either by changing the focus of the optical trap or by using piezo-mirrors to change the beam pointing. 

\subsubsection{Absorption imaging}
The most important experimental parameters of the bath are the atom number, density and temperature, which are commonly obtained by absorption imaging. An absorption image is taken by shining light resonant to an optical transition on the atoms for a fixed duration. The atoms absorb the light and this leads to a depletion
of the intensity of the probe beam depending on the optical density OD of the cloud at each specific spatial position. Subtracting images of the recorded light intensities with and without the atoms present gives the absorption image and information on the optical density of the cloud. For low intensities, resonant light, and a closed optical transition, the optical density is related to the column density $n(x,y)$ by the resonance
cross section $\sigma$, i.e. $OD = n(x,y)\sigma$. In this example, the imaging happens along the $z$-direction. The absorption images are then directly related to the column density of the atomic cloud and by integrating along one direction a 1D density distribution is obtained. This density profile can then be fitted with the
appropriate distribution function~\citep{Inguscio2008ufg,Ketterle1999mpa}, which gives the atom number and width of the cloud. The 1D density profile of a a thermal cloud is fitted with a Gaussian distribution function, for a BEC a parabola or bimodal distribution is used and a DFG requires a polylogarithmic function.  Integrating another time, the total atom number can be found. 

%Time-of-flight
To obtain the temperature, a time-of-flight measurement is taken. Here the atomic cloud is imaged at various times after it is released from the optical trap. The expansion of the cloud in the absence of any confining potential is related to its temperature. From the time-of-flight curves, assuming free expansion, the temperature $T_x$ and $T_y$ as well as the in-situ width of the cloud, $\sigma^0_x$ and $\sigma^0_y$,  in both x- and y- direction can be obtained. This follows from fitting the expansion of the width of the cloud as a function of expansion time using
\begin{equation}
\sigma_\alpha (t)=\sqrt{(\sigma^0_\alpha)^2 + \frac{k_B T_a}{m_a}\,t^2}.
\end{equation}
For determining the width, it is important to measure the magnification of the camera.

%Discuss Time-of-flight
The density of the cloud can also be inferred from the time-of-flight measurements as $n=N/V=N/\left((2\pi)^{3/2}\sigma^0_x\, \sigma^0_y\, \sigma^0_z\right)$. Comparing this to Eq.~\ref{peakdensity}, the trapfrequencies can also be obtained as $\sigma^0_\alpha=\left( \frac{k T}{m \omega_\alpha^2}\right)^{1/2}$. They can also be measured independently by exciting and measuring the frequency of the collective modes in the trap~\citep{Grimm2008ufg}.

\subsection{Ion-atom mixtures}
\label{sec:AImixtures}
%Introduce table and different approaches
Several ion-atom experiments currently exist that look into the interplay between an ion and an ultracold ($\sim\mu$K) cloud of atoms. See table~\ref{AImixtures} for an overview including their latest experimental results. Here, we focus on the setups with ultracold atoms beyond the MOT phase, where the atom cloud is already in the regime where only the $s$-wave scattering length characterizes the atom-atom interactions~\footnote{Ion-MOT experiments exist with the combinations of $^{17}$OH$^-$-$^{87}$Rb~\citep{Hassan2022adi},
$^{28}$N$_2^-$-$^{87}$Rb and
$^{32}$O$_2^-$-$^{87}$Rb~\citep{Doerfler2019lrv},
$^{174}$Yb$^+$-$^{87}$Rb~\citep{Bahrami2019ooa},
$^{173}$BaCl$+$-$^{87}$Ca~\citep{Puri2019rbi},
$^{174}$Yb$^+$-$^{40}$Ca~\citep{Mills2019ees,Li2019ean},
$^{40}$Ca$^+$-$^{28}$Na~\citep{Kwolek2019moc}, 
$^{40}$Ca$^+$-$^{39}$K~\citep{Hui2020pmc} and
$^{55}$Cs-$^{55}$Cs~\citep{Dutta2020moc}}.
%Yb+: Olmschenk2007mad , Ba+: Dietrich2010hao, Ca+: Roos2000edo, Sr+, Rb+
Most setups prepare the atoms away from the ion and then move the atoms on top of the ion or vice versa, e.g.~\citep{Hirzler2020esf, Perego2020eoi, Meir:2017, Schmid2012aaf}. The transport can be done optically by changing the position of the atoms' optical dipole trap or using compensation electric fields for displacing the ion.   The separate preparation requires either a two-chamber or two-stage design of the experimental setup, which makes it easier to have good optical access facilitating large beams for the atomic MOT-stage. While the ion can be reinitialized, usually the atom bath is discarded and is reloaded for each new measurement.

\begin{table}[tp]
\centering
\setlength\extrarowheight{3pt}
\caption{\label{AImixtures} Overview of experimental ion-atom mixtures with ultracold atoms and their characteristics. Given are the ion-atom mass ratio $\xi$, the $C_4$-coefficient with definition $V_\text{ia} (r)=-\frac{C_4}{r^4}$, the $s$-wave energy $E_s$ and the most up-to-date experimental reference. The ion can be either trapped in a Paul trap (PT), in an ion-only optical dipole trap (IDT) or together with the atoms in a Bi-chromatic optical dipole trap (Bi-ODT). The ion can also be unconfined, which is typically for creation via the ionisation of a Rydberg excitation (Ry). Moreover the ion can be created from the atom bath by three-body recombination (TBR) and subsequent chemical reactions.}
\begin{tabular}{|l|l|l|l|l|l|l|l|}
\hline\hline
Ion & Atom & $\xi=\frac{m_a}{m_i}$ &$C_4$ & $E_s/k_\text{B}$  &Type& Reference\\ 
&&&$(E_ha_0^4)$&($\mu$K)& &\\\hline\hline
$^{171/174}$Yb$^+$&$^6$Li & 0.035 & 82 & 8.6 &PT  &\cite{Hirzler2022ooc}\\\hline
$^{138}$Ba$^+$& $^6$Li & 0.044 & 82 & 8.7 &PT+IDT&\cite{Weckesser2021oof}\\\hline
$^{40}$Ca$^+$& $^6$Li & 0.15 & 82& 10.6 &PT&\cite{Saito:2017}\\\hline
$^{138}$Ba$^+$& $^{87}$Rb & 0.63& 160 & 0.052 &PT &\cite{Mohammadi:2021}\\\hline
$^{138}$Ba$^+$& $^{87}$Rb & 0.63& 160 & 0.052 &PT + Bi-ODT&\cite{Schmidt2020otf}\\\hline
$^{88}$Sr$^+$& $^{87}$Rb & 0.99 & 160 &0.078  &PT&\cite{Katz2022qld}\\\hline
$^{87}$Rb$^+$& $^{87}$Rb & 1.0 & 160 & 0.078 &PT&\cite{Mahdian2021doo}\\\hline
$^{87}$Rb$^+$& $^{87}$Rb & 1.0 & 160 & 0.078 &Ry&\cite{Dieterle2021toa}\\\hline
$^{87}$Rb$^+$& $^{87}$Rb & 1.0 & 160 & 0.078 &PT + TBR&~\cite{Mahdian2021doo}\\\hline\hline
%$^{17}$OH$^-$& $^{87}$Rb (MOT) & 5.1 & 160 & 0.73 &\cite{Hassan2022adi}\\\hline
%$^{28}$N$_2^-$& $^{87}$Rb (MOT) & 3.1 & 160 & 0.33 &\cite{Doerfler2019lrv}\\\hline
%$^{32}$O$_2^-$& $^{87}$Rb (MOT) & 2.7 & 160 & 0.27 &\cite{Doerfler2019lrv}\\\hline
%$^{174}$Yb$^+$&$^{40}$Ca (MOT)& 0.23 && &\cite{Mills2019ees}\\\hline
%$^{40}$Ca$^+$&$^{28}$Na (MOT)& 0.23 && &\cite{Kwolek2019moc}\\\hline
%$^{171/174}$Yb$^+$& $^{161}$Dy & 0.94 &  &&129&\\\hline\hline
\end{tabular}
\end{table} 

% Other traps
In ion-atom mixtures, the ion is typically trapped in a radio-frequency Paul trap~\citep{Paul:1990}, which relies on a time-dependent potential (see Sec.~\ref{paultrap}). However, ions by themselves can also be trapped in a multipole trap~\citep{Wester2009RMT} or in an optical dipole trap~\citep{Schaetz2017tia, Karpa2019tsi}. Combining any of these traps with an ultracold cloud of atoms leads to some difficulties. For multipole traps, reaching very low temperatures for the ion-atom collision energy is in principle easier compared to a Paul trap as the ion can be confined in a region with little micromotion. But on the other hand, this makes the ion more prone to stray electric fields that cause excess micromotion~\citep{Trimby2022bgc, Niranjan2021aom}. Multipole traps have been used for studying ion-atom collisions and buffer gas cooling at K-mK temperatures~~\citep{Wester2009RMT,Asvany2009NSK,Notzold2020TMI}. They allow the exploration of ion-atom systems beyond the critical mass ratio of $\xi=m_\text{a}/m_\text{i}$ of about 1~\citep{Holtkemeier2016bgc}. 

Optical trapping of the ion requires relatively high optical powers to create a deep enough trap to catch the ion, which is typically laser-cooled to the Doppler limit\footnote{Ground-state cooling of an ion in an optical lattice is feasible and was demonstrated by~\cite{Karpa2013soi}}. The optical trap depth  of the ion is given by the optical potential and reduced by stray electric fields and static defocusing. Moreover, the atoms also react to this optical potential, which is typically very deep compared to what is normally used for optical atom trapping. This causes a very tight confinement of the cloud, where the high density leads to enhanced three-body losses~\citep{Harter2012sia} as well as a high heating rate of the atoms by the ion's trapping light. Precooling the ion to sub-Doppler temperatures, would reduce the optical potential needed and thus the effect on the atoms. Additionally, a second light field with an anti-confining frequency can be used to compensate for the detrimental effects of the ion optical trap. This bi-chromatic trap creates a milder potential for the atoms, while still deeply trapping the ion~\citep{Karpa2021ioi}. For Ba$^+$-Rb mixture a bi-chromatic optical trap of 532 and 1064\,nm was demonstrated in which both atoms and ions could be confined~\citep{Schmidt2020otf}. Here, the 532\,nm light is actually anti-trapping for the atoms and this is compensated by the 1064\,nm light. In Ba$^+$-Li an ion-only optical trap at 532\,nm was used to capture the ion after precooling in the Paul trap and by the atom buffer gas. Both setups still require precooling of the ion in the Paul trap, to reduce its energy to trappable regimes. Developments for an electro-optical trap which combines the optical trapping with an electrostatic field~\citep{Karpa2013soi, Karpa2019tsi} are also under way~\citep{Perego2020eoi}.  
Alternatively, an ion can be obtained following three-body recombination of the atoms into a molecular ion which subsequently photo-ionizes and dissociates, creating a single ion as was demonstrated for Rb~\citep{Mahdian2021doo}.

% free ions/Rydberg ions
Furthermore, ion-atom mixtures can be created by exciting atoms to a Rydberg state and ionizing the Rydberg atom. These ions are formed from the ultracold bath and lead to homonuclear ion-atom mixtures. These ions could then again be trapped, yet at the moment one studies how they move as free ions through the cloud. Their interaction with the bath can be probed for the duration of tens of microseconds and is limited by three-body recombination. For the Rb$^+$-Rb mixture an ion with an energy of about $k_\text{B}\times 50\,\mu$K was created with a lifetime of $\sim 20\,\mu$s~\citep{Dieterle2021toa}. The Rydberg blockade takes care that only a single ionic impurity is created~\citep{Balewski2013cas, Kleinbach2018iii}.
Rydberg atoms can also be coupled to either free or trapped ions~\citep{Engel2018oor,Ewald:2019,Haze:2019,Deiss2021lra,Zuber2021sio}. These Rydberg atoms have a stronger polarization and provide a strong interaction with the ion. Moreover, a trapped ion itself can be excited to a Rydberg state~\citep{Mokhberi2020tri}, however its influence on the interaction with the bath has not yet been studied.

% quantum regime & mass ratio
Reaching the quantum regime is more accommodating for ion-atom combinations with a heavy ion and light atoms. There, $\xi\ll 1$ and the reduced mass $\mu\approx m_\text{a}$. This leads to $s$-wave energies in the $\mu$K-regime, as $E_{s}=\frac{\hbar^4}{4\mu^2C_4}$. This energy needs to be compared to the collision energy which also depends on the atom and ion masses as well as their energies (see Eq.~\ref{Ecol}). Ultracold atoms can be routinely cooled to the few $\mu$K regime, whereas ions after Doppler cooling are typically in the several hundreds of $\mu$K regime. Therefore, reducing the collision energy is most profitable when cooling the ion further, either by buffer gas cooling or using sub-Doppler cooling techniques such as resolved side-band or EIT-cooling~\cite{Diedrich:1989,Roos:2000a}. The mass ratio plays not only a role  for cooling, but also in the energy distribution of the ion. For a heavy ion with a light atom, the shape of the distribution remains largely unaffected. However, for equal masses a non-thermal Tsallis-distribution was seen~\citep{Meir2016doa} and the mass-ratio was predicted to influence the exact nature of this nonthermal energy distribution that appears after multiple collisions~\citep{Rouse2019ted, Pinkas2020eoi}. Large ion to atom mass ratios result in energy distributions that are indistinguishable from a thermal distribution~\citep{Feldker2020bgc}. 

% Rf-induced heating vs cooling rate
For ions in a Paul trap, the rf-induced heating, caused by the coupling of the micromotion with the attractive ion-atom potential is a limiting factor. This heating sets a lower bound on the ion temperatures that can be reached and thus on the attainability of the quantum regime. For one dimension, the ion heating in a single collision between an ion and atom at rest was found~\citep{Cetina2012flt}, which translate to a 3D heating of 
\begin{equation}
    W^{3D}_0=\frac{8}{3\pi}\left(\frac{m_\text{a}}{m_\text{i}+m_\text{a}}\right)^{5/3}\left(\frac{m_\text{i}^2\omega^4C_4}{q^2}\right)^{1/3}.
\end{equation}

This equation has been confirmed by numerical simulations~\citep{Cetina2012flt,Pinkas2020eoi}. We can use it to compare the prospects for various ion-atom combinations in ideal Paul traps.

\begin{table}[tp]
\centering
\setlength\extrarowheight{3pt}
\caption{\label{AImixturesheating} Estimation of the effect of rf-induced heating for various ion-atom combinations. Table shows the maximum values $\Omega_\text{rf}^\text{max}$ such that $W^{3D}_0=E_s$ and $q=0.2$, the maximal ion energy $E_i^\text{max}=\frac{m_i}{\mu}E_s$ and maximal radial offset field $\mathcal{E}_\text{offs}^\text{max}$ (Eq.~\ref{Eoffset}) as well as the reduced mass $\mu$ in atomic mass units.}
\begin{tabular}{|l|l||l|l|l|l|}
\hline\hline
Ion & Atom & $\mu$ & $\Omega_\text{rf}^\text{max}/2\pi$ &$E_\text{i}^\text{max}/k_\text{B}$ & $\mathcal{E}_\text{offs}^\text{max}$ \\
& & (u) & (MHz)&($\mu$K)& (V/m)\\\hline\hline
$^{171}$Yb$^+$&$^6$Li & 5.81 & 22.8 & 253 & 2.82\\\hline
$^{138}$Ba$^+$& $^6$Li & 5.76 & 19.9 & 209 & 2.00\\\hline
$^{88}$Sr$^+$& $^6$Li & 5.63 & 15.2 & 143 & 0.35\\\hline
$^{40}$Ca$^+$& $^6$Li & 5.23 & 10.3 & 81.2& 1.01\\\hline
$^{171}$Yb$^+$& $^{23}$Na & 20.3 & 0.745 & 6.02 & 0.014\\\hline
$^{138}$Ba$^+$& $^{87}$Rb & 53.3 & 0.0225 & 0.135&  $5.76\times 10^{-5}$\\\hline
$^{88}$Sr$^+$& $^{87}$Rb & 43.7 & 0.0277 & 0.156& $6.09\times 10^{-5}$\\\hline
$^{87}$Rb$^+$& $^{87}$Rb & 43.5 & 0.0279& 0.157 & $6.12\times 10^{-5}$\\\hline\hline
\end{tabular}
\end{table} 

We can equate $W^{3D}_0$ with the $s$-wave energy $E_S=\hbar^4/(2\mu^2 C_4)$ to find out what restrictions we have in choosing $\Omega$ and $q$ such that the energy released in the first collision between a ground state cooled ion and ultracold atom remains in the $s$-wave regime. Setting $q=0.2$ and in the limit $|q_j^2|\gg |a_j|$ such that $\omega_j\sim |q_j|\Omega_\text{rf}/\sqrt{8}$,  we show the results in table~\ref{AImixturesheating}. We see that combinations with a large ion-to-atom mass ratio allow for using Paul traps with $\Omega_\text{rf}/(2\pi) \lesssim$~20~MHz. On the other hand, smaller mass ratios require smaller drive frequencies with already $^{23}$Na demanding $\Omega/(2\pi) \lesssim$~1~MHz. For such small drive frequencies, radial excess micromotion is expected to be the dominant limiting factor. 

Equating the $s$-wave limit with the contribution of the ion's micromotion energy to the collision energy allows us to estimate the requirements on static offset field compensation. The collision energy (Eq.~\ref{Ecol}) of an ion in a bath of atoms in the limit $T_\text{a}\rightarrow 0$, is given by $\frac{\mu}{m_\text{i}}E_\text{s}$. This means that the ion can have a maximal energy of $E_\text{i}^\text{max}=\frac{m_\text{i}}{\mu}E_s$, in order for the ion-atom collision to be in the quantum regime i.e $E_\text{col} < E_s$. Using Eq.~\ref{Eoffset}, we can further calculate the maximal radial offset field that can be present in the setup. Both values are listed in table~\ref{AImixturesheating} for various ion-atom combinations. For Li and $^{23}$Na combined with heavy ions these values are experimentally challenging but in reach, while alkali atoms heavier than $^{23}$Na require forbidding micromotion compensation and seem out of reach.  Note that the present discussion only includes the first collision and cannot be used to predict the thermalization of the ion in a gas or indeed whether it thermalizes at all.

\begin{table}[tp]
\label{sims}
\centering
\setlength\extrarowheight{3pt}
\caption{\label{AImixturesheatinglit} Heating and the final energy for realistic ion-atom systems in an RF-Paul trap with the static and dynamic stability parameters $a, q$ and trap driving frequency $\Omega$ given by literature. The thermal equilibrium energy of the ion is obtained from numerical simulations using the code of~\citep{Trimby2022bgc} taking into account an radial stray field of 0.05V/m and an atom bath of $2\,\mu$K.}
\begin{tabular}{|l|l||l|l|l|l|l|l|}
\hline\hline
Ion & Atom & $a_x$ & $q$ & $\Omega_\text{rf}/(2\pi)$&$W^{3D}_0/k_\text{B}$  & $E^\text{ion}_\infty/k_\text{B}$& Refs.\\
& & & & (MHz) & ($\mu$K) & ($\mu$K)&  \\\hline\hline
$^{171}$Yb$^+$&$^6$Li & -0.00073 & 0.22 & 2& 0.31 & $28.0 $&\cite{Hirzler2020esf}\\\hline
$^{138}$Ba$^+$& $^6$Li & -0.00011 &0.24 & 1.433 & 0.30 & 35.5 &\cite{Weckesser2021tsa}\\\hline
%wsec=2Pi*(123,122,7.6) kHZ, R0=9mm, Vrfmax=1140V 
$^{40}$Ca$^+$& $^6$Li & -0.0016 &0.16 & 4.8 & 3.29 & 87.1 &\cite{Saito:2017} \\\hline
%wsec=2Pi*(252, 95)khZ, Vrf=44V 
$^{138}$Ba$^+$& $^{87}$Rb  & -0.00093 &0.13 & 5.24 & 56.3 & $1.47\times 10^3$ &\cite{Schmid2012aaf}\\\hline
%wsec=2Pi*(250, 80)khZ, r0=2.1mm, q=0.13 
$^{138}$Ba$^+$& $^{87}$Rb  & -0.00032 &0.40 & 1.4 & 20.7 & 667 &\cite{Schmidt2020otf}\\\hline
% wsec=2Pi*(200, 12.5)khZ 
$^{88}$Sr$^+$& $^{87}$Rb  & -0.052 &0.58 & 4.22 & 104 &  $313\times 10^3$&\cite{Katz2022qld}\\\hline
% wsec = 2*Pi*(1000, 400, 480) kHz , Omega = 26.5 = 2*Pi*4.218 &\cite{Katz2022qld} 
$^{87}$Rb$^+$& $^{87}$Rb  & -0.0015 &0.21 & 5.24 & 87.5 & $552\times 10^3$ &\cite{Schmid2012aaf}\\\hline\hline
%%wsec=2Pi*(390, 100)khZ, q=0.21 &\cite{Schmid:2012aaf}
\end{tabular}
\end{table} 

Apart from heating the ion, atom baths also buffer gas cool the ion as their temperature is typically 2-3 orders of magnitude lower than that of the ion (see Sec.~\ref{bgcool}). The competition between heating and cooling, determines the final temperature reachable in ion-atom mixtures. Using numerical simulations based on the code published by~\cite{Trimby2022bgc}, we now proceed to obtain this equilibrium temperature of the ion for the various ion-atom systems discussed, in the limit of many collisions. We do this for a realistic experimental scenario. We assume an atom bath of $2\,\mu$k and the ion initially at rest and a radial stray field of 0.05V/m. The latter reflects the stray field compensation achievable in the ion-atom experiments with Paul traps\footnote{For optical trapping of ions, electric field compensations below 0.01V/m have been reported~\citep{Huber:2014, Weckesser2021oof}}. After several thousands of collisions of the ion with the atoms, we extract the final secular ion temperature given the realistic values of $q$ and $\Omega_\text{rf}$. The total equilibrium kinetic energy of the ion is then obtained using $E^\text{ion}_\infty=\frac{5}{2}k_\text{B} T_\text{sec}$ and is shown in table~\ref{AImixturesheatinglit}. Here, we assume all three directions to contribute equally to the kinetic energy and the intrinsic micromotion to add kinetic energy on the order of $k_\text{B}T$ to the ion~\citep{Berkeland:1998}. The mixtures with a heavy ion and a light atom show the lowest final energies. By comparing the two Ba$^+$-Rb setups, the influence of the trap parameters becomes clear.

\section{Experimental Results}
\label{results}
The recent experiments with hybrid ion-atom systems especially focus on cooling to the quantum regime, understanding the interactions and losses of the system, and observing how the ion behaves in a many-body environment of atoms. The colder the system is, the more likely it is that quantum effects play a role. This requires the interactions to be treated quantum mechanically and shed light on what happens beyond the typical Langevin picture of collisions. Here we discuss these recent experiments and the results that were obtained.

\subsection{Buffer gas cooling}
\label{bgcool}
%general buffer gas cooling
The idea of cooling one system by immersing it into another system which is much colder is a general concept. This buffer gas cooling or sympathetic cooling relies on the energy exchange between the hot system and the coolant. In ion-atom systems this method is used to reduce the energy of the typically much hotter ion by the atom bath. As it purely relies on collisions between the two components, it is a very versatile technique applicable to many atomic and ionic species. Unlike laser-cooling or magnetic evaporation, it does not require the atom or ion to have a particular resonant transition or dipole moment. Atom baths can be readily prepared at sub-Doppler temperatures in the 10-0.1$\,\mu$K regime, while  the ion's temperature is in the $10-0.1\,$mK regime. Buffer gas cooling was also proposed as an application in trapped ion quantum computing~\citep{Daley2004sac}, where the superfluid coolant, e.g. a Bose-Einstein condensate, protects the ion-qubit.  Besides reducing the ion's energy, buffer gas cooling is the way to reduce the collision energy and reach the quantum regime. 

%ion thermometry + detection + Tsallis
To determine the effect of buffer gas cooling ion thermometry is used. Depending on the temperature regime different methods are applied. For instance, in a first experiment with an $^{174}$Yb$^+$ ion in a Rb-BEC, the technique of Doppler recooling~\citep{Epstein:2007, Wesenberg:2007} showed the cooling possibilities of atomic baths to below the K regime~\citep{Zipkes2010ats}. With the same technique, in the 10\,K-mk regime, buffer gas cooling in ion-atom systems was demonstrated with Ca$^+$~\citep{Haze2018cdo} by a Li-bath of about 4.5$\,\mu$K. For detecting lower temperatures,  thermometry using a narrow-line transition can be applied. For instance, carrier Rabi spectroscopy was used to detect the heating of a groundstate cooled $^{88}$Sr$^+$ ion in a bath of 5$\,\mu$K $^{87}$Rb atoms~\citep{Meir2016doa}, below the mK regime. For this equal mass ion-atom mixture, the ion's energy distribution was demonstrated to evolve from a thermal (Maxwell-Boltzmann) to a power-law distribution best described by the Tsallis function, by interacting with the atom cloud. This is caused by the interplay between the rf fields and the ion-atom interaction, which depends on the mass-ratio~\citep{Rouse2019ted}. Alternatively, the ion can be captured in a shallow optical potential, which sets an upper bound on the energy of the ion~\citep{Schneider2010oto, Weckesser2021oof}. 

%2020: Ba-Rb bichromatic trap https://journals.aps.org/prl/pdf/10.1103/PhysRevLett.124.053402 
Recently, for Ba$^+$-Rb in optical traps, buffer gas cooling was measured to reduce the ion energy by 100$\,\mu$K~\citep{Schmidt2020otf} in a single collision. This was shown using a bi-chromatic optical trap, which captured both Ba$^+$ and Rb and could cool Ba$^+$ to below its Doppler temperature of $370\,\mu$K. About 10 collisions between the ion and atom cloud would be needed to reach thermal equilibrium. However, extending the measurements to further cool down the ion to about the temperature of the bath was not possible due to three-body losses and the experimental stability. Especially the good relative alignment of the two beams of the optical trap and the presence of parasitic ions triggered by multi-photon ionization and three-body recombination, remain ongoing challenges~\citep{Karpa2021ioi}. Loading the atoms not as an ensemble, but into individual lattice sites of an optical lattice, could be a possible way out.  

% BGC to quantum regime https://www.nature.com/articles/s41567-019-0772-5 
Moreover, buffer gas cooling to the $s$-wave regime was observed in the Yb$^+$-Li mixture~\citep{Feldker2020bgc}. For cold collisions one can decompose the total scattering wave function into contributions of different partial waves, i.e. different angular momenta of the rotational motion. For $E_\text{col}<E_s$, the ion-atom collisions can be called ultracold and are entirely determined by a single partial-wave scattering in the incoming collisional channel. For the temperatures of Yb$^+$-Li mixture, most partial waves are frozen out and the quantization of interaction leads to only contributions of the $s$-wave and $p$-wave molecular potentials to the collisions. When studying collisions, this system should therefore show deviations from the classical treatment of collisions and, indeed, quantum effects in the spin-exchange rate as a function of collision energy were seen (see Section\ref{QE}).  

%Collision energy
By precise thermometry and taking into account the full energy budget of the ion a collision energy of $E_\text{col}=1.15(\pm 0.23) E_s$ was found after 1\,s of interaction with a ~2\,$\mu$K cloud of Li. The Yb$^+$-Li has an $s$-wave energy of $k_\text{B} \times 8.6\mu$K. The collision energy depends on the atom kinetic energy ($\frac{3}{2}k_\text{B} T_\text{a}$) and the total kinetic energy of the ion, which is made up of several components. For Yb$^+$-Li, besides the secular temperature in axial and radial directions, the intrinsic micromotion and excess micromotion in all three directions was determined to deduce the final collision energy.  Furthermore, care was taken to compensate the excess micromotion as much as possible~\footnote{These commonly used micromotion compensation and thermometry methods are described in Section~\ref{micromotion} and ~\ref{Iondetection}.}. 

\begin{figure}[!t]
\center
\includegraphics[height=0.6\textwidth]{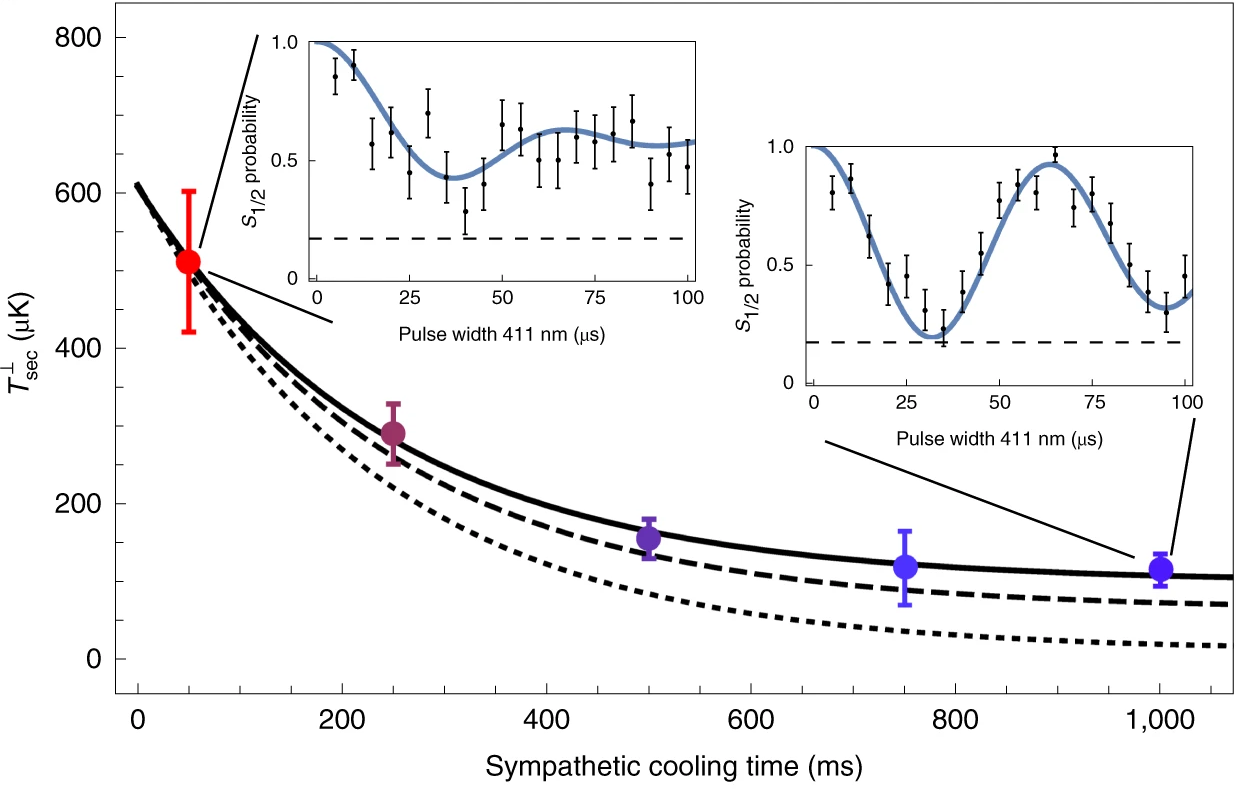}
\caption{Buffer gas cooling to the $s$-wave regime in the Yb$^+$-Li mixture. The data shows the measured secular radial temperature of the ion for various ion-atom interaction time. The solid line represents and exponential fit to the data, while dotted (dashed) lines are molecular dynamics simulations without (with) the time-dependence of the Paul trap. The insets show the Rabi oscillations on the 411\,nm $S_{1/2}\rightarrow$ $D_{5/2}$ transition used to determine the temperature. From~\cite{Feldker2020bgc}}.
\label{BGCnatphys}
%\vspace{-5cm}
\end{figure}
%cooling
The buffer gas cooling of Yb$^+$ outperformed Doppler cooling by a factor of five. The cooling curve of the ion by a 10\,$\mu$K $^6$Li bath with a peak density $n_\text{a}=3.1 (\pm 1.5)\times 10^{10}\,$cm$^{-3}$ is shown in Figure~\ref{BGCnatphys}. The secular radial temperature was obtained by measuring the laser excitation on the 411\,nm $S_{1/2}\rightarrow$ $D_{5/2}$ transition as a function of pulse width. These Rabi oscillations are shown as insets for both the hot and cold ions. Both the frequency and damping of the oscillations change with temperature and by fitting these oscillations to a model assuming a thermal distribution, the mean motional quantum number and subsequently the secular temperature are obtained~\citep{Leibfried:1996, Meir2016doa}. The ion was cooled from $T_\text{sec}^\perp = 600\,\mu$K, which is close to the Doppler temperature of $0.63\,$mK, to a final temperature of $98 (\pm11)\,\mu$K. With the experimental uncertainties, the presence of the ion was not found to influence the atom temperature or number, as determined by time-of-flight analysis and spin-selective absorption imaging.    

% Tion > Tatom
The ion temperature equilibrates to a value which is higher by an order of magnitude than the temperature of the atom bath. This energy difference is confirmed by molecular dynamics simulations~\citep{Fuerst2018por} as shown by the dotted and dashed lines in Figure~\ref{BGCnatphys}. Taking only the secular description of the Paul trap into account (dotted line) a final temperature equal to that of the atom is found. However, when including the time-dependence (micromotion) of the Paul trap, the realistic excess micromotion of the setup and background heating of the ion without the bath, agreement with the data is found as shown by the dashed line. The remaining mismatch in final temperature may be partially explained by temperature overestimation. It was assumed that the dephasing in the Rabi flops was fully caused by the motion of the ion, while other possible effects such as laser frequency and intensity noise were not considered. Another, more tantalizing reason may be the onset of quantum effects~\citep{Oghittu2021doa}. Nonetheless the discrepancy between the ion's and atom's final temperatures has its origin mostly in the micromotion-induced heating caused by the ion-atom interaction, which is typical for ion-atom systems in Paul traps~\citep{Meir2016doa, DeVoe2009pld, Zipkes2011koa, Chen2014ngs, Rouse2017sed, Holtkemeier2016bgc}. Using the same type of simulations, one could show that the final ion temperature reachable by buffer gas cooling could even be a factor of two lower, through trap parameter optimization~\citep{Trimby2022bgc}

\subsection{Collisions and chemistry}
%why ion-atom systems (good control & cold) contribute to this topic and difference collisions vs chemistry
The excellent control and read-out of the ion, makes it a good reaction center with which collisions and chemical reactions can be studied. Collisions can lead to state or momentum changes, whereas chemistry leads to a change in the nature of the components for instance by combining the reactants into a molecule or exchanging an electron between them. Moreover, because ion-atom systems are ultracold, they facilitate the exploration of chemistry and collisions at temperatures well below the mK regime. There, the classical picture of colliding particles and barriers needs to be replaced by a full quantum mechanical treatment and measurements are necessary to shed light at the chemistry that goes on there~\citep{Heazlewood2021tca}. 

% what you need to observe collisions and chemistry
Observation of collisions and chemistry relies on the controlled preparation of the reactants and the detection of the energy, state and/or kind of the products. In ion-atom experiments, the initial state of the atom and ion reactants can be readily prepared and the final state of the products can be read-out through both state-selective atom and ion imaging (see Sec.~\ref{setup}). Moreover the gain or loss in kinetic energy can be measured with ion thermometry or by time-of-flight analysis of the atom cloud. Another method is to look at atom or ion loss from the trap. Particles with a higher energy than the trapping depth of the potential, leave the trap and this can be detected as loss either by measuring the atom number or observing the loss of ion-fluorescence. As ion traps are very deep, the large energy released during an exothermic reaction will commonly not excite the ion out of the trap. However it will give the ion enough kinetic energy which will make it off-resonant with the fluorescence light due to the Doppler shift. See Fig.\ref{MIDimer}a and b for an example fluorescence measurement of a bright and dark ion.  

% formation of new particles
Loss of fluorescence can also point to the creation of a different ion species that can not fluoresce at the given wavelength. For instance a molecular ion is formed or charge exchange leads to the atom becoming ionized and the ion becoming neutral. As long as the mass of the new product is still fulfilling the trapping criteria for the ion-trap, it will remain trapped after the reaction, however the optical control and read-out is lost. The trappability of this other ion depends mainly on its mass and whether that matches the stability criteria for the trap. The outcome of such a reaction can be detected by mass spectrometry (e.g. ~\citep{Harter2013pdo, Schmidt2020msr}), which reveals the mass of the product or through quantum logic spectroscopy~\citep{Schmidt:2005} when co-trapping another ion. 

%Quantum logic spectroscopy  https://arxiv.org/pdf/2107.08441.pdf
Quantum logic spectroscopy~\citep{Wolf:2016, Chou:2017, Sinhal:2020} provides a way to detect product ions of chemical reactions of whom the optical control and read-out is inaccessible, either because the setup does not have the relevant lasers available or the optical detection of this particular product has not yet been shown. The power of this method was demonstrated for isotopes of the Sr$^+$ ion inside a Rb-cloud~\citep{Katz2022qld}. Here the $^{88}$Sr$^+$ was used as the read-out ion. This logic ion was then co-trapped with another ion, called the chemistry ion which is either $^{84}$Sr$^+$,$^{86}$Sr$^+$, $^{87}$Sr$^+$ or another $^{88}$Sr$^+$. The latter is used for calibration of the measurements. Isotopes can be selectively loaded by adjusting the photo-ionization frequency and are sympathetically cooled by the logic ion. The loading of the isotopes can be detected via mass spectrometry~\citep{Drewsen2004nio}. 

%example of Rb-Sr+ 
By preparing the Rb atoms in an excited hyperfine state ($F=2$), the exothermic hyperfine and spin changing reactions between these atoms and various Sr ion isotopes could be studied with quantum logic spectroscopy~\citep{Katz2022qld}. Through an ion-atom collision a Rb atom can relax to the lower hyperfine state ($F=1$) and the energy of the atom decreases by the amount of the hyperfine energy, which is about $k_\text{B}\times$328\,mK for Rb. Because of conservation of spin, the Sr ion will also change its state and energy by about $k_\text{B}\times$1\,mK. The total internal energy release during the reaction $\Delta E=\Delta E_a\pm \Delta E_i$, depends on whether the ion gains or reduces its internal energy. As energy is conserved, the released energy is distributed among the atom and ion as kinetic energy. The ratio of distribution depends on their mass and, in the center-of-mass frame, the ion takes up $\frac{\mu}{m_i} \Delta E$. This energy gets divided over the six motional modes of the trap and corresponds to about $k_\text{B}\times$27\,mK per mode.

%how the logic ion senses the chemistry ion
The increase in the motion of the chemistry ion is sensed by the logic ion as the two are coupled by the Coulomb force. Any motional change in one thus influences the other. The logic ion can than be readily read-out using ion thermometry (see section~\ref{Iondetection}) and in the Sr$^+$-Rb setup this is done by electron shelving. As a result the hyperfine changing collision rate for all isotopes could be determined and interestingly the odd isotope showed a twice as low rate. The precise determination of spin exchange and spin relaxation rates aids the \textit{ab initio} calculations of the molecular potentials of intermediate complexes that can form during the reaction of Rb with Sr$^+$ and gives insight into the dynamics of the system. 

\subsubsection{Molecular ions} 
In ion-atom systems molecular ions can be formed through spontaneous radiative association, photoassociation, magnetoassociation as well as through three-body recombination~\citep{Mohammadi:2021} or chemical reactions between an ion and Feshbach dimers~\citep{Hirzler2022ooc}. Typically these molecular ions are still weakly bound and in highly excited states. This makes them very reactive. Once a molecular ion is formed, they too can collide with the atom cloud and their survival depends on how hard it is to dissociate them either by collisions or assisted by light present in the setup. Especially the far off-resonant light of the optical dipole trap for the atoms can be detrimental to the lifetime of molecular ions. Moreover collisions can lead to electronic and vibrational state changes of the molecular ion. Here we discuss the recent studies on molecular ions performed using ion-atom mixtures with ~$\mu$K atoms in optical dipole traps. However, molecular ions are widely studied e.g. in the field of trapped molecular ions~\citep{Najafian2020iom, Sinhal:2020} as cold molecular ions promise applications in quantum information and precision spectroscopy~\cite{MurPetit:2012, Khanyile2015, Wolf:2016, Chou:2017, Sinhal:2020, Katz2022qld}. 

%BaRb+ paper  https://journals.aps.org/prresearch/abstract/10.1103/PhysRevResearch.3.013196 
For BaRb$^+$, the question of how the molecular ion collides, relaxes, or dissociates within the cloud of atoms was measured by studying the elastic, inelastic and reactive processes~\citep{Mohammadi:2021}. They measured the rates for collisional and radiative relaxation as well as photodissociation, spin-flip collisions, and chemical reactions between this alkaline earth and alkali species. Directly after formation, the collisions between BaRb$^+$ and Rb lead mainly to vibrational relaxation. As the molecular ion gets deeper bound, the dominant mechanisms shifts to radiative relaxation. Other collisional and dissociation processes found, led to the detection of Ba$^+$, Rb$^+$ and Rb$_2^+$. This shows the possible pathways along the lifetime of a molecular ion. 
    
\begin{figure}[!t]
\center
\includegraphics[height=0.6\textwidth]{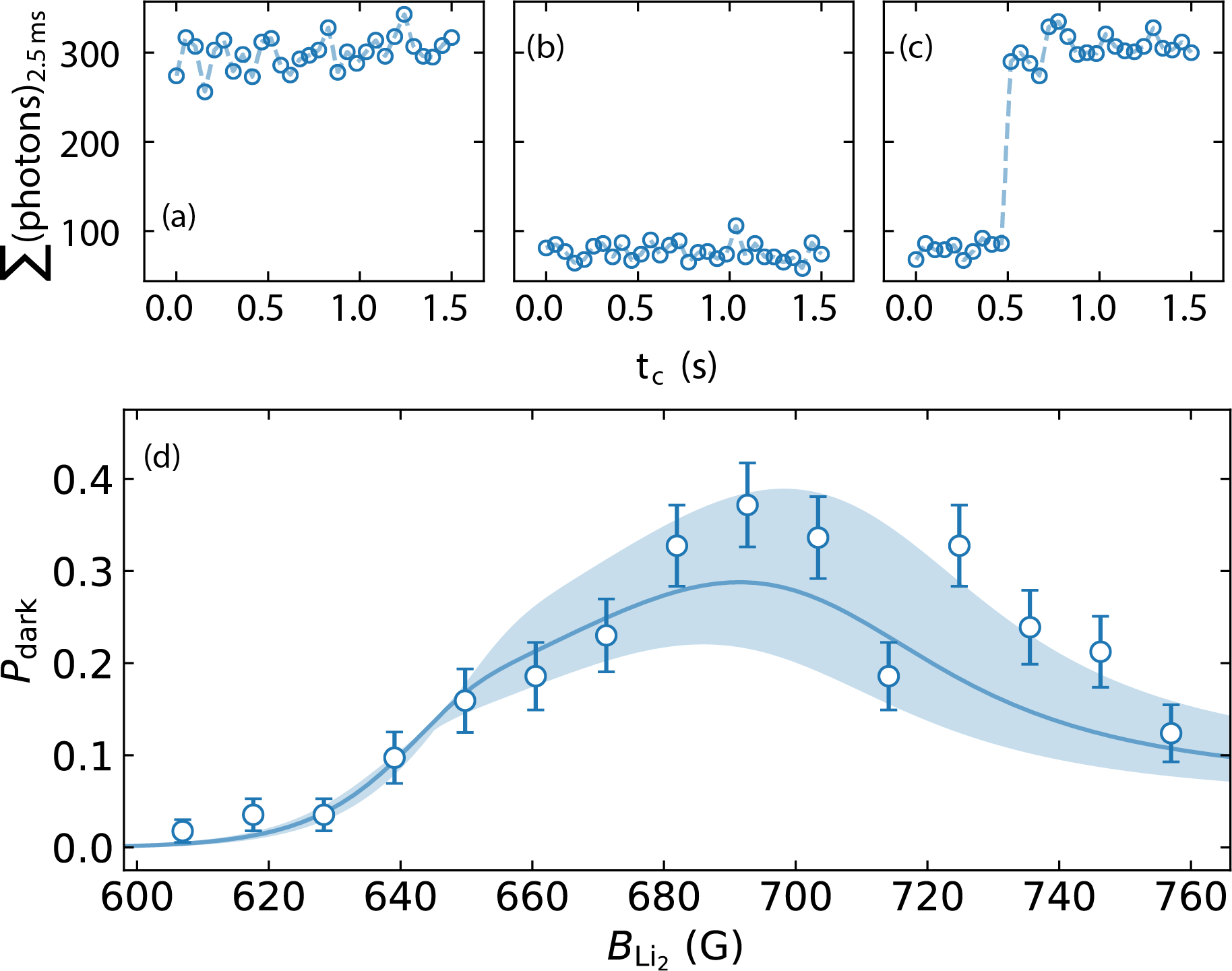}
\caption{Ion interacting with ultracold molecules. (a-c) Ion fluorescence measurements as a function of recooling time, after the ion has interacted for 500\,ms with the atom-dimer bath. The fluorescence is collected for 2.5\,ms. The ion appears bright(a), dark (b) or shows a recooling event(c) where the ion turns bright again. (d) Dark ion probability (events c and d) as a function of the evaporation field $B_{\textrm{Li}_2}$ which varies the numbers of Feshbach dimers. The blue line is a numerical solution to rate equations describing independently the number of dimers based on the atomic evaporation ramp which creates the dimers, the atom density, temperature and $B_{\textrm{Li}_2}$. The shaded regions correspond to a 20\% error in the atom temperature. Adapted from~\citep{Hirzler2022ooc}.}
\label{MIDimer}
%\vspace{-5cm}
\end{figure}

%Recooling events and what we can learn from them. 
The dissociation of the molecular ion into an atom and ion typically leads to the observation of so-called recooling events. See Fig. \ref{MIDimer}c for an example. Here, the fluorescence returns after a certain cooling time $t_\text{c}$. As fluorescence is collected by shining resonant Doppler light in the setup, a `hot' ion can be cooled back to a temperature regime where it will fluoresce again. Because of their kinetic energy, `hot' ions see a Doppler shift with respect to the imaging transition and remain dark until they are cold enough to be resonant with the light. Despite the Doppler shift, the light still Doppler cools the ion. For Doppler cooling being red-detuned with respect to the transition is a prerequisite. The recooling time depends on the energy of the ions and could be traced back to the energy of the molecular ion. However, this requires a good knowledge of all contributions to the (molecular) ion's energy. Nevertheless, these kind of events are used to measure the energy of the ion in so-called single-shot Doppler cooling thermometry~\citep{Meir:2017}.

% Lidimer-MI PRL paper. %Ion-molecule collisions 
A novel way to form a molecular ion is through chemical reactions between the ion and Feshbach dimers~\citep{Hirzler2020ctn}, which was recently observed with Yb$^+$-Li~\citep{Hirzler2022ooc}. Using Feshbach resonances, weakly bound ultracold molecules in high vibrational states, so-called Feshbach dimers, can be created from an ultracold gas. In the $^{174}$Yb$^+$-$^6$Li systems, the reaction Li$_2 + $ Yb$^+ \rightarrow$ LiYb$^+ + $Li was studied by looking at the probability of the ion going dark after interaction with the atom-dimer bath (see Fig.~\ref{MIDimer}). The dimer density was maximally 10\% of the atom density and the ion was prepared in its ground state ($S_{1/2}$). The dark-probability showed an anti-correlation with the atom density, excluding atoms to be taking part in the reaction. This was further confirmed by exciting the ion to the $P_{1/2}$ state and measuring the charge exchange rate, which is a good probe for the local density of the atomic cloud~\citep{Joger:2017}. The creation of the molecular ion was confirmed with mass spectrometry. Similarly as to the BaRb$^+$, recooling events were seen as is depicted in Fig.~\ref{MIDimer}c. The most likely cause here was photodissociation of the molecular ion by the 1064\,nm light of the optical trap holding the atoms and dimers.   

% sensing properties of ion.
The molecular ion formation through ion-molecule collisions was confirmed by varying the number of dimers and demonstrating the sensing properties of an ion probe. Figure~\ref{MIDimer}d shows the remarkable agreement between the probability of molecular ion formation and the number of dimers. The latter can be varied by varying the magnetic field at which the Feshbach dimers are formed through three-body recombination of the atoms. The blue shaded area shows the calculated probability based on atom-data alone and relies upon the assumption that each ion-molecule collision leads to a molecular ion, which agrees with the ion measurements. For a $P_\text{dark}=0.2$ about 50 dimers in a bath of $20 000$ atoms could still be detected. Thus, the ion was used as a probe for trace amounts of Li$_2$ molecules. This shows the sensing capabilities of an ion interacting with a many-body system. 

%next step
The next step for molecular ions in hybrid ion-atom systems would be to resolve the rotational and vibrational states of the molecular ion and gain control on the preparation of specific molecular states. For instance by using quantum logic spectroscopy~\citep{Wolf:2016} with a single ion as the logic ion and the molecular ion as the chemistry ion. These techniques are being developed in the field of trapped molecular ions, e.g. Ca$^+$ and N$_2^+$~\citep{Najafian2020iom, Sinhal:2020}, and could also be applied for molecular ions in an atomic bath. Although the dissociation by the trapping light of the atoms might limit the lifetime, this can be prevented by directly discarding the bath after the molecular ion was formed. A way to control the energy of the molecular ion is by controlling the binding energy of the Feshbach dimers that play a role in its creation~\citep{Hirzler2020ctn}. The cold controlled molecular ions could be used in combination with precision spectroscopy in the search of new physics~\citep{Safronova2018sfn} or for laboratory astrochemistry.  Furthermore, the ion-molecule collisions which can be studied within hybrid ion-atom systems give access to a new regime, as the molecules created from the ultracold gas are in the $\mu$K regime. Thereby they can complement the ion-molecule research done with Coulomb crystals~\citep{Heazlewood2015ltk, Heazlewood2019cic} and molecular beams~\citep{Meyer2017imr}.  

\subsubsection{Other reactions}
%Excitation exchange 2020 https://journals.aps.org/pra/pdf/10.1103/PhysRevA.102.031301 and https://journals.aps.org/pra/abstract/10.1103/PhysRevA.103.032805
When the ion is prepared in an excited state, for most ion-atom mixtures, charge-exchange reactions dominate the inelastic collisions. However, for Sr$^+$-Rb two other mechanisms were measured to play a more important role. These were electronic excitation exchange and spin-orbit change~\citep{Benshlomi2020doo}. The reason lies in the lack of avoided crossings between the potential energy curves for Sr$^+$-Rb and Rb$^+$-Sr. Distinction between the two mechanisms could be made by using single-shot Doppler cooling thermometry~\citep{Meir:2017} on the ion and measuring the ion's energy after a few collisions with the atoms. An excited Sr$^+$($D_{5/2}$ or $D_{3/2}$) was measured to rapidly decay to Sr$^+$($S_{1/2}$) by simultaneously exciting Rb from the $S_{1/2}$ to the $P_{1/2}$ state and releasing the remaining energy into motion. Looking at the energy distribution of the ion, they could distinguish between the two mechanisms, as they led to distinguishable amounts of energy being released. The cross section of these processes for varying collision energy, was subsequently measured by shuttling the atoms with an optical lattice through the ion. Depending on the speed with which the lattice was moved the collision energy could be tuned from 0.2-12\,mK$\times k_\text{B}$ with a resolution of 200$\,\mu$K$\times k_\text{B}$~\citep{BenShlomi2021her}. The cross section was shown to follow the expected Langevin scaling of $E_\text{col}^{-1/2}$. By improving the long-term stability of this measurement method, significant deviations from this scaling could be observed, which would point towards quantum resonances (see Sec.~\ref{QE}). 

%Swap Cooling 2021 https://iopscience.iop.org/article/10.1088/1367-2630/ac0575
Charge exchange reactions are especially interesting when discussing homo-nuclear ion-atom mixtures, where the charge gets swapped between the ion and atom, i.e. $A+A^+\rightarrow A^++A$. In these systems, as a bonus, the charge exchange leads to swap cooling and provides a way to create a cold ion in a single step~\citep{Ravi2012cas}. This was directly observed by comparing cooling in a mixture of Rb-Rb$^+$ to Rb-Ba$^+$~\citep{Mahdian2021doo}, where the atom bath had a temperature of $T_\text{a} = 600\,$nK. In the homo-nuclear mixture a fast cooling of the ion was seen, whereas in the heteronuclear mixture this was absent. The mechanism behind the so-called swap cooling is resonant charge exchange and happens through glancing collisions, limiting the application to ions with temperatures above the 200\,K. Below the 200\,K, the buffer gas cooling due to Langevin collisions dominates and determines the cooling dynamics. 

\subsection{Quantum effects}
\label{QE}
Although attaining ion-atom collision energies close to the $s$-wave regime remains challenging, it is possible. In this regime, only a single partial wave describes the interaction. Therefore, the interaction can be described by a single parameter, i.e. the ion-atom scattering length $a$. Furthermore, $s$-wave Feshbach resonances (FRs) can then be used to tune the ion-atom interaction~\citep{Idziaszek2011mqd, Gacesa2017cti, Tomza2015cib}, in analogy to ultracold quantum gases where FRs are the workhorse of the field~\citep{Chin2010fri}.  

The characteristic energy below which the $s$-wave regime is reached is given by $E_s=\frac{\hbar^4}{4\mu^2C_4}$. To know that the quantum regime is attained one can either look for quantum effects or measure the collision energy. The latter requires a well-rounded measurement of all energy contributions to the total ion energy (secular, IMM, EMM) and the atom bath. Subsequently the collision energy can be calculated and compared to the characteristic energy scale $E_s$.

With a favorable ion-atom mass ratio and good control over stray electric field the quantum regime can be accessed and quantum effects have been seen. The first observation saw a deviation from the classical Langevin rate for spin-exchange for low collision energies with Yb$^+$/Li. In Ba$^+$/Li Feshbach resonances were seen when approaching the $s$-wave limit. We discuss these discoveries below together with the effects that can still be observed in the mK regime.  

\subsubsection{In the mK regime}
Reminiscent features of the quantum regime can be observed in the mK regime when looking at the spin dependence of ion-atom collisions.  This effect is called partial-wave phase locking and was observed for both Sr$^+$-Rb and Yb$^+$-Li~\citep{Fuerst2018doa, Sikorsky2018plb, Cote2018sot}. When ions and atoms collide, spin exchange or spin relaxation can happen and the rate of these types of collisions depends on the initial state of the ion and the atom cloud. The spin exchange collision dynamics is given by the singlet and triplet scattering lengths as well as the atomic polarizability because of the short-range nature of the spin exchange interaction. Even at high temperatures, i.e. when multiple partial waves are present, it is insensitive to the centrifugal barrier~\citep{Sikorsky2018plb, Cote2018sot}. 

The spin-exchange measurements allow for an estimation of the difference between the singlet and triplet scattering lengths, which can be used to predict the magnetic fields at which the ion-atom Feshbach resonances are expected to occur~\citep{Chin2010fri}. Extending the measured spin dependent rates to different isotope combinations, enables a good comparison with coupled-channel scattering calculations. From matching theory and experiment, then an estimate for the triplet ($a_T$) and singlet ($a_S$) scattering lengths can be inferred. For Yb$^+$-Li, this resulted in the prediction of the scattering lengths being large and opposite in sign, e.g. $a_T\sim -a_S \sim R_4$~\citep{Fuerst2018doa}. This is in good agreement with the later measured values of $a_S=1.2(0.3)R_4$ and $a_T=-1.5(0.7)R_4$~\citep{Feldker2020bgc}, when the $s$-wave regime was probed.   

\subsubsection{Spin exchange} % https://www.nature.com/articles/s41567-019-0772-5#ref-CR2
With the coldest system of Yb$^+$-Li, quantum effects in ion-atom collisions were observed by studying the spin-exchange dynamics as a function of collision energy. The latter was tuned by deliberately adding excess micromotion to the ion using compensation electrodes. This increases the ion's energy and thus the collision energy. In the classical regime, the spin exchange rate is expected to be proportional to the Langevin collision rate $\Gamma= n_\text{a} \sigma v$, which depends on the atom density $n_\text{a}$, the relative velocity $v$ and the collision cross-section $\sigma$ and is independent of the collision energy (see section~\ref{Langevin}). However in the quantum regime, deviations are expected. Here, the wave-nature of the interactions matters. This leads to quantization of the angular momemtum $l$ and the scattering of the particles is described by partial waves with $l=0,1,2,...$, called $s,p,d,...$-waves. With each partial wave of $l>0$, a centrifugal barrier can be associated. Particles can both tunnel through this barrier or be reflected from it and the likelihood that this happens depends on the collision energy. Therefore, shape resonances and structure are expected to show up in the collision energy dependence of the spin-exchange rate. 

\begin{figure}[!t]
\center
\includegraphics[height=0.6\textwidth]{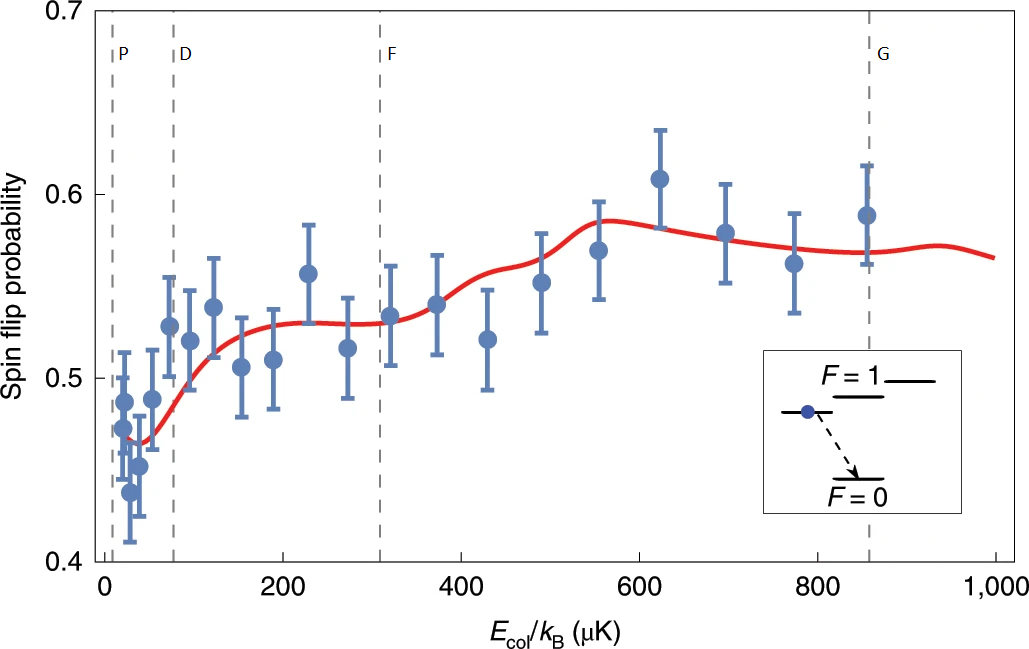}
\caption{Deviation from energy-independent classical Langevin rate in spin exchange measurements. Shown is the probability to detect the Ytterbium ion spin in the $\ket{F=0, m_F=0}$ state after preparing it in the \ket{1,-1} state (see inset) and an interaction time of ~10\,ms with a cold Lithium atomic cloud. The error bars reflect the quantum projection noise. The red line is the theoretical prediction from quantum scattering calculations taking into account the energy distribution of the ion. The energy barriers of $l=1,2,3,4$ angular momentum states are indicated by the vertical dashed line. From~\cite{Feldker2020bgc}.}
\label{Spinexchange}
%\vspace{-5cm}
\end{figure}

The measured spin-exchange probability versus collision energy is shown in figure~\ref{Spinexchange} after a 10-ms interaction with an atom cloud of $n_\text{a}= 21(\pm)\times10^{15}$\,m$^{-3}$. A clear deviation from an energy-independent Langevin-rate can be seen. The location of the centrifugal barriers for the various partial waves are depicted by the dashed vertical lines. The red theory line represents a best fit to the data of multichannel quantum scattering calculations which take into account the complete description of molecular and hyperfine structures as well as the collision energy distribution of the ion. As fit parameters the triplet and singlet scattering lengths, i.e. $a_s=1.2(0.3)$ and $a_T=-1.5(0.7)R_4$ with $R_4=70$\,nm, and an average number of 1.2 Langevin collisions were found. 

The observation of this quantum effect, demonstrated that the Yb$^+$-Li mixture has access to the quantum regime. A next step is the search for Feshbach resonances, which have been predicted~\citep{Idziaszek2011mqd, Tomza2015cib}. These studies would be feasible when increasing the density of the atom cloud, which decreases the cycling time of the measurements and makes the detection of Feshbach resonances possible. 

\subsubsection{Feshbach resonances} %https://www.nature.com/articles/s41586-021-04112-y
With Ba$^+$-Li, ion-atom Feshbach resonances were detected~\citep{Weckesser2021oof} by using a high atomic density spin-polarized Li gas with $n_\text{a}=33\times 10^{17}$\,m$^{-3}$. These resonances occur due to coupling between a molecular potential and a background continuum of states. Together with only a single partial wave contributing to the collision, a Feshbach resonance gives the prospects of controlled magnetic-tuning of ion-atom interactions. In neutral-neutral atom mixtures~\citep{Chin2010fri}, this control and tunability of interactions already led to a wide exploration of both many-body~\citep{Bloch2008mbp} and few-body physics~\citep{Wang2013ufb}. Around the Feshbach resonance's center, the interaction increases and changes sign. Therefore the signatures are an increase of the collision rate between the atom and ion, resulting in ion loss or an increase in cooling capacity of the atom buffer gas.  

\begin{figure}[!t]
\center
\includegraphics[height=0.4\textwidth]{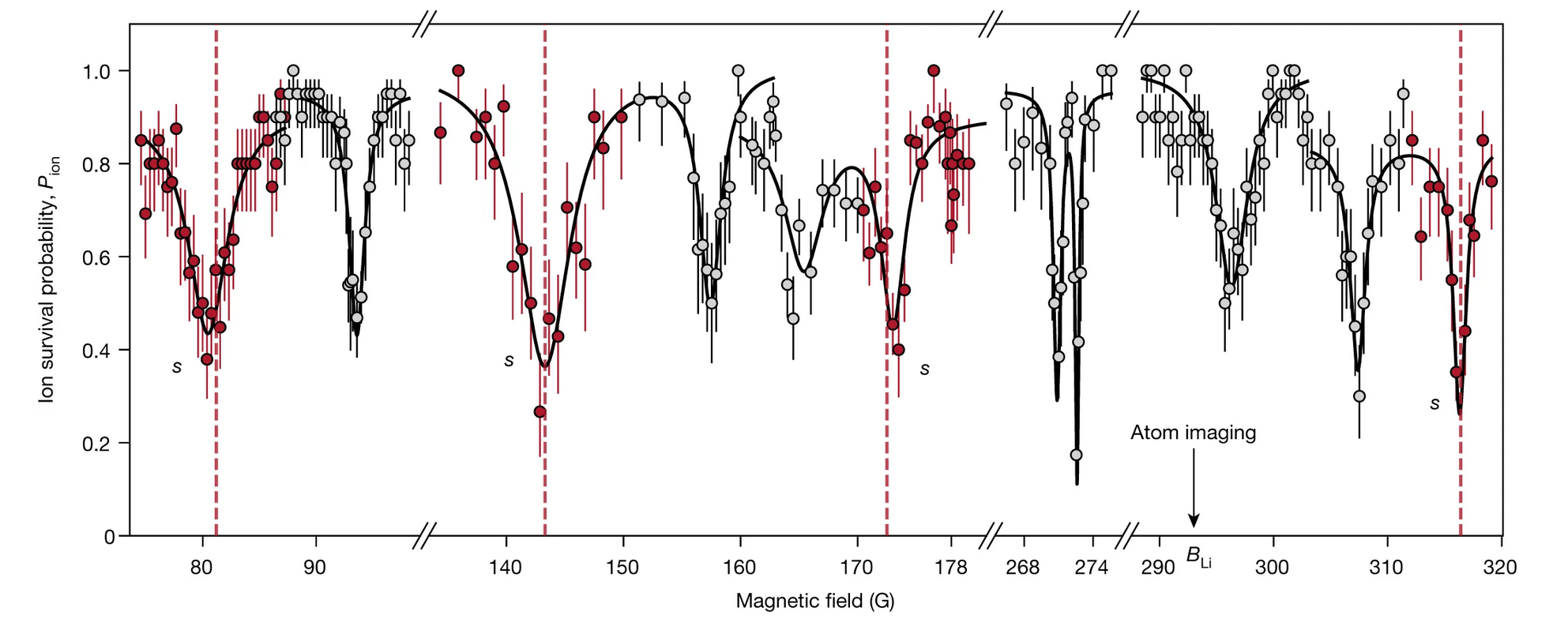}
\caption{Feshbach resonances in the Ba$^+$-Li mixture. The ion survival probability as a function of magnetic field is shown. The loss features are fit with Lorentzians and in red the $s$-wave resonances are shown. The vertical dashed lines correspond to the theoretical prediction of the Feshbach resonance center. From~\cite{Weckesser2021oof}.} 

\label{FR}
%\vspace{-5cm}
\end{figure}

The measurements looked at the ion survival probability after interaction with the atoms for various magnetic fields. They are shown in Figure~\ref{FR}. At a Feshbach resonance,  three-body recombination is greatly enhanced, which leads to the observed ion-loss. In the magnetic field range of 705-330\,G, 11 loss features where seen. From fitting a Lorentzian to the features, the center and full-width-half-maximum was determined. The overall collision energy of the system was below 90$\,\mu$K, which allows for both $s$-, $p$- and $d$-wave resonances to be detected. An $l$-wave resonances, is a Feshbach resonance where an $l$-wave molecular level couples resonantly to the entrance collision channel.

Four resonances, shown in red, could be attributed to $s$-wave Feshbach resonances through comparison with theoretical predictions (dashed vertical lines). The theory is based on ab initio electronic structure and multichannel quantum scattering calculations. In total, only 5 resonances were predicted when considering spin-projection conserving electronic interaction ($\delta m_F=0$). However, the $s$-wave resonances turned out to all be associated with $\delta m_F=1$ interactions. This shows that coupling terms like second-order spin-orbit coupling need to be considered, increasing the number of expected resonances~\citep{Ticknor2004mso}. The theory can predict the number of resonances, the distance between them and their width. However, it does require experimental input to determine the exact location of the Feshbach resonances. Combining the data with the theory, the singlet and triplet scattering length was obtained, which pins down the Feshbach resonance locations. For Ba$^+$-Li, $a_S=0.236R_4$ and $a_T=-0.053R_4$, with $R_4=69$\,nm was found~\citep{Weckesser2021oof}. The assignment of the other resonances, requires more theory investigation. 

To further explore the found Feshbach resonances, the three-body recombination and the cooling properties of the gas were probed. Close to the $s$-wave FR at 296.13\,G, the loss rate of the ion was determined for varying atom densities. It was found to follow $\Gamma_\text{Loss}\propto n_\text{a}^2$, which confirmed that the observed ion-loss was caused by three-body recombination. Moreover, the increased cooling of the ion by the atoms was observed by determining the trapping probability of the ion in an optical trap for varying magnetic fields around the FR center. The shallow potential of the optical trap only holds the ion if it is cold enough, compared to its trap depth. By holding the ion for a fixed interaction time in the cold atom bath, a different trapping probability was found for varying magnetic fields. For these observations the stray electric fields were compensated down to 0.003 V/m.

The next steps would be to use the tunability of the ion-atom interaction through their Feshbach resonance to explore various interaction regimes or even fully switch off the interaction. Furthermore, a Feshbach resonance has a bound molecular state on the repulsive side, which by ramping across the resonance can be used for the association of molecular ions from the ion and the atoms themselves. Detecting these molecular ions would be an important step forward too. For Ba$^+$-Li, these studies would be enabled by freezing out the other partial waves contributions and reaching the $s$-wave regime.

\subsection{Ion transport}
% Motivation; why transport in a BEC?
Ion-atom collisions, both in the classical and quantum regime, are especially of interest when looking at charge transport. An ion moving through a dense environment will interact with its surroundings, which affects the ion's motion and direction. Therefore to understand the dynamics of an ion travelling through a medium, insight into the collisions behavior is important. Here, experiments in hybrid ion-atom systems can contribute, especially when combined with a good spatial read-out of the ion's location in the cloud. In a radio-frequency or optical trap, the ion's motion is controlled by the trapping potential. However a free ion in an atom cloud can be created by photo-ionizing an atom of the cloud via a Rydberg excitation. 

% how to get the ion % exp. setup and measurements
Using a Rb$^+$-Rb mixture, the transport of a single ion inside a Bose-Einstein condensate (BEC) could be measured~\citep{Dieterle2021toa} for the timescale of tens of microseconds.  A single rubidium ion was created by applying a laser pulse, leading to a Rydberg excitation of a Rubidium atom. Subsequently, a small dc electric field pulse ionized the Rydberg atom. The two-step process leads to the deterministic creation of a single ion, because of a strong Rydberg blockade~\citep{Balewski2013cas, Kleinbach2018iii, Engel2018oor}. The free ion was not confined by any external electric or optical potential and had an initial energy of about $50\,\mu$K, coming from the small electric field that creates it. 
To study the transport, the ion was subject to an external force coming from a tunable homogenous electric field of 1-6\,mV/cm. For a period of 1-20\,$\mu$s, the free Rb$^+$ ion was transported through the Rb BEC ($n_\text{a}=4\times10^{14}$cm$^{-3}$). The pulling of the field accelerates the ion for a given time after which the location of the ion is detected. The detection happens by turning on a strong electric field and accelerating the ion towards a multichannel plate through an ion lens. By measuring the time it took the ion to arrive at the detector after turning on the read-out fields, the location of the ion in the cloud was found. In comparison, the ion was transported through a thermal gas with a 40 times lower density.   

% Results
The ion transport in the BEC showed diffusive behavior, even though frictionless behavior could have been expected because of the superfluidity of the BEC. For the 40 times sparser gas, the transport was ballistic. This shows that frequent ion-atom collisions lead to the diffusive transport observed in the BEC, which was further confirmed by numerical simulations. Taking into account the variable density that the ion encounters when travelling throught the BEC, the ion's motion was modeled by subsequent elastic Langevin collisions with ballistic motion in between~\citep{Zipkes2011koa, Chen2014ngs}. The ion-atom Langevin scattering rate $\gamma_L= 2\pi n_\text{a}\sqrt{C_4/\mu}$ set the time in-between collisions events. Furthermore, glancing collisions were not included and given the good agreement between data and simulations, were found to have a negligible role in the transport dynamics.

% Ion-dimers 
Another mechanism encountered in the Rb$^+$-Rb measurements, was three-body recombination via Rb$^+$ + Rb + Rb $\rightarrow$ Rb$_2^+$ + Rb~\citep{Dieterle2020icd}. Because of the heavier mass, Rb$_2^+$ arrives later at the detector than Rb$^+$ and the two ions can be discriminated. The electric field could also be used to destroy the ion dimer. The amount of electric field needed for the dissociation reaction (Rb$_2^+$ + $\mathcal{E} \rightarrow$ Rb$^+$ + Rb) to happen, was found to be a good measure of the binding energy of the dimer. The binding energy of the dimer could be read-out, by measuring the ratio of ion and ion-dimers detected for varying strengths of the electric field~\citep{Dieterle2020icd}. The longer the ion-dimer resided within the atom cloud, the more it became deeply bound. This points to inelastic secondary collisions with the Rb atoms as the likely cause of the relaxation. 

%Future
Looking at sub-microsecond timescales might provide access to the regime where quantum effects dominate and only a few partial waves contribute to the collisions that describe the transport of the ion. By improving the control of the electric fields and the spatial resolution of the detection scheme, single collisions could be detectable. This opens the possibility to probe the quantum regime of ion impurity transport~\citep{Cote2000fcm}, formation of mesoscopic molecular ions~\citep{Cote2002mmi} and more specifically charged polaron physics~\citep{Casteels2011ppo, Astrakharchik2021ipi, Christensen2021cpa, Oghittu2021doa}. Here, high resolution microscope techniques and momentum spectroscopy as developed for quantum gases could be useful, e.g.~\citep{Veit2021pim, Geppert2021adi}.

\section{Prospects}
With the observation of quantum effects in the hybrid ion-atom systems, exciting possibilities for further exploration lie ahead. For large mass-imbalanced ion-atom systems the tunability of ion-atom interactions is within reach, which enables full control of the ion-atom collisional properties and creates a versatile platform for state-to-state quantum chemistry, molecular ion creation and detection as well as for quantum simulation. However, there is still room to gain, as the colder the ion as well as the atomic gas, the better the control over the inter and intra-species interactions. Moreover, for the more equal mass ion-atom systems the quest remains on how the quantum regime can be attained and probed. Here, we briefly outline the experimental directions and prospects of what might be next. 

%Exp. setup requirements
From the experimental perspective of probing the quantum regime, two angles are important. The development of new techniques to better control and probe the current systems at hand as well as the progress in new designs which facilitate other types of ion-atom systems to be studied at colder temperatures. Although for now the Paul trap offers the most direct way to the quantum regime, other trap designs are important for bringing more ion-atom systems to the quantum regime. Here, combinations of optical traps, tweezers and combinations of static fields with optical fields are all on the table, e.g.~\citep{Schaetz2017tia, Schmidt2020otf, Perego2020eoi, Karpa2021ioi}. Each ion-atom combination provides it own set of promises and challenges for quantum simulation. In current setups the experimental challenge lies in simplifying and standardizing the overall cooling, trapping and detection sequences, in order to improve their stability and shorten the run time of the experiments. This provides room for extending the measurement cycle with additional cooling or trapping steps to further decrease the collision energy and increase the density to reach deeper in the quantum regime. There, the ion-atom interaction can be tuned independently and becomes an experimental knob to turn.  

% Quantum simulation, impurity physics
With control come the opportunities of using the ion-atom mixtures as test beds for quantum simulation of charged impurity physics. Of special interest is the ionic polaron, which unlike its neutral atom counterpart, is not short-ranged. This causes the typical polaron picture to break down and measuring the ionic polarons properties would shed light on this quasiparticle of intermediate range and its transport properties~\citep{Casteels2011ppo, Astrakharchik2021ipi, Christensen2021cpa, Oghittu2021doa, Christensen2022mii}. For the transport studies improving the time-resolution to look at shorter timescales is an important premise, as here the quantum effects play a role. The ionic impurity-bath studies are also interesting in the light of quantum information. Especially to see what happens to qubits in quantum baths as for example, the measurement of decoherence of a spin qubit in spin-polarized Bose-Einstein condensate~\citep{Ratschbacher2013doa}. It would be interesting to see what happens in a degenerate Fermi gas and for varying ion-atom interaction strengths. Moreover, by going for large systems of a few ions inside a bath, the prospects appear to look into induced-interactions mediated by the bath~\citep{ding2022mib}

%ultracold quantum chemistry and molecular ions
Not only from the many-body aspect, but also from the few-body perspective the impurity in a bath is of interest~\citep{PerezRios2021cca}. There is much more to explore on the front of ultracold chemistry. Especially when looking into the creation of molecular ions, promises lie ahead on getting a better grip on the product-state distribution. For instance, by resolving the rotational and vibrational states of the molecular ion and gaining control on the preparation of specific molecular states. Probing the energy of the molecular ion right after creation by e.g. quantum logic spectroscopy, would give insights into the reaction path and if there are possible ways to influence it. 
Here the better read-out and protection of molecular ions or in-between product states would be greatly beneficial.

%More complex baths. 
Moreover to extend the ion-bath studies to different regimes, enhancing the complexity of the ion-atom system under study is another direction worth pursuing. For instance, an ion in a bath of dipolar atoms or molecules, would open the possibility to look into baths which have other controllable interactions beyond the van der Waals interaction. The ion can be used as a sensor of the many-body properties of the complex bath. The power of ion sensing was already demonstrated by measurements involving an ion in BEC, which revealed that the ion could measure the density profile of the atom cloud~\citep{Schmid2010doa, Zipkes2010ats} or the atom-dimer creation by studying chemical reactions~\citep{Hirzler2022ooc}. Here, one can already use the tunability of the bath, to sense different many-body environments with the ion. However, simultaneous control over the ion-atom interaction would further increase the flexibility of the system. So far most ion-atom systems with an ultracold bath only employ single-atomic ions which are singly-charged. This can be extended to probe the interactions between a bath and multiple-charged ions or molecular ions. Several ion-MOT experiments exist in this direction~\citep{Hassan2022adi, Doerfler2019lrv, Puri2019rbi} and they could advance to the few $\mu$K regime for the atom bath by implementing an optical dipole trap and evaporative cooling. Also the various Rydberg-bath hybrid systems~\citep{Engel2018oor,Ewald:2019,Haze:2019,Deiss2021lra,Zuber2021sio, Mokhberi2020tri} come to mind to shed light on the interaction of ions within a complex bath. 

All together these prospects which rely on the experimental control, design and ingenuity yet to come, will benefit our knowledge of ion-atom interactions from both the many-body and few-body perspectives. This is requisite to define the possibilities of ion-atom hybrid systems not only as quantum simulators but also for their applications for quantum technologies.

\section*{Acknowledgements}
\addcontentsline{toc}{section}{Acknowledgements}
We are grateful to Eleanor Trimby for running the simulations of Table~\ref{sims} and careful reading of the manuscript. We thank Susanne Yelin, Leon Karpa, Jes\'us~P\'erez-R\'{i}os and Tobias Sch\"atz, Michał Tomza and Florian Meinert for comments on the manuscript. This work is supported by the Netherlands Organization for Scientific Research (Start-up grant 740.018.008, and Vrije Programma 680.92.18.05 (R.G.)) and by the funding from the European Unions Horizon 2020 research and innovation programme under the Marie Sklodowska-Curie grant agreement No 895473 (R.S.L.).

%\section*{References}
\addcontentsline{toc}{section}{References}
\bibliographystyle{elsarticle-harv.bst}

\bibliography{reviewbib}

\begin{thebibliography}{200}
\expandafter\ifx\csname natexlab\endcsname\relax\def\natexlab#1{#1}\fi
\providecommand{\url}[1]{\texttt{#1}}
\providecommand{\href}[2]{#2}
\providecommand{\path}[1]{#1}
\providecommand{\DOIprefix}{doi:}
\providecommand{\ArXivprefix}{arXiv:}
\providecommand{\URLprefix}{URL: }
\providecommand{\Pubmedprefix}{pmid:}
\providecommand{\doi}[1]{\href{http://dx.doi.org/#1}{\path{#1}}}
\providecommand{\Pubmed}[1]{\href{pmid:#1}{\path{#1}}}
\providecommand{\bibinfo}[2]{#2}
\ifx\xfnm\relax \def\xfnm[#1]{\unskip,\space#1}\fi
%Type = Article
\bibitem[{Alheit et~al.(1995)Alheit, Hennig, Morgenstern, Vedel and
  Werth}]{Alheit1995OIP}
\bibinfo{author}{Alheit, R.}, \bibinfo{author}{Hennig, C.},
  \bibinfo{author}{Morgenstern, R.}, \bibinfo{author}{Vedel, F.},
  \bibinfo{author}{Werth, G.}, \bibinfo{year}{1995}.
\newblock \bibinfo{journal}{Applied Physics B} \bibinfo{volume}{61}.
\newblock \URLprefix \url{https://doi.org/10.1007/BF01082047}.
%Type = Article
\bibitem[{Astrakharchik et~al.(2021)Astrakharchik, Ardila, Schmidt, Jachymski
  and Negretti}]{Astrakharchik2021ipi}
\bibinfo{author}{Astrakharchik, G.E.}, \bibinfo{author}{Ardila, L.A.P.},
  \bibinfo{author}{Schmidt, R.}, \bibinfo{author}{Jachymski, K.},
  \bibinfo{author}{Negretti, A.}, \bibinfo{year}{2021}.
\newblock \bibinfo{title}{Ionic polaron in a {B}ose-{E}instein condensate}.
\newblock \bibinfo{journal}{Comm. Phys.} \bibinfo{volume}{4},
  \bibinfo{pages}{94}.
\newblock \URLprefix \url{https://doi.org/10.1038/s42005-021-00597-1}.
%Type = Article
\bibitem[{Asvany and Schlemmer(2009)}]{Asvany2009NSK}
\bibinfo{author}{Asvany, O.}, \bibinfo{author}{Schlemmer, S.},
  \bibinfo{year}{2009}.
\newblock \bibinfo{journal}{International Journal of Mass Spectrometry}
  \bibinfo{volume}{279}.
\newblock \URLprefix \url{https://doi.org/10.1016/j.ijms.2008.10.022}.
%Type = Article
\bibitem[{Bahrami et~al.(2019)Bahrami, Müller, Drechsler, Joger, Gerritsma and
  Schmidt-Kaler}]{Bahrami2019ooa}
\bibinfo{author}{Bahrami, A.}, \bibinfo{author}{Müller, M.},
  \bibinfo{author}{Drechsler, M.}, \bibinfo{author}{Joger, J.},
  \bibinfo{author}{Gerritsma, R.}, \bibinfo{author}{Schmidt-Kaler, F.},
  \bibinfo{year}{2019}.
\newblock \bibinfo{title}{Operation of a microfabricated planar ion-trap for
  studies of a {Y}b$^+$–{R}b hybrid quantum system}.
\newblock \bibinfo{journal}{physica status solidi (b)} \bibinfo{volume}{256},
  \bibinfo{pages}{1800647}.
\newblock \URLprefix
  \url{https://onlinelibrary.wiley.com/doi/abs/10.1002/pssb.201800647},
  \DOIprefix\doi{https://doi.org/10.1002/pssb.201800647}.
%Type = Article
\bibitem[{Balewski et~al.(2013)Balewski, Krupp, Gaj, Peter, B\"uchler, L\"ow,
  Hofferberth and Pfau}]{Balewski2013cas}
\bibinfo{author}{Balewski, J.B.}, \bibinfo{author}{Krupp, A.T.},
  \bibinfo{author}{Gaj, A.}, \bibinfo{author}{Peter, D.},
  \bibinfo{author}{B\"uchler, H.P.}, \bibinfo{author}{L\"ow, R.},
  \bibinfo{author}{Hofferberth, S.}, \bibinfo{author}{Pfau, T.},
  \bibinfo{year}{2013}.
\newblock \bibinfo{title}{Coupling a single electron to a {B}ose-{E}instein
  condensate}.
\newblock \bibinfo{journal}{Nature} \bibinfo{volume}{502},
  \bibinfo{pages}{664}.
\newblock \DOIprefix\doi{https://doi.org/10.1038/nature12592}.
%Type = Misc
\bibitem[{Barakhshan et~al.()Barakhshan, Marrs, Bhosale, Arora, Eigenmann and
  Safronova}]{UDportal}
\bibinfo{author}{Barakhshan, P.}, \bibinfo{author}{Marrs, A.},
  \bibinfo{author}{Bhosale, A.}, \bibinfo{author}{Arora, B.},
  \bibinfo{author}{Eigenmann, R.}, \bibinfo{author}{Safronova, M.S.}, .
\newblock \bibinfo{howpublished}{{\textit{Portal for High-Precision Atomic Data
  and Computation}} (version 2.0). University of Delaware, Newark, DE, USA.
  URL: {https://www.udel.edu/atom} [February 2022].}
\newblock \URLprefix \url{https://www.udel.edu/atom}.
%Type = Article
\bibitem[{Ben-shlomi et~al.(2021)Ben-shlomi, Pinkas, Meir, Sikorsky, Katz,
  Akerman and Ozeri}]{BenShlomi2021her}
\bibinfo{author}{Ben-shlomi, R.}, \bibinfo{author}{Pinkas, M.},
  \bibinfo{author}{Meir, Z.}, \bibinfo{author}{Sikorsky, T.},
  \bibinfo{author}{Katz, O.}, \bibinfo{author}{Akerman, N.},
  \bibinfo{author}{Ozeri, R.}, \bibinfo{year}{2021}.
\newblock \bibinfo{title}{High-energy-resolution measurements of an
  ultracold-atom--ion collisional cross section}.
\newblock \bibinfo{journal}{Phys. Rev. A} \bibinfo{volume}{103},
  \bibinfo{pages}{032805}.
\newblock \URLprefix
  \url{https://link.aps.org/doi/10.1103/PhysRevA.103.032805},
  \DOIprefix\doi{10.1103/PhysRevA.103.032805}.
%Type = Article
\bibitem[{Ben-shlomi et~al.(2020)Ben-shlomi, Vexiau, Meir, Sikorsky, Akerman,
  Pinkas, Dulieu and Ozeri}]{Benshlomi2020doo}
\bibinfo{author}{Ben-shlomi, R.}, \bibinfo{author}{Vexiau, R.},
  \bibinfo{author}{Meir, Z.}, \bibinfo{author}{Sikorsky, T.},
  \bibinfo{author}{Akerman, N.}, \bibinfo{author}{Pinkas, M.},
  \bibinfo{author}{Dulieu, O.}, \bibinfo{author}{Ozeri, R.},
  \bibinfo{year}{2020}.
\newblock \bibinfo{title}{Direct observation of ultracold atom-ion excitation
  exchange}.
\newblock \bibinfo{journal}{Phys. Rev. A} \bibinfo{volume}{102},
  \bibinfo{pages}{031301}.
\newblock \URLprefix
  \url{https://link.aps.org/doi/10.1103/PhysRevA.102.031301},
  \DOIprefix\doi{10.1103/PhysRevA.102.031301}.
%Type = Article
\bibitem[{Berkeland et~al.(1998)Berkeland, Miller, Bergquist, Itano and
  Wineland}]{Berkeland:1998}
\bibinfo{author}{Berkeland, D.J.}, \bibinfo{author}{Miller, J.D.},
  \bibinfo{author}{Bergquist, J.C.}, \bibinfo{author}{Itano, W.M.},
  \bibinfo{author}{Wineland, D.J.}, \bibinfo{year}{1998}.
\newblock \bibinfo{title}{Minimization of ion micromotion in a {P}aul trap}.
\newblock \bibinfo{journal}{J.~App.~Phys.} \bibinfo{volume}{83},
  \bibinfo{pages}{5025--5033}.
\newblock \DOIprefix\doi{10.1063/1.367318}.
%Type = Article
\bibitem[{Bissbort et~al.(2013)Bissbort, Cocks, Negretti, Idziaszek, Calarco,
  Schmidt-Kaler, Hofstetter and Gerritsma}]{Bissbort:2013}
\bibinfo{author}{Bissbort, U.}, \bibinfo{author}{Cocks, D.},
  \bibinfo{author}{Negretti, A.}, \bibinfo{author}{Idziaszek, Z.},
  \bibinfo{author}{Calarco, T.}, \bibinfo{author}{Schmidt-Kaler, F.},
  \bibinfo{author}{Hofstetter, W.}, \bibinfo{author}{Gerritsma, R.},
  \bibinfo{year}{2013}.
\newblock \bibinfo{title}{Emulating solid-state physics with a hybrid system of
  ultracold ions and atoms}.
\newblock \bibinfo{journal}{Phys.~Rev.~Lett.} \bibinfo{volume}{111},
  \bibinfo{pages}{080501}.
\newblock \DOIprefix\doi{10.1103/PhysRevLett.111.080501}.
%Type = Article
\bibitem[{Bloch et~al.(2008)Bloch, Dalibard and Zwerger}]{Bloch2008mbp}
\bibinfo{author}{Bloch, I.}, \bibinfo{author}{Dalibard, J.},
  \bibinfo{author}{Zwerger, W.}, \bibinfo{year}{2008}.
\newblock \bibinfo{title}{Many-body physics with ultracold gases}.
\newblock \bibinfo{journal}{Rev. Mod. Phys.} \bibinfo{volume}{80},
  \bibinfo{pages}{885}.
\newblock \DOIprefix\doi{10.1103/RevModPhys.80.885}.
%Type = Article
\bibitem[{Bosworth et~al.(2021)Bosworth, Pyzh and Schmelcher}]{Bosworth2021spo}
\bibinfo{author}{Bosworth, D.J.}, \bibinfo{author}{Pyzh, M.},
  \bibinfo{author}{Schmelcher, P.}, \bibinfo{year}{2021}.
\newblock \bibinfo{title}{Spectral properties of a three-body atom-ion hybrid
  system}.
\newblock \bibinfo{journal}{Phys. Rev. A} \bibinfo{volume}{103},
  \bibinfo{pages}{033303}.
\newblock \URLprefix
  \url{https://link.aps.org/doi/10.1103/PhysRevA.103.033303},
  \DOIprefix\doi{10.1103/PhysRevA.103.033303}.
%Type = Article
\bibitem[{Brownnutt et~al.(2015)Brownnutt, Kumph, Rabl and
  Blatt}]{Brownnutt:2015}
\bibinfo{author}{Brownnutt, M.}, \bibinfo{author}{Kumph, M.},
  \bibinfo{author}{Rabl, P.}, \bibinfo{author}{Blatt, R.},
  \bibinfo{year}{2015}.
\newblock \bibinfo{title}{Ion-trap measurements of electric-field noise near
  surfaces}.
\newblock \bibinfo{journal}{Rev.~Mod.~Phys.} \bibinfo{volume}{87},
  \bibinfo{pages}{1419}.
\newblock \DOIprefix\doi{10.1103/RevModPhys.87.1419}.
%Type = Article
\bibitem[{Bruzewicz et~al.(2019)Bruzewicz, Chiaverini, McConnell and
  Sage}]{Bruzewicz2019tiq}
\bibinfo{author}{Bruzewicz, C.D.}, \bibinfo{author}{Chiaverini, J.},
  \bibinfo{author}{McConnell, R.}, \bibinfo{author}{Sage, J.M.},
  \bibinfo{year}{2019}.
\newblock \bibinfo{title}{Trapped-ion quantum computing: Progress and
  challenges}.
\newblock \bibinfo{journal}{Applied Physics Reviews} \bibinfo{volume}{6},
  \bibinfo{pages}{021314}.
\newblock \URLprefix \url{https://doi.org/10.1063/1.5088164},
  \DOIprefix\doi{10.1063/1.5088164},
  \href{http://arxiv.org/abs/https://doi.org/10.1063/1.5088164}{{\tt
  arXiv:https://doi.org/10.1063/1.5088164}}.
%Type = Article
\bibitem[{Casteels et~al.(2011)Casteels, Tempere and
  Devreese}]{Casteels2011ppo}
\bibinfo{author}{Casteels, W.}, \bibinfo{author}{Tempere, J.},
  \bibinfo{author}{Devreese, J.}, \bibinfo{year}{2011}.
\newblock \bibinfo{title}{Polaronic properties of an ion in a {B}ose-{E}instein
  condensate in the strong-coupling limit}.
\newblock \bibinfo{journal}{J. Low Temp. Phys.} \bibinfo{volume}{162},
  \bibinfo{pages}{266}.
\newblock \DOIprefix\doi{10.1007/s10909-010-0286-0}.
%Type = Article
\bibitem[{Cetina et~al.(2012)Cetina, Grier and Vuleti{\'c}}]{Cetina2012flt}
\bibinfo{author}{Cetina, M.}, \bibinfo{author}{Grier, A.T.},
  \bibinfo{author}{Vuleti{\'c}, V.}, \bibinfo{year}{2012}.
\newblock \bibinfo{title}{Fundamental limit to atom-ion sympathetic cooling in
  paul traps}.
\newblock \bibinfo{journal}{Phys. Rev. Lett.} \bibinfo{volume}{109},
  \bibinfo{pages}{253201}.
\newblock \DOIprefix\doi{10.1103/PhysRevLett.109.253201}.
%Type = Article
\bibitem[{Chen et~al.(2014)Chen, Sullivan and Hudson}]{Chen2014ngs}
\bibinfo{author}{Chen, K.}, \bibinfo{author}{Sullivan, S.T.},
  \bibinfo{author}{Hudson, E.R.}, \bibinfo{year}{2014}.
\newblock \bibinfo{title}{Neutral gas sympathetic cooling of an ion in a {P}aul
  trap}.
\newblock \bibinfo{journal}{Phys.~Rev.~Lett.} \bibinfo{volume}{112},
  \bibinfo{pages}{143009}.
\newblock \DOIprefix\doi{10.1103/PhysRevLett.112.143009}.
%Type = Article
\bibitem[{Chin et~al.(2010)Chin, Grimm, Julienne and Tiesinga}]{Chin2010fri}
\bibinfo{author}{Chin, C.}, \bibinfo{author}{Grimm, R.},
  \bibinfo{author}{Julienne, P.S.}, \bibinfo{author}{Tiesinga, E.},
  \bibinfo{year}{2010}.
\newblock \bibinfo{title}{Feshbach resonances in ultracold gases}.
\newblock \bibinfo{journal}{Rev. Mod. Phys.} \bibinfo{volume}{82},
  \bibinfo{pages}{1225}.
\newblock \DOIprefix\doi{10.1103/RevModPhys.82.1225}.
%Type = Article
\bibitem[{wan Chou et~al.(2017)wan Chou, Kurz, Hume, Plessow, Leibrandt and
  Leibfried}]{Chou:2017}
\bibinfo{author}{wan Chou, C.}, \bibinfo{author}{Kurz, C.},
  \bibinfo{author}{Hume, D.B.}, \bibinfo{author}{Plessow, P.N.},
  \bibinfo{author}{Leibrandt, D.R.}, \bibinfo{author}{Leibfried, D.},
  \bibinfo{year}{2017}.
\newblock \bibinfo{title}{Preparation and coherent manipulation of pure quantum
  states of a single molecular ion}.
\newblock \bibinfo{journal}{Nature} \bibinfo{volume}{545},
  \bibinfo{pages}{203--207}.
\newblock \DOIprefix\doi{10.1038/nature22338}.
%Type = Article
\bibitem[{Christensen et~al.(2021)Christensen, Camacho-Guardian and
  Bruun}]{Christensen2021cpa}
\bibinfo{author}{Christensen, E.R.}, \bibinfo{author}{Camacho-Guardian, A.},
  \bibinfo{author}{Bruun, G.M.}, \bibinfo{year}{2021}.
\newblock \bibinfo{title}{Charged polarons and molecules in a {B}ose-{E}instein
  condensate}.
\newblock \bibinfo{journal}{Phys. Rev. Lett.} \bibinfo{volume}{126},
  \bibinfo{pages}{243001}.
\newblock \URLprefix
  \url{https://link.aps.org/doi/10.1103/PhysRevLett.126.243001},
  \DOIprefix\doi{10.1103/PhysRevLett.126.243001}.
%Type = Article
\bibitem[{Christensen et~al.(2022)Christensen, Camacho-Guardian and
  Bruun}]{Christensen2022mii}
\bibinfo{author}{Christensen, E.R.}, \bibinfo{author}{Camacho-Guardian, A.},
  \bibinfo{author}{Bruun, G.M.}, \bibinfo{year}{2022}.
\newblock \bibinfo{title}{Mobile ion in a {F}ermi sea}.
\newblock \bibinfo{journal}{Phys. Rev. A} \bibinfo{volume}{105},
  \bibinfo{pages}{023309}.
\newblock \URLprefix
  \url{https://link.aps.org/doi/10.1103/PhysRevA.105.023309},
  \DOIprefix\doi{10.1103/PhysRevA.105.023309}.
%Type = Article
\bibitem[{C\^ot\'e(2000)}]{Cote2000fcm}
\bibinfo{author}{C\^ot\'e, R.}, \bibinfo{year}{2000}.
\newblock \bibinfo{title}{From classical mobility to hopping conductivity:
  Charge hopping in an ultracold gas}.
\newblock \bibinfo{journal}{Phys. Rev. Lett.} \bibinfo{volume}{85},
  \bibinfo{pages}{5316--5319}.
\newblock \URLprefix
  \url{https://link.aps.org/doi/10.1103/PhysRevLett.85.5316},
  \DOIprefix\doi{10.1103/PhysRevLett.85.5316}.
%Type = Incollection
\bibitem[{C\^ot\'e(2016)}]{Cote2016uha}
\bibinfo{author}{C\^ot\'e, R.}, \bibinfo{year}{2016}.
\newblock \bibinfo{title}{Chapter two - ultracold hybrid atom–ion systems},
  \bibinfo{publisher}{Academic Press}. volume~\bibinfo{volume}{65} of
  \textit{\bibinfo{series}{Advances In Atomic, Molecular, and Optical
  Physics}}, pp. \bibinfo{pages}{67--126}.
\newblock \URLprefix
  \url{https://www.sciencedirect.com/science/article/pii/S1049250X16300088},
  \DOIprefix\doi{https://doi.org/10.1016/bs.aamop.2016.04.004}.
%Type = Article
\bibitem[{C\^ot\'e and Dalgarno(2000)}]{Cote2000uai}
\bibinfo{author}{C\^ot\'e, R.}, \bibinfo{author}{Dalgarno, A.},
  \bibinfo{year}{2000}.
\newblock \bibinfo{title}{Ultracold atom-ion collisions}.
\newblock \bibinfo{journal}{Phys. Rev. A} \bibinfo{volume}{62},
  \bibinfo{pages}{012709}.
\newblock \URLprefix \url{https://link.aps.org/doi/10.1103/PhysRevA.62.012709},
  \DOIprefix\doi{10.1103/PhysRevA.62.012709}.
%Type = Article
\bibitem[{C\^ot\'e et~al.(2002)C\^ot\'e, Kharchenko and Lukin}]{Cote2002mmi}
\bibinfo{author}{C\^ot\'e, R.}, \bibinfo{author}{Kharchenko, V.},
  \bibinfo{author}{Lukin, M.D.}, \bibinfo{year}{2002}.
\newblock \bibinfo{title}{Mesoscopic molecular ions in {B}ose-{E}instein
  condensates}.
\newblock \bibinfo{journal}{Phys. Rev. Lett.} \bibinfo{volume}{89},
  \bibinfo{pages}{093001}.
\newblock \URLprefix
  \url{https://link.aps.org/doi/10.1103/PhysRevLett.89.093001},
  \DOIprefix\doi{10.1103/PhysRevLett.89.093001}.
%Type = Article
\bibitem[{{C{\^o}t{\'e}} and {Simbotin}(2018)}]{Cote2018sot}
\bibinfo{author}{{C{\^o}t{\'e}}, R.}, \bibinfo{author}{{Simbotin}, I.},
  \bibinfo{year}{2018}.
\newblock \bibinfo{title}{{Signature of the $s$-wave regime high above ultralow
  temperatures}}.
\newblock \bibinfo{journal}{Phys.~Rev.~Lett.} \bibinfo{volume}{121},
  \bibinfo{pages}{173401}.
\newblock \DOIprefix\doi{10.1103/PhysRevLett.121.173401}.
%Type = Article
\bibitem[{Daley et~al.(2004)Daley, Fedichev and Zoller}]{Daley2004sac}
\bibinfo{author}{Daley, A.J.}, \bibinfo{author}{Fedichev, P.O.},
  \bibinfo{author}{Zoller, P.}, \bibinfo{year}{2004}.
\newblock \bibinfo{title}{Single-atom cooling by superfluid immersion: {A}
  nondestructive method for qubits}.
\newblock \bibinfo{journal}{Phys.~Rev.~A} \bibinfo{volume}{69},
  \bibinfo{pages}{022306}.
\newblock \DOIprefix\doi{10.1103/PhysRevA.69.022306}.
%Type = Article
\bibitem[{Dehmelt(1975)}]{Dehmelt:1975}
\bibinfo{author}{Dehmelt, H.G.}, \bibinfo{year}{1975}.
\newblock \bibinfo{title}{Proposed $10^{14}\:\delta\nu < \nu$ laser
  fluorescence spectroscopy on {Tl}$^+$ mono--ion oscillator {II} (spontaneous
  quantum jumps)}.
\newblock \bibinfo{journal}{Bull.~Am.~Phys.~Soc.} \bibinfo{volume}{20},
  \bibinfo{pages}{60}.
%Type = Article
\bibitem[{Deiß et~al.(2021)Deiß, Haze and Hecker~Denschlag}]{Deiss2021lra}
\bibinfo{author}{Deiß, M.}, \bibinfo{author}{Haze, S.},
  \bibinfo{author}{Hecker~Denschlag, J.}, \bibinfo{year}{2021}.
\newblock \bibinfo{title}{Long-range atom–ion {R}ydberg molecule: A novel
  molecular binding mechanism}.
\newblock \bibinfo{journal}{Atoms} \bibinfo{volume}{9}.
\newblock \URLprefix \url{https://www.mdpi.com/2218-2004/9/2/34},
  \DOIprefix\doi{10.3390/atoms9020034}.
%Type = Article
\bibitem[{Deng et~al.(2015)Deng, Che, Lan, Ge, Xu, Yuan, Zhang and
  Lu}]{DengDBS2015}
\bibinfo{author}{Deng, K.}, \bibinfo{author}{Che, H.}, \bibinfo{author}{Lan,
  Y.}, \bibinfo{author}{Ge, Y.P.}, \bibinfo{author}{Xu, Z.T.},
  \bibinfo{author}{Yuan, W.H.}, \bibinfo{author}{Zhang, J.},
  \bibinfo{author}{Lu, Z.H.}, \bibinfo{year}{2015}.
\newblock \bibinfo{journal}{J. Appl. Phys.} \bibinfo{volume}{118}.
\newblock \URLprefix \url{https://doi.org/10.1063/1.4931420}.
%Type = Article
\bibitem[{DeVoe(2009)}]{DeVoe2009pld}
\bibinfo{author}{DeVoe, R.G.}, \bibinfo{year}{2009}.
\newblock \bibinfo{title}{Power-law distributions for a trapped ion interacting
  with a classical buffer gas}.
\newblock \bibinfo{journal}{Phys.~Rev.~Lett.} \bibinfo{volume}{102},
  \bibinfo{pages}{063001}.
\newblock \DOIprefix\doi{10.1103/PhysRevLett.102.063001}.
%Type = Article
\bibitem[{Diedrich et~al.(1989)Diedrich, Bergquist, Itano and
  Wineland}]{Diedrich:1989}
\bibinfo{author}{Diedrich, F.}, \bibinfo{author}{Bergquist, J.C.},
  \bibinfo{author}{Itano, W.M.}, \bibinfo{author}{Wineland, D.J.},
  \bibinfo{year}{1989}.
\newblock \bibinfo{title}{{L}aser cooling to the zero-point energy of motion.}
\newblock \bibinfo{journal}{Phys.~Rev.~Lett.} \bibinfo{volume}{62},
  \bibinfo{pages}{403--406}.
\newblock \URLprefix
  \url{https://journals.aps.org/prl/abstract/10.1103/PhysRevLett.62.403},
  \DOIprefix\doi{10.1103/PhysRevLett.62.403}.
%Type = Article
\bibitem[{Dieterle et~al.(2020)Dieterle, Berngruber, H\"olzl, L\"ow, Jachymski,
  Pfau and Meinert}]{Dieterle2020icd}
\bibinfo{author}{Dieterle, T.}, \bibinfo{author}{Berngruber, M.},
  \bibinfo{author}{H\"olzl, C.}, \bibinfo{author}{L\"ow, R.},
  \bibinfo{author}{Jachymski, K.}, \bibinfo{author}{Pfau, T.},
  \bibinfo{author}{Meinert, F.}, \bibinfo{year}{2020}.
\newblock \bibinfo{title}{Inelastic collision dynamics of a single cold ion
  immersed in a bose-einstein condensate}.
\newblock \bibinfo{journal}{Phys. Rev. A} \bibinfo{volume}{102},
  \bibinfo{pages}{041301}.
\newblock \URLprefix
  \url{https://link.aps.org/doi/10.1103/PhysRevA.102.041301},
  \DOIprefix\doi{10.1103/PhysRevA.102.041301}.
%Type = Article
\bibitem[{Dieterle et~al.(2021)Dieterle, Berngruber, H\"olzl, L\"ow, Jachymski,
  Pfau and Meinert}]{Dieterle2021toa}
\bibinfo{author}{Dieterle, T.}, \bibinfo{author}{Berngruber, M.},
  \bibinfo{author}{H\"olzl, C.}, \bibinfo{author}{L\"ow, R.},
  \bibinfo{author}{Jachymski, K.}, \bibinfo{author}{Pfau, T.},
  \bibinfo{author}{Meinert, F.}, \bibinfo{year}{2021}.
\newblock \bibinfo{title}{Transport of a single cold ion immersed in a
  {B}ose-{E}instein condensate}.
\newblock \bibinfo{journal}{Phys. Rev. Lett.} \bibinfo{volume}{126},
  \bibinfo{pages}{033401}.
\newblock \URLprefix
  \url{https://link.aps.org/doi/10.1103/PhysRevLett.126.033401},
  \DOIprefix\doi{10.1103/PhysRevLett.126.033401}.
%Type = Article
\bibitem[{Ding et~al.(2022)Ding, Drewsen, Arlt and Bruun}]{ding2022mib}
\bibinfo{author}{Ding, S.}, \bibinfo{author}{Drewsen, M.},
  \bibinfo{author}{Arlt, J.J.}, \bibinfo{author}{Bruun, G.},
  \bibinfo{year}{2022}.
\newblock \bibinfo{title}{Mediated interactions between ions in quantum
  degenerate gases}.
\newblock \bibinfo{journal}{arXiv preprint arXiv:2203.02768} \URLprefix
  \url{https://arxiv.org/abs/2203.02768}.
%Type = Article
\bibitem[{Doerk et~al.(2010)Doerk, Idziaszek and Calarco}]{Doerk:2010}
\bibinfo{author}{Doerk, H.}, \bibinfo{author}{Idziaszek, Z.},
  \bibinfo{author}{Calarco, T.}, \bibinfo{year}{2010}.
\newblock \bibinfo{title}{Atom-ion quantum gate}.
\newblock \bibinfo{journal}{Phys.~Rev.~A} \bibinfo{volume}{81},
  \bibinfo{pages}{012708}.
\newblock \DOIprefix\doi{10.1103/PhysRevA.81.012708}.
%Type = Article
\bibitem[{D\"orfler et~al.(2020)D\"orfler, Yurtsever, Villarreal,
  Gonz\'alez-Lezana, Gianturco and Willitsch}]{Doerfler2020rsc}
\bibinfo{author}{D\"orfler, A.D.}, \bibinfo{author}{Yurtsever, E.},
  \bibinfo{author}{Villarreal, P.}, \bibinfo{author}{Gonz\'alez-Lezana, T.},
  \bibinfo{author}{Gianturco, F.A.}, \bibinfo{author}{Willitsch, S.},
  \bibinfo{year}{2020}.
\newblock \bibinfo{title}{Rotational-state-changing collisions between
  {N}$_{2}^{+}$ and {R}b at low energies}.
\newblock \bibinfo{journal}{Phys. Rev. A} \bibinfo{volume}{101},
  \bibinfo{pages}{012706}.
\newblock \URLprefix
  \url{https://link.aps.org/doi/10.1103/PhysRevA.101.012706},
  \DOIprefix\doi{10.1103/PhysRevA.101.012706}.
%Type = Article
\bibitem[{Drewsen et~al.(2004)Drewsen, Mortensen, Martinussen, Staanum and
  S\o{}rensen}]{Drewsen2004nio}
\bibinfo{author}{Drewsen, M.}, \bibinfo{author}{Mortensen, A.},
  \bibinfo{author}{Martinussen, R.}, \bibinfo{author}{Staanum, P.},
  \bibinfo{author}{S\o{}rensen, J.L.}, \bibinfo{year}{2004}.
\newblock \bibinfo{title}{Nondestructive identification of cold and extremely
  localized single molecular ions}.
\newblock \bibinfo{journal}{Phys. Rev. Lett.} \bibinfo{volume}{93},
  \bibinfo{pages}{243201}.
\newblock \URLprefix
  \url{https://link.aps.org/doi/10.1103/PhysRevLett.93.243201},
  \DOIprefix\doi{10.1103/PhysRevLett.93.243201}.
%Type = Article
\bibitem[{Dutta and Rangwala(2020)}]{Dutta2020moc}
\bibinfo{author}{Dutta, S.}, \bibinfo{author}{Rangwala, S.A.},
  \bibinfo{year}{2020}.
\newblock \bibinfo{title}{Measurement of collisions between laser-cooled cesium
  atoms and trapped cesium ions}.
\newblock \bibinfo{journal}{Phys. Rev. A} \bibinfo{volume}{102},
  \bibinfo{pages}{033309}.
\newblock \URLprefix
  \url{https://link.aps.org/doi/10.1103/PhysRevA.102.033309},
  \DOIprefix\doi{10.1103/PhysRevA.102.033309}.
%Type = Article
\bibitem[{Dörfler et~al.(2019)Dörfler, Eberle, Koner, Tomza, Meuwly and
  Willitsch}]{Doerfler2019lrv}
\bibinfo{author}{Dörfler, A.D.}, \bibinfo{author}{Eberle, P.},
  \bibinfo{author}{Koner, D.}, \bibinfo{author}{Tomza, M.},
  \bibinfo{author}{Meuwly, M.}, \bibinfo{author}{Willitsch, S.},
  \bibinfo{year}{2019}.
\newblock \bibinfo{title}{Long-range versus short-range effects in cold
  molecular ion-neutral collisions}.
\newblock \bibinfo{journal}{Nature Communications} \bibinfo{volume}{10}.
\newblock \DOIprefix\doi{10.1038/s41467-019-13218-x}.
%Type = Article
\bibitem[{{Earnshaw}(1842)}]{Earnshaw:1842}
\bibinfo{author}{{Earnshaw}, S.}, \bibinfo{year}{1842}.
\newblock \bibinfo{title}{{On the nature of the molecular forces which regulate
  the constitution of the luminferous ether}}.
\newblock \bibinfo{journal}{Transactions of the Cambridge Philosophical
  Society} \bibinfo{volume}{7}, \bibinfo{pages}{97–112}.
%Type = Article
\bibitem[{Ebgha et~al.(2019)Ebgha, Saeidian, Schmelcher and
  Negretti}]{Ebgha2019cai}
\bibinfo{author}{Ebgha, M.R.}, \bibinfo{author}{Saeidian, S.},
  \bibinfo{author}{Schmelcher, P.}, \bibinfo{author}{Negretti, A.},
  \bibinfo{year}{2019}.
\newblock \bibinfo{title}{Compound atom-ion {J}osephson junction: Effects of
  finite temperature and ion motion}.
\newblock \bibinfo{journal}{Phys. Rev. A} \bibinfo{volume}{100},
  \bibinfo{pages}{033616}.
\newblock \URLprefix
  \url{https://link.aps.org/doi/10.1103/PhysRevA.100.033616},
  \DOIprefix\doi{10.1103/PhysRevA.100.033616}.
%Type = Article
\bibitem[{Engel et~al.(2018)Engel, Dieterle, Schmid, Tomschitz, Veit, Zuber,
  L\"ow, Pfau and Meinert}]{Engel2018oor}
\bibinfo{author}{Engel, F.}, \bibinfo{author}{Dieterle, T.},
  \bibinfo{author}{Schmid, T.}, \bibinfo{author}{Tomschitz, C.},
  \bibinfo{author}{Veit, C.}, \bibinfo{author}{Zuber, N.},
  \bibinfo{author}{L\"ow, R.}, \bibinfo{author}{Pfau, T.},
  \bibinfo{author}{Meinert, F.}, \bibinfo{year}{2018}.
\newblock \bibinfo{title}{Observation of {R}ydberg blockade induced by a single
  ion}.
\newblock \bibinfo{journal}{Phys. Rev. Lett.} \bibinfo{volume}{121},
  \bibinfo{pages}{193401}.
\newblock \URLprefix
  \url{https://link.aps.org/doi/10.1103/PhysRevLett.121.193401},
  \DOIprefix\doi{10.1103/PhysRevLett.121.193401}.
%Type = Article
\bibitem[{Epstein et~al.(2007)Epstein, Seidelin, Leibfried, Wesenberg,
  Bollinger, Amini, Blakestad, Britton, Home, Itano, Jost, Knill, Langer,
  Ozeri, Shiga and Wineland}]{Epstein:2007}
\bibinfo{author}{Epstein, R.J.}, \bibinfo{author}{Seidelin, S.},
  \bibinfo{author}{Leibfried, D.}, \bibinfo{author}{Wesenberg, J.H.},
  \bibinfo{author}{Bollinger, J.J.}, \bibinfo{author}{Amini, J.M.},
  \bibinfo{author}{Blakestad, R.B.}, \bibinfo{author}{Britton, J.},
  \bibinfo{author}{Home, J.P.}, \bibinfo{author}{Itano, W.M.},
  \bibinfo{author}{Jost, J.D.}, \bibinfo{author}{Knill, E.},
  \bibinfo{author}{Langer, C.}, \bibinfo{author}{Ozeri, R.},
  \bibinfo{author}{Shiga, N.}, \bibinfo{author}{Wineland, D.J.},
  \bibinfo{year}{2007}.
\newblock \bibinfo{title}{Simplified motional heating rate measurements of
  trapped ions}.
\newblock \bibinfo{journal}{Phys.~Rev.~A} \bibinfo{volume}{76},
  \bibinfo{pages}{033411}.
\newblock \URLprefix
  \url{https://journals.aps.org/pra/references/10.1103/PhysRevA.76.033411},
  \DOIprefix\doi{10.1103/PhysRevA.76.033411}.
%Type = Article
\bibitem[{Ewald et~al.(2019)Ewald, Feldker, Hirzler, F\"urst and
  Gerritsma}]{Ewald:2019}
\bibinfo{author}{Ewald, N.V.}, \bibinfo{author}{Feldker, T.},
  \bibinfo{author}{Hirzler, H.}, \bibinfo{author}{F\"urst, H.},
  \bibinfo{author}{Gerritsma, R.}, \bibinfo{year}{2019}.
\newblock \bibinfo{title}{Observation of interactions between trapped ions and
  ultracold {R}ydberg atoms}.
\newblock \bibinfo{journal}{Phys. Rev. Lett.} \bibinfo{volume}{122},
  \bibinfo{pages}{253401}.
\newblock \DOIprefix\doi{10.1103/PhysRevLett.122.253401}.
%Type = Article
\bibitem[{Feldker et~al.(2020)Feldker, F\"urst, Hirzler, Ewald, Mazzanti,
  Wiater, Tomza and Gerritsma}]{Feldker2020bgc}
\bibinfo{author}{Feldker, T.}, \bibinfo{author}{F\"urst, H.},
  \bibinfo{author}{Hirzler, H.}, \bibinfo{author}{Ewald, N.V.},
  \bibinfo{author}{Mazzanti, M.}, \bibinfo{author}{Wiater, D.},
  \bibinfo{author}{Tomza, M.}, \bibinfo{author}{Gerritsma, R.},
  \bibinfo{year}{2020}.
\newblock \bibinfo{title}{Buffer gas cooling of a trapped ion to the quantum
  regime}.
\newblock \bibinfo{journal}{Nat. Phys.} \bibinfo{volume}{16},
  \bibinfo{pages}{413--416}.
\newblock \DOIprefix\doi{10.1038/s41567-019-0772-5}.
%Type = Incollection
\bibitem[{Ferlaino et~al.(2009)Ferlaino, Knoop and Grimm}]{Ferlaino2009ufm}
\bibinfo{author}{Ferlaino, F.}, \bibinfo{author}{Knoop, S.},
  \bibinfo{author}{Grimm, R.}, \bibinfo{year}{2009}.
\newblock \bibinfo{title}{Ultracold feshbach molecules}, in:
  \bibinfo{editor}{Krems, R.}, \bibinfo{editor}{Friedrich, B.},
  \bibinfo{editor}{Stwalley, W.C.} (Eds.), \bibinfo{booktitle}{Cold Molecules}.
  \bibinfo{publisher}{{CRC} Press}.
\newblock \DOIprefix\doi{10.1201/9781420059045}.
%Type = Article
\bibitem[{Fl\"uhmann and Home(2020)}]{Fluehmann:2020}
\bibinfo{author}{Fl\"uhmann, C.}, \bibinfo{author}{Home, J.},
  \bibinfo{year}{2020}.
\newblock \bibinfo{title}{Direct characteristic-function tomography of quantum
  states of the trapped-ion motional oscillator}.
\newblock \bibinfo{journal}{Phys.~Rev.~Lett.} \bibinfo{volume}{125},
  \bibinfo{pages}{043602}.
\newblock \URLprefix
  \url{https://journals.aps.org/prl/abstract/10.1103/PhysRevLett.125.043602},
  \DOIprefix\doi{10.1103/PhysRevLett.125.043602}.
%Type = Article
\bibitem[{F{\"u}rst et~al.(2018)F{\"u}rst, Ewald, Secker, Joger, Feldker and
  Gerritsma}]{Fuerst2018por}
\bibinfo{author}{F{\"u}rst, H.}, \bibinfo{author}{Ewald, N.V.},
  \bibinfo{author}{Secker, T.}, \bibinfo{author}{Joger, J.},
  \bibinfo{author}{Feldker, T.}, \bibinfo{author}{Gerritsma, R.},
  \bibinfo{year}{2018}.
\newblock \bibinfo{title}{Prospects of reaching the quantum regime in
  {L}i-{Y}b$^+$ mixtures}.
\newblock \bibinfo{journal}{J. Phys. B} \bibinfo{volume}{51},
  \bibinfo{pages}{195001}.
\newblock \URLprefix \url{http://iopscience.iop.org/10.1088/1361-6455/aadd7d}.
%Type = Article
\bibitem[{F\"urst et~al.(2018)F\"urst, Feldker, Ewald, Joger, Tomza and
  Gerritsma}]{Fuerst2018doa}
\bibinfo{author}{F\"urst, H.}, \bibinfo{author}{Feldker, T.},
  \bibinfo{author}{Ewald, N.V.}, \bibinfo{author}{Joger, J.},
  \bibinfo{author}{Tomza, M.}, \bibinfo{author}{Gerritsma, R.},
  \bibinfo{year}{2018}.
\newblock \bibinfo{title}{Dynamics of a single ion-spin impurity in a
  spin-polarized atomic bath}.
\newblock \bibinfo{journal}{Phys. Rev. A} \bibinfo{volume}{98},
  \bibinfo{pages}{012713}.
\newblock \URLprefix \url{https://link.aps.org/doi/10.1103/PhysRevA.98.012713},
  \DOIprefix\doi{10.1103/PhysRevA.98.012713}.
%Type = Article
\bibitem[{Gacesa and C{\^o}t{\'e}(2017)}]{Gacesa2017cti}
\bibinfo{author}{Gacesa, M.}, \bibinfo{author}{C{\^o}t{\'e}, R.},
  \bibinfo{year}{2017}.
\newblock \bibinfo{title}{Charge transfer in ultracold gases via {F}eshbach
  resonances}.
\newblock \bibinfo{journal}{Phys. Rev. A} \bibinfo{volume}{95},
  \bibinfo{pages}{062704}.
\newblock \DOIprefix\doi{10.1103/PhysRevA.95.062704}.
%Type = Article
\bibitem[{Geppert et~al.(2021)Geppert, Althön, Fichtner and
  Ott}]{Geppert2021adi}
\bibinfo{author}{Geppert, P.}, \bibinfo{author}{Althön, M.},
  \bibinfo{author}{Fichtner, D.}, \bibinfo{author}{Ott, H.},
  \bibinfo{year}{2021}.
\newblock \bibinfo{title}{Diffusive-like redistribution in state-changing
  collisions between rydberg atoms and ground state atoms}.
\newblock \bibinfo{journal}{Nat. Commun.} \bibinfo{volume}{12},
  \bibinfo{pages}{3900}.
\newblock \DOIprefix\doi{10.1038/s41467-021-24146-0}.
%Type = Article
\bibitem[{Gerlich(1995)}]{Gerlich:1995}
\bibinfo{author}{Gerlich, D.}, \bibinfo{year}{1995}.
\newblock \bibinfo{title}{Ion-neutral collisions in a 22-pole trap at very low
  energies}.
\newblock \bibinfo{journal}{Phys. Scr.} \bibinfo{volume}{T59},
  \bibinfo{pages}{256}.
\newblock \DOIprefix\doi{10.1088/0031-8949/1995/T59/035}.
%Type = Incollection
\bibitem[{Gerlich(2008)}]{Gerlich2008tso}
\bibinfo{author}{Gerlich, D.}, \bibinfo{year}{2008}.
\newblock \bibinfo{title}{The study of cold collisions using ion guides and
  traps}, in: \bibinfo{editor}{Smith, I.W.M.} (Ed.), \bibinfo{booktitle}{Low
  Temperatures and Cold Molecules}. \bibinfo{publisher}{PUBLISHED BY IMPERIAL
  COLLEGE PRESS AND DISTRIBUTED BY WORLD SCIENTIFIC PUBLISHING CO.}.
  chapter~\bibinfo{chapter}{3}, pp. \bibinfo{pages}{121--174}.
\newblock \URLprefix
  \url{https://www.worldscientific.com/doi/abs/10.1142/9781848162105_0003},
  \DOIprefix\doi{10.1142/9781848162105_0003}.
%Type = Article
\bibitem[{Gerritsma et~al.(2010)Gerritsma, Kirchmair, Z{\"a}hringer, Solano,
  Blatt and Roos}]{Gerritsma:2010}
\bibinfo{author}{Gerritsma, R.}, \bibinfo{author}{Kirchmair, G.},
  \bibinfo{author}{Z{\"a}hringer, F.}, \bibinfo{author}{Solano, E.},
  \bibinfo{author}{Blatt, R.}, \bibinfo{author}{Roos, C.F.},
  \bibinfo{year}{2010}.
\newblock \bibinfo{title}{Quantum simulation of the {D}irac equation.}
\newblock \bibinfo{journal}{Nature} \bibinfo{volume}{463}, \bibinfo{pages}{68}.
\newblock \URLprefix \url{https://doi.org/10.1038/nature08688},
  \DOIprefix\doi{10.1038/nature08688}.
%Type = Book
\bibitem[{Ghosh(1995)}]{Ghosh:1995}
\bibinfo{author}{Ghosh, P.K.}, \bibinfo{year}{1995}.
\newblock \bibinfo{title}{Ion Traps}.
\newblock \bibinfo{publisher}{Clarendon Press}.
\newblock \URLprefix
  \url{https://global.oup.com/academic/product/ion-traps-9780198539957?cc=nl&lang=en&}.
%Type = Article
\bibitem[{Goham and Britton(2021)}]{Goham:2021}
\bibinfo{author}{Goham, C.J.B.}, \bibinfo{author}{Britton, J.W.},
  \bibinfo{year}{2021}.
\newblock \bibinfo{title}{Resolved-sideband micromotion sensing in {Y}b$^{+}$
  on the 935 nm repump transition}.
\newblock \bibinfo{journal}{arXiv:2111.11504} \URLprefix
  \url{https://arxiv.org/abs/2111.11504}.
%Type = Article
\bibitem[{Grier et~al.(2009)Grier, Cetina, Oru\ifmmode \check{c}\else
  \v{c}\fi{}evi\ifmmode~\acute{c}\else \'{c}\fi{} and
  Vuleti\ifmmode~\acute{c}\else \'{c}\fi{}}]{Grier2009ooc}
\bibinfo{author}{Grier, A.T.}, \bibinfo{author}{Cetina, M.},
  \bibinfo{author}{Oru\ifmmode \check{c}\else
  \v{c}\fi{}evi\ifmmode~\acute{c}\else \'{c}\fi{}, F.},
  \bibinfo{author}{Vuleti\ifmmode~\acute{c}\else \'{c}\fi{}, V.},
  \bibinfo{year}{2009}.
\newblock \bibinfo{title}{Observation of cold collisions between trapped ions
  and trapped atoms}.
\newblock \bibinfo{journal}{Phys. Rev. Lett.} \bibinfo{volume}{102},
  \bibinfo{pages}{223201}.
\newblock \URLprefix
  \url{https://link.aps.org/doi/10.1103/PhysRevLett.102.223201},
  \DOIprefix\doi{10.1103/PhysRevLett.102.223201}.
%Type = Incollection
\bibitem[{Grimm(2008)}]{Grimm2008ufg}
\bibinfo{author}{Grimm, R.}, \bibinfo{year}{2008}.
\newblock
  \bibinfo{title}{\href{https://arxiv.org/abs/cond-mat/0703091}{Ultracold Fermi
  gases in the BEC-BCS crossover: a review from the Innsbruck perspective}},
  in: \bibinfo{editor}{Inguscio, M.}, \bibinfo{editor}{Ketterle, W.},
  \bibinfo{editor}{Salomon, C.} (Eds.), \bibinfo{booktitle}{Ultra-cold Fermi
  Gases}.
\newblock \bibinfo{note}{{P}roceedings of the International School of Physics
  ``Enrico Fermi'', Course CLXIV, Varenna, 20-30 June 2006}.
%Type = Article
\bibitem[{Grimm et~al.(2000)Grimm, Weidem{\"u}ller and
  Ovchinnikov}]{Grimm2000odt}
\bibinfo{author}{Grimm, R.}, \bibinfo{author}{Weidem{\"u}ller, M.},
  \bibinfo{author}{Ovchinnikov, Y.B.}, \bibinfo{year}{2000}.
\newblock \bibinfo{title}{Optical dipole traps for neutral atoms}.
\newblock \bibinfo{journal}{Adv. At. Mol. Opt. Phys.} \bibinfo{volume}{42},
  \bibinfo{pages}{95}.
\newblock \DOIprefix\doi{10.1016/S1049-250X(08)60186-X}.
%Type = Article
\bibitem[{Harlander et~al.(2010)Harlander, Brownnutt, H{\"a}nsel and
  Blatt}]{Harlander:2010}
\bibinfo{author}{Harlander, M.}, \bibinfo{author}{Brownnutt, M.},
  \bibinfo{author}{H{\"a}nsel, W.}, \bibinfo{author}{Blatt, R.},
  \bibinfo{year}{2010}.
\newblock \bibinfo{title}{Trapped-ion probing of light-induced charging effects
  on dielectrics}.
\newblock \bibinfo{journal}{New J.~Phys.} \bibinfo{volume}{12},
  \bibinfo{pages}{093035}.
\newblock \URLprefix \url{https://doi.org/10.1088/1367-2630/12/9/093035},
  \DOIprefix\doi{10.1088/1367-2630/12/9/093035}.
%Type = Article
\bibitem[{H\"arter and Denschlag(2014)}]{Harter2014cai}
\bibinfo{author}{H\"arter, A.}, \bibinfo{author}{Denschlag, J.H.},
  \bibinfo{year}{2014}.
\newblock \bibinfo{title}{Cold atom-ion experiments in hybrid traps}.
\newblock \bibinfo{journal}{Contemp. Phys.} \bibinfo{volume}{55},
  \bibinfo{pages}{33--45}.
\newblock \DOIprefix\doi{10.1080/00107514.2013.854618}.
%Type = Article
\bibitem[{H\"arter et~al.(2012)H\"arter, Kr\"ukow, Brunner, Schnitzler, Schmid
  and Denschlag}]{Harter2012sia}
\bibinfo{author}{H\"arter, A.}, \bibinfo{author}{Kr\"ukow, A.},
  \bibinfo{author}{Brunner, A.}, \bibinfo{author}{Schnitzler, W.},
  \bibinfo{author}{Schmid, S.}, \bibinfo{author}{Denschlag, J.H.},
  \bibinfo{year}{2012}.
\newblock \bibinfo{title}{Single ion as a three-body reaction center in an
  ultracold atomic gas}.
\newblock \bibinfo{journal}{Phys. Rev. Lett.} \bibinfo{volume}{109},
  \bibinfo{pages}{123201}.
\newblock \URLprefix
  \url{https://link.aps.org/doi/10.1103/PhysRevLett.109.123201},
  \DOIprefix\doi{10.1103/PhysRevLett.109.123201}.
%Type = Article
\bibitem[{H\"arter et~al.(2013)H\"arter, Kr\"ukow, Dei{\ss}, Tiemann and
  Hecker~Denschlag}]{Harter2013pdo}
\bibinfo{author}{H\"arter, A.}, \bibinfo{author}{Kr\"ukow, A.},
  \bibinfo{author}{Dei{\ss}, B.}, \bibinfo{author}{Tiemann, E.},
  \bibinfo{author}{Hecker~Denschlag, J.}, \bibinfo{year}{2013}.
\newblock \bibinfo{title}{Population distribution of product states following
  three-body recombination in an ultracold atomic gas}.
\newblock \bibinfo{journal}{Nat. Phys.} \bibinfo{volume}{9},
  \bibinfo{pages}{512--517}.
\newblock \DOIprefix\doi{10.1038/nphys2661}.
%Type = Article
\bibitem[{Hassan et~al.(2022)Hassan, Tauch, Kas, Nötzold, Carrera, Endres,
  Wester and Weidemüller}]{Hassan2022adi}
\bibinfo{author}{Hassan, S.Z.}, \bibinfo{author}{Tauch, J.},
  \bibinfo{author}{Kas, M.}, \bibinfo{author}{Nötzold, M.},
  \bibinfo{author}{Carrera, H.L.}, \bibinfo{author}{Endres, E.S.},
  \bibinfo{author}{Wester, R.}, \bibinfo{author}{Weidemüller, M.},
  \bibinfo{year}{2022}.
\newblock \bibinfo{title}{Associative detachment in anion-atom reactions
  involving a dipole-bound electron}.
\newblock \bibinfo{journal}{Nat. Commun.} \bibinfo{volume}{13},
  \bibinfo{pages}{818}.
\newblock \DOIprefix\doi{10.1038/s41467-022-28382-w}.
%Type = Article
\bibitem[{Haze et~al.(2015)Haze, Saito, Fujinaga and Mukaiyama}]{Haze:2015}
\bibinfo{author}{Haze, S.}, \bibinfo{author}{Saito, R.},
  \bibinfo{author}{Fujinaga, M.}, \bibinfo{author}{Mukaiyama, T.},
  \bibinfo{year}{2015}.
\newblock \bibinfo{title}{Charge-exchange collisions between ultracold
  fermionic lithium atoms and calcium ions}.
\newblock \bibinfo{journal}{Phys.~Rev.~A} \bibinfo{volume}{91},
  \bibinfo{pages}{032709}.
\newblock \DOIprefix\doi{10.1103/PhysRevA.91.032709}.
%Type = Article
\bibitem[{Haze et~al.(2018)Haze, Sasakawa, Saito, Nakai and
  Mukaiyama}]{Haze2018cdo}
\bibinfo{author}{Haze, S.}, \bibinfo{author}{Sasakawa, M.},
  \bibinfo{author}{Saito, R.}, \bibinfo{author}{Nakai, R.},
  \bibinfo{author}{Mukaiyama, T.}, \bibinfo{year}{2018}.
\newblock \bibinfo{title}{Cooling dynamics of a single trapped ion via elastic
  collisions with small-mass atoms}.
\newblock \bibinfo{journal}{Phys.~Rev.~Lett.} \bibinfo{volume}{120},
  \bibinfo{pages}{043401}.
\newblock \DOIprefix\doi{10.1103/PhysRevLett.120.043401}.
%Type = Article
\bibitem[{Haze et~al.(2019)Haze, Wolf, Deiß, Wang, Raithel and
  Denschlag}]{Haze:2019}
\bibinfo{author}{Haze, S.}, \bibinfo{author}{Wolf, J.}, \bibinfo{author}{Deiß,
  M.}, \bibinfo{author}{Wang, L.}, \bibinfo{author}{Raithel, G.},
  \bibinfo{author}{Denschlag, J.H.}, \bibinfo{year}{2019}.
\newblock \bibinfo{title}{Stark spectroscopy of {R}ydberg atoms in an atom-ion
  hybrid trap}.
\newblock \bibinfo{journal}{arXiv:1901.11069} \URLprefix
  \url{https://arxiv.org/abs/1901.11069}.
%Type = Article
\bibitem[{Heazlewood(2019)}]{Heazlewood2019cic}
\bibinfo{author}{Heazlewood, B.R.}, \bibinfo{year}{2019}.
\newblock \bibinfo{title}{Cold ion chemistry within coulomb crystals}.
\newblock \bibinfo{journal}{Molecular Physics} \bibinfo{volume}{117},
  \bibinfo{pages}{1934--1941}.
\newblock \URLprefix \url{https://doi.org/10.1080/00268976.2018.1564850},
  \DOIprefix\doi{10.1080/00268976.2018.1564850},
  \href{http://arxiv.org/abs/https://doi.org/10.1080/00268976.2018.1564850}{{\tt
  arXiv:https://doi.org/10.1080/00268976.2018.1564850}}.
%Type = Article
\bibitem[{Heazlewood and Softley(2015)}]{Heazlewood2015ltk}
\bibinfo{author}{Heazlewood, B.R.}, \bibinfo{author}{Softley, T.P.},
  \bibinfo{year}{2015}.
\newblock \bibinfo{title}{Low-temperature kinetics and dynamics with {C}oulomb
  crystals}.
\newblock \bibinfo{journal}{Annual Review of Physical Chemistry}
  \bibinfo{volume}{66}, \bibinfo{pages}{475--495}.
\newblock \URLprefix
  \url{https://doi.org/10.1146/annurev-physchem-040214-121527},
  \DOIprefix\doi{10.1146/annurev-physchem-040214-121527},
  \href{http://arxiv.org/abs/https://doi.org/10.1146/annurev-physchem-040214-121527}{{\tt
  arXiv:https://doi.org/10.1146/annurev-physchem-040214-121527}}.
  \bibinfo{note}{pMID: 25594853}.
%Type = Article
\bibitem[{Heazlewood and Softley(2021)}]{Heazlewood2021tca}
\bibinfo{author}{Heazlewood, B.R.}, \bibinfo{author}{Softley, T.P.},
  \bibinfo{year}{2021}.
\newblock \bibinfo{title}{Towards chemistry at absolute zero}.
\newblock \bibinfo{journal}{Nat. Rev. Chem.} \bibinfo{volume}{5},
  \bibinfo{pages}{125--140}.
\newblock \URLprefix \url{https://doi.org/10.1038/s41570-020-00239-0},
  \DOIprefix\doi{10.1038/s41570-020-00239-0}.
%Type = Article
\bibitem[{Higgins et~al.(2021)Higgins, Salim, Zhang, Parke, Pokorny and
  Hennrich}]{Higgins:2021}
\bibinfo{author}{Higgins, G.}, \bibinfo{author}{Salim, S.},
  \bibinfo{author}{Zhang, C.}, \bibinfo{author}{Parke, H.},
  \bibinfo{author}{Pokorny, F.}, \bibinfo{author}{Hennrich, M.},
  \bibinfo{year}{2021}.
\newblock \bibinfo{title}{Micromotion minimization using {R}amsey
  interferometry}.
\newblock \bibinfo{journal}{New Journal of Physics} \bibinfo{volume}{23},
  \bibinfo{pages}{123028}.
\newblock \URLprefix \url{https://doi.org/10.1088/1367-2630/ac3db6},
  \DOIprefix\doi{10.1088/1367-2630/ac3db6}.
%Type = Article
\bibitem[{Hirzler et~al.(2020a)Hirzler, Feldker, F\"urst, Ewald, Trimby, Lous,
  Arias~Espinoza, Mazzanti, Joger and Gerritsma}]{Hirzler2020esf}
\bibinfo{author}{Hirzler, H.}, \bibinfo{author}{Feldker, T.},
  \bibinfo{author}{F\"urst, H.}, \bibinfo{author}{Ewald, N.V.},
  \bibinfo{author}{Trimby, E.}, \bibinfo{author}{Lous, R.S.},
  \bibinfo{author}{Arias~Espinoza, J.D.}, \bibinfo{author}{Mazzanti, M.},
  \bibinfo{author}{Joger, J.}, \bibinfo{author}{Gerritsma, R.},
  \bibinfo{year}{2020}a.
\newblock \bibinfo{title}{Experimental setup for studying an ultracold mixture
  of trapped {Y}b$^{+}\text{--}^{6}${L}i}.
\newblock \bibinfo{journal}{Phys. Rev. A} \bibinfo{volume}{102},
  \bibinfo{pages}{033109}.
\newblock \URLprefix
  \url{https://link.aps.org/doi/10.1103/PhysRevA.102.033109},
  \DOIprefix\doi{10.1103/PhysRevA.102.033109}.
%Type = Article
\bibitem[{Hirzler et~al.(2022)Hirzler, Lous, Trimby, P\'erez-R\'{\i}os,
  Safavi-Naini and Gerritsma}]{Hirzler2022ooc}
\bibinfo{author}{Hirzler, H.}, \bibinfo{author}{Lous, R.S.},
  \bibinfo{author}{Trimby, E.}, \bibinfo{author}{P\'erez-R\'{\i}os, J.},
  \bibinfo{author}{Safavi-Naini, A.}, \bibinfo{author}{Gerritsma, R.},
  \bibinfo{year}{2022}.
\newblock \bibinfo{title}{Observation of chemical reactions between a trapped
  ion and ultracold {F}eshbach dimers}.
\newblock \bibinfo{journal}{Phys. Rev. Lett.} \bibinfo{volume}{128},
  \bibinfo{pages}{103401}.
\newblock \URLprefix
  \url{https://link.aps.org/doi/10.1103/PhysRevLett.128.103401},
  \DOIprefix\doi{10.1103/PhysRevLett.128.103401}.
%Type = Article
\bibitem[{Hirzler et~al.(2020b)Hirzler, Trimby, Lous, Groenenboom, Gerritsma
  and P{\'{e}}rez-R{\'{\i}}os}]{Hirzler2020ctn}
\bibinfo{author}{Hirzler, H.}, \bibinfo{author}{Trimby, E.},
  \bibinfo{author}{Lous, R.S.}, \bibinfo{author}{Groenenboom, G.C.},
  \bibinfo{author}{Gerritsma, R.}, \bibinfo{author}{P{\'{e}}rez-R{\'{\i}}os,
  J.}, \bibinfo{year}{2020}b.
\newblock \bibinfo{title}{Controlling the nature of a charged impurity in a
  bath of {F}eshbach dimers}.
\newblock \bibinfo{journal}{Phys. Rev. Research} \bibinfo{volume}{2},
  \bibinfo{pages}{033232}.
\newblock \URLprefix
  \url{https://journals.aps.org/prresearch/abstract/10.1103/PhysRevResearch.2.033232}.
%Type = Article
\bibitem[{H{\"o}ltkemeier et~al.(2016)H{\"o}ltkemeier, Weckesser,
  L{\'o}pez-Carrera and Weidem{\"u}ller}]{Holtkemeier2016bgc}
\bibinfo{author}{H{\"o}ltkemeier, B.}, \bibinfo{author}{Weckesser, P.},
  \bibinfo{author}{L{\'o}pez-Carrera, H.}, \bibinfo{author}{Weidem{\"u}ller,
  M.}, \bibinfo{year}{2016}.
\newblock \bibinfo{title}{Buffer-gas cooling of ions in a multipole radio
  frequency trap beyond the critical mass ratio}.
\newblock \bibinfo{journal}{Phys.~Rev.~Lett.} \bibinfo{volume}{116},
  \bibinfo{pages}{233003}.
\newblock \DOIprefix\doi{10.1103/PhysRevLett.116.233003}.
%Type = Article
\bibitem[{Home(2013)}]{Home:2013}
\bibinfo{author}{Home, J.P.}, \bibinfo{year}{2013}.
\newblock \bibinfo{title}{Chapter 4 - quantum science and metrology with
  mixed-species ion chains}.
\newblock \bibinfo{journal}{Advances in Atomic, Molecular, and Optical Physics}
  \bibinfo{volume}{62}, \bibinfo{pages}{231--277}.
\newblock \URLprefix \url{https://doi.org/10.1016/B978-0-12-408090-4.00004-9},
  \DOIprefix\doi{10.1016/B978-0-12-408090-4.00004-9}.
%Type = Article
\bibitem[{Huber et~al.(2014)Huber, Lambrecht, Schmidt, Karpa and
  Schaetz}]{Huber:2014}
\bibinfo{author}{Huber, T.}, \bibinfo{author}{Lambrecht, A.},
  \bibinfo{author}{Schmidt, J.}, \bibinfo{author}{Karpa, L.},
  \bibinfo{author}{Schaetz, T.}, \bibinfo{year}{2014}.
\newblock \bibinfo{title}{A far-off-resonance optical trap for a {B}a$^+$ ion}.
\newblock \bibinfo{journal}{Nature Communications} \bibinfo{volume}{5},
  \bibinfo{pages}{5587}.
\newblock \URLprefix \url{https://doi.org/10.1038/ncomms6587},
  \DOIprefix\doi{10.1038/ncomms6587}.
%Type = Article
\bibitem[{Ibaraki et~al.(2011)Ibaraki, Tanaka and Urabe}]{Ibaraki:2011}
\bibinfo{author}{Ibaraki, Y.}, \bibinfo{author}{Tanaka, U.},
  \bibinfo{author}{Urabe, S.}, \bibinfo{year}{2011}.
\newblock \bibinfo{title}{Detection of parametric resonance of trapped ions for
  micromotion compensation}.
\newblock \bibinfo{journal}{{Applied Physics B}} \bibinfo{volume}{105},
  \bibinfo{pages}{219--223}.
\newblock \URLprefix
  \url{https://link.springer.com/article/10.1007/s00340-011-4463-x},
  \DOIprefix\doi{10.1007/s00340-011-4463-x}.
%Type = Article
\bibitem[{Idziaszek et~al.(2011)Idziaszek, Simoni, Calarco and
  Julienne}]{Idziaszek2011mqd}
\bibinfo{author}{Idziaszek, Z.}, \bibinfo{author}{Simoni, A.},
  \bibinfo{author}{Calarco, T.}, \bibinfo{author}{Julienne, P.S.},
  \bibinfo{year}{2011}.
\newblock \bibinfo{title}{Multichannel quantum-defect theory for ultracold
  atom-ion collisions}.
\newblock \bibinfo{journal}{New J.~Phys.} \bibinfo{volume}{13},
  \bibinfo{pages}{083005}.
\newblock \DOIprefix\doi{10.1088/1367-2630/13/8/083005}.
%Type = Book
\bibitem[{Inguscio et~al.(2008)Inguscio, Ketterle and
  Salomon}]{Inguscio2008ufg}
\bibinfo{editor}{Inguscio, M.}, \bibinfo{editor}{Ketterle, W.},
  \bibinfo{editor}{Salomon, C.} (Eds.), \bibinfo{year}{2008}.
\newblock
  \bibinfo{title}{\href{https://www.iospress.nl/book/ultra-cold-fermi-gases/}{Ultra-cold
  Fermi Gases}}.
\newblock \bibinfo{publisher}{IOS Press, Amsterdam}.
\newblock \bibinfo{note}{{P}roceedings of the International School of Physics
  ``Enrico Fermi'', Course CLXIV, Varenna, 20-30 June 2006}.
%Type = Article
\bibitem[{Itano et~al.(1995)Itano, Bergquist, Bollinger and
  Wineland}]{Itano:1995}
\bibinfo{author}{Itano, W.M.}, \bibinfo{author}{Bergquist, J.C.},
  \bibinfo{author}{Bollinger, J.J.}, \bibinfo{author}{Wineland, D.J.},
  \bibinfo{year}{1995}.
\newblock \bibinfo{title}{Cooling methods in ion traps}.
\newblock \bibinfo{journal}{Phys. Scr.} \bibinfo{volume}{T59},
  \bibinfo{pages}{106}.
\newblock \DOIprefix\doi{10.1088/0031-8949/1995/T59/013}.
%Type = Article
\bibitem[{James(1998)}]{James:1998}
\bibinfo{author}{James, D.F.V.}, \bibinfo{year}{1998}.
\newblock \bibinfo{title}{Quantum dynamics of cold trapped ions with
  application to quantum computation}.
\newblock \bibinfo{journal}{Appl.~Phys.~B} \bibinfo{volume}{66},
  \bibinfo{pages}{181}.
\newblock \URLprefix \url{https://doi.org/10.1007/s003400050373},
  \DOIprefix\doi{10.1007/s003400050373}.
%Type = Incollection
\bibitem[{Jervis and Thywissen(2014)}]{Jervis2014mau}
\bibinfo{author}{Jervis, D.}, \bibinfo{author}{Thywissen, J.H.},
  \bibinfo{year}{2014}.
\newblock
  \bibinfo{title}{\href{https://www.worldscientific.com/doi/abs/10.1142/9781783264766_0002}{Making
  an Ultracold Gas}}, in: \bibinfo{editor}{T{\"{o}}rm{\"{a}}, P.},
  \bibinfo{editor}{Sengstock, K.} (Eds.), \bibinfo{booktitle}{Quantum Gas
  Experiments: Exploring Many-Body States}. \bibinfo{publisher}{Imperial
  College Press}. chapter~\bibinfo{chapter}{2}.
\newblock \URLprefix
  \url{https://www.worldscientific.com/doi/abs/10.1142/9781783264766_0002},
  \DOIprefix\doi{10.1142/9781783264766_0002}.
%Type = Article
\bibitem[{Joger et~al.(2017)Joger, F\"urst, Ewald, Feldker, Tomza and
  Gerritsma}]{Joger:2017}
\bibinfo{author}{Joger, J.}, \bibinfo{author}{F\"urst, H.},
  \bibinfo{author}{Ewald, N.}, \bibinfo{author}{Feldker, T.},
  \bibinfo{author}{Tomza, M.}, \bibinfo{author}{Gerritsma, R.},
  \bibinfo{year}{2017}.
\newblock \bibinfo{title}{Observation of collisions between cold {L}i atoms and
  {Y}b$^{+}$ ions}.
\newblock \bibinfo{journal}{Phys. Rev. A} \bibinfo{volume}{96},
  \bibinfo{pages}{030703(R)}.
\newblock \DOIprefix\doi{10.1103/PhysRevA.96.030703}.
%Type = Book
\bibitem[{Karpa(2019)}]{Karpa2019tsi}
\bibinfo{author}{Karpa, L.}, \bibinfo{year}{2019}.
\newblock \bibinfo{publisher}{Springer International Publishing},
  \bibinfo{address}{Cham}.
\newblock \URLprefix \url{https://doi.org/10.1007/978-3-030-27716-1},
  \DOIprefix\doi{10.1007/978-3-030-27716-1}.
%Type = Article
\bibitem[{Karpa(2021)}]{Karpa2021ioi}
\bibinfo{author}{Karpa, L.}, \bibinfo{year}{2021}.
\newblock \bibinfo{title}{Interactions of ions and ultracold neutral atom
  ensembles in composite optical dipole traps: Developments and perspectives}.
\newblock \bibinfo{journal}{Atoms} \bibinfo{volume}{9}.
\newblock \URLprefix \url{https://www.mdpi.com/2218-2004/9/3/39},
  \DOIprefix\doi{10.3390/atoms9030039}.
%Type = Article
\bibitem[{Karpa et~al.(2013)Karpa, Bylinskii, Gangloff, Cetina and
  Vuleti\ifmmode~\acute{c}\else \'{c}\fi{}}]{Karpa2013soi}
\bibinfo{author}{Karpa, L.}, \bibinfo{author}{Bylinskii, A.},
  \bibinfo{author}{Gangloff, D.}, \bibinfo{author}{Cetina, M.},
  \bibinfo{author}{Vuleti\ifmmode~\acute{c}\else \'{c}\fi{}, V.},
  \bibinfo{year}{2013}.
\newblock \bibinfo{title}{Suppression of ion transport due to long-lived
  subwavelength localization by an optical lattice}.
\newblock \bibinfo{journal}{Phys. Rev. Lett.} \bibinfo{volume}{111},
  \bibinfo{pages}{163002}.
\newblock \URLprefix
  \url{https://link.aps.org/doi/10.1103/PhysRevLett.111.163002},
  \DOIprefix\doi{10.1103/PhysRevLett.111.163002}.
%Type = Article
\bibitem[{Katz et~al.(2022)Katz, Pinkas, Akerman and Ozeri}]{Katz2022qld}
\bibinfo{author}{Katz, O.}, \bibinfo{author}{Pinkas, M.},
  \bibinfo{author}{Akerman, N.}, \bibinfo{author}{Ozeri, R.},
  \bibinfo{year}{2022}.
\newblock \bibinfo{title}{Quantum logic detection of collisions between single
  atom–ion pairs}.
\newblock \bibinfo{journal}{Nat. Phys.} \bibinfo{volume}{78},
  \bibinfo{pages}{1745--2481}.
\newblock \DOIprefix\doi{10.1038/s41567-022-01517-y}.
%Type = Article
\bibitem[{Keller et~al.(2015)Keller, Partnera, Burgermeister and
  Mehlst\"aubler}]{Keller:2015}
\bibinfo{author}{Keller, J.}, \bibinfo{author}{Partnera, H.L.},
  \bibinfo{author}{Burgermeister, T.}, \bibinfo{author}{Mehlst\"aubler, T.E.},
  \bibinfo{year}{2015}.
\newblock \bibinfo{title}{Precise determination of micromotion for trapped-ion
  optical clocks}.
\newblock \bibinfo{journal}{J.~App.~Phys.} \bibinfo{volume}{118},
  \bibinfo{pages}{104501}.
\newblock \URLprefix \url{https://aip.scitation.org/doi/10.1063/1.4930037},
  \DOIprefix\doi{https://doi.org/10.1063/1.4930037}.
%Type = Incollection
\bibitem[{Ketterle et~al.(1999)Ketterle, Durfee and
  Stamper-Kurn}]{Ketterle1999mpa}
\bibinfo{author}{Ketterle, W.}, \bibinfo{author}{Durfee, D.S.},
  \bibinfo{author}{Stamper-Kurn, D.M.}, \bibinfo{year}{1999}.
\newblock
  \bibinfo{title}{\href{http://ebooks.iospress.nl/volumearticle/33326}{Making,
  probing and understanding Bose-Einstein condensates}}, in:
  \bibinfo{editor}{Inguscio, M.}, \bibinfo{editor}{Stringari, S.},
  \bibinfo{editor}{Wieman, C.E.} (Eds.), \bibinfo{booktitle}{Bose-Einstein
  Condensation in Atomic Gases}. \bibinfo{publisher}{IOS Press}. volume
  \bibinfo{volume}{140}.
\newblock \bibinfo{note}{{P}roceedings of the International School of Physics
  ``Enrico Fermi'', Course CXL, Varenna, 7-17 July 1998}.
%Type = Article
\bibitem[{Ketterle and {van Druten}(1996)}]{Ketterle1996eco}
\bibinfo{author}{Ketterle, W.}, \bibinfo{author}{{van Druten}, N.J.},
  \bibinfo{year}{1996}.
\newblock \bibinfo{title}{Evaporative cooling of trapped atoms}.
\newblock \bibinfo{journal}{Adv. At. Mol. Opt. Phys.} \bibinfo{volume}{37},
  \bibinfo{pages}{181}.
\newblock \DOIprefix\doi{10.1016/S1049-250X(08)60101-9}.
%Type = Article
\bibitem[{Khanyile et~al.(2015)Khanyile, Shu and Brown}]{Khanyile2015}
\bibinfo{author}{Khanyile, N.B.}, \bibinfo{author}{Shu, G.},
  \bibinfo{author}{Brown, K.R.}, \bibinfo{year}{2015}.
\newblock \bibinfo{title}{Observation of vibrational overtones by
  single-molecule resonant photodissociation}.
\newblock \bibinfo{journal}{Nat. Comm.} \bibinfo{volume}{6},
  \bibinfo{pages}{7825}.
\newblock \URLprefix \url{https://www.nature.com/articles/ncomms8825}.
%Type = Article
\bibitem[{Kleinbach et~al.(2018)Kleinbach, Engel, Dieterle, L\"ow, Pfau and
  Meinert}]{Kleinbach2018iii}
\bibinfo{author}{Kleinbach, K.S.}, \bibinfo{author}{Engel, F.},
  \bibinfo{author}{Dieterle, T.}, \bibinfo{author}{L\"ow, R.},
  \bibinfo{author}{Pfau, T.}, \bibinfo{author}{Meinert, F.},
  \bibinfo{year}{2018}.
\newblock \bibinfo{title}{Ionic impurity in a {B}ose-{E}instein condensate at
  submicrokelvin temperatures}.
\newblock \bibinfo{journal}{Phys. Rev. Lett.} \bibinfo{volume}{120},
  \bibinfo{pages}{193401}.
\newblock \URLprefix
  \url{https://link.aps.org/doi/10.1103/PhysRevLett.120.193401},
  \DOIprefix\doi{10.1103/PhysRevLett.120.193401}.
%Type = Article
\bibitem[{K\"ohler et~al.(2006)K\"ohler, G\'oral and Julienne}]{Koehler2006poc}
\bibinfo{author}{K\"ohler, T.}, \bibinfo{author}{G\'oral, K.},
  \bibinfo{author}{Julienne, P.S.}, \bibinfo{year}{2006}.
\newblock \bibinfo{title}{Production of cold molecules via magnetically tunable
  {F}eshbach resonances}.
\newblock \bibinfo{journal}{Rev. Mod. Phys.} \bibinfo{volume}{78},
  \bibinfo{pages}{1311--1361}.
\newblock \URLprefix \url{https://link.aps.org/doi/10.1103/RevModPhys.78.1311},
  \DOIprefix\doi{10.1103/RevModPhys.78.1311}.
%Type = Article
\bibitem[{Krükow et~al.(2016)Krükow, Mohammadi, Härter and
  Denschlag}]{Kruekow2016}
\bibinfo{author}{Krükow, A.}, \bibinfo{author}{Mohammadi, A.},
  \bibinfo{author}{Härter, A.}, \bibinfo{author}{Denschlag, J.H.},
  \bibinfo{year}{2016}.
\newblock \bibinfo{title}{Reactive two-body and three-body collisions {of
  Ba}$^+$in an ultracold rb gas}.
\newblock \bibinfo{journal}{Phys. Rev. A} \bibinfo{volume}{94},
  \bibinfo{pages}{030701(R)}.
\newblock \URLprefix
  \url{https://journals.aps.org/pra/abstract/10.1103/PhysRevA.94.030701}.
%Type = Article
\bibitem[{Kwolek et~al.(2019)Kwolek, Goodman, Slayton, Bl\"umel, Wells,
  Narducci and Smith}]{Kwolek2019moc}
\bibinfo{author}{Kwolek, J.M.}, \bibinfo{author}{Goodman, D.S.},
  \bibinfo{author}{Slayton, B.}, \bibinfo{author}{Bl\"umel, R.},
  \bibinfo{author}{Wells, J.E.}, \bibinfo{author}{Narducci, F.A.},
  \bibinfo{author}{Smith, W.W.}, \bibinfo{year}{2019}.
\newblock \bibinfo{title}{Measurement of charge exchange between na and
  {C}a$^{+}$ in a hybrid trap}.
\newblock \bibinfo{journal}{Phys. Rev. A} \bibinfo{volume}{99},
  \bibinfo{pages}{052703}.
\newblock \URLprefix \url{https://link.aps.org/doi/10.1103/PhysRevA.99.052703},
  \DOIprefix\doi{10.1103/PhysRevA.99.052703}.
%Type = Article
\bibitem[{Lamata et~al.(2007)Lamata, Le\'{o}n, Sch\"{a}tz and
  Solano}]{Lamata:2007}
\bibinfo{author}{Lamata, L.}, \bibinfo{author}{Le\'{o}n, J.},
  \bibinfo{author}{Sch\"{a}tz, T.}, \bibinfo{author}{Solano, E.},
  \bibinfo{year}{2007}.
\newblock \bibinfo{title}{Dirac equation and quantum relativistic effects in a
  single trapped ion}.
\newblock \bibinfo{journal}{Phys.~Rev.~Lett.} \bibinfo{volume}{98},
  \bibinfo{pages}{253005}.
\newblock \URLprefix
  \url{https://journals.aps.org/prl/abstract/10.1103/PhysRevLett.98.253005},
  \DOIprefix\doi{10.1103/PhysRevLett.98.253005}.
%Type = Article
\bibitem[{Lange et~al.(2021)Lange, Peshkov, Huntemann, Tamm, Surzhykov and
  Peik}]{Lange:2021}
\bibinfo{author}{Lange, R.}, \bibinfo{author}{Peshkov, A.A.},
  \bibinfo{author}{Huntemann, N.}, \bibinfo{author}{Tamm, C.},
  \bibinfo{author}{Surzhykov, A.}, \bibinfo{author}{Peik, E.},
  \bibinfo{year}{2021}.
\newblock \bibinfo{title}{Lifetime of the $^{2}{F}_{7/2}$ level in {Y}b$^{+}$
  for spontaneous emission of electric octupole radiation}.
\newblock \bibinfo{journal}{Phys. Rev. Lett.} \bibinfo{volume}{127},
  \bibinfo{pages}{213001}.
\newblock \URLprefix
  \url{https://link.aps.org/doi/10.1103/PhysRevLett.127.213001},
  \DOIprefix\doi{10.1103/PhysRevLett.127.213001}.
%Type = Article
\bibitem[{Leibfried et~al.(2003)Leibfried, Blatt, Monroe and
  Wineland}]{Leibfried:2003}
\bibinfo{author}{Leibfried, D.}, \bibinfo{author}{Blatt, R.},
  \bibinfo{author}{Monroe, C.}, \bibinfo{author}{Wineland, D.},
  \bibinfo{year}{2003}.
\newblock \bibinfo{title}{Quantum dynamics of single trapped ions}.
\newblock \bibinfo{journal}{Rev.~Mod.~Phys.} \bibinfo{volume}{75},
  \bibinfo{pages}{281}.
\newblock \URLprefix \url{https://link.aps.org/doi/10.1103/RevModPhys.75.281},
  \DOIprefix\doi{10.1103/RevModPhys.75.281}.
%Type = Article
\bibitem[{Leibfried et~al.(1996)Leibfried, Meekhof, King, Monroe, Itano and
  Wineland}]{Leibfried:1996}
\bibinfo{author}{Leibfried, D.}, \bibinfo{author}{Meekhof, D.M.},
  \bibinfo{author}{King, B.E.}, \bibinfo{author}{Monroe, C.},
  \bibinfo{author}{Itano, W.M.}, \bibinfo{author}{Wineland, D.J.},
  \bibinfo{year}{1996}.
\newblock \bibinfo{title}{Experimental determination of the motional quantum
  state of a trapped atom}.
\newblock \bibinfo{journal}{Phys.~Rev.~Lett.} \bibinfo{volume}{77},
  \bibinfo{pages}{4281--4285}.
\newblock \URLprefix
  \url{https://journals.aps.org/prl/abstract/10.1103/PhysRevLett.77.4281},
  \DOIprefix\doi{10.1103/PhysRevLett.77.4281}.
%Type = Article
\bibitem[{Leibfried et~al.(1998)Leibfried, Pfau and Monroe}]{Leibfried:1998}
\bibinfo{author}{Leibfried, D.}, \bibinfo{author}{Pfau, T.},
  \bibinfo{author}{Monroe, C.R.}, \bibinfo{year}{1998}.
\newblock \bibinfo{title}{Shadows and mirrors: Reconstructing quantum states of
  atom motion}.
\newblock \bibinfo{journal}{Phys. Today} \bibinfo{volume}{51},
  \bibinfo{pages}{22}.
\newblock \URLprefix
  \url{https://physicstoday.scitation.org/doi/10.1063/1.882256},
  \DOIprefix\doi{10.1063/1.882256}.
%Type = Article
\bibitem[{Li et~al.(2020)Li, Jyothi, Li, Kłos, Petrov, Brown and
  Kotochigova}]{Hui2020pmc}
\bibinfo{author}{Li, H.}, \bibinfo{author}{Jyothi, S.}, \bibinfo{author}{Li,
  M.}, \bibinfo{author}{Kłos, J.}, \bibinfo{author}{Petrov, A.},
  \bibinfo{author}{Brown, K.R.}, \bibinfo{author}{Kotochigova, S.},
  \bibinfo{year}{2020}.
\newblock \bibinfo{title}{Photon-mediated charge exchange reactions between
  $^{39}${K} atoms and $^{40}${C}a$^+$ ions in a hybrid trap}.
\newblock \bibinfo{journal}{Phys. Chem. Chem. Phys.} \bibinfo{volume}{22},
  \bibinfo{pages}{10870--10881}.
\newblock \URLprefix \url{http://dx.doi.org/10.1039/D0CP01131B},
  \DOIprefix\doi{10.1039/D0CP01131B}.
%Type = Article
\bibitem[{Li et~al.(2019)Li, Mills, Puri, Petrov, Hudson and
  Kotochigova}]{Li2019ean}
\bibinfo{author}{Li, M.}, \bibinfo{author}{Mills, M.}, \bibinfo{author}{Puri,
  P.}, \bibinfo{author}{Petrov, A.}, \bibinfo{author}{Hudson, E.R.},
  \bibinfo{author}{Kotochigova, S.}, \bibinfo{year}{2019}.
\newblock \bibinfo{title}{Excitation-assisted nonadiabatic charge-transfer
  reaction in a mixed atom-ion system}.
\newblock \bibinfo{journal}{Phys. Rev. A} \bibinfo{volume}{99},
  \bibinfo{pages}{062706}.
\newblock \URLprefix \url{https://link.aps.org/doi/10.1103/PhysRevA.99.062706},
  \DOIprefix\doi{10.1103/PhysRevA.99.062706}.
%Type = Article
\bibitem[{Lougovski et~al.(2006)Lougovski, Walther and Solano}]{Lougovski:2006}
\bibinfo{author}{Lougovski, P.}, \bibinfo{author}{Walther, H.},
  \bibinfo{author}{Solano, E.}, \bibinfo{year}{2006}.
\newblock \bibinfo{title}{Instantaneous measurement of field quadrature moments
  and entanglement}.
\newblock \bibinfo{journal}{Eur.~Phys.~J.~D} \bibinfo{volume}{38},
  \bibinfo{pages}{423--426}.
\newblock \URLprefix
  \url{https://link.springer.com/article/10.1140/epjd/e2006-00085-3},
  \DOIprefix\doi{10.1140/epjd/e2006-00085-3}.
%Type = Article
\bibitem[{Mahdian et~al.(2021)Mahdian, Krükow and Denschlag}]{Mahdian2021doo}
\bibinfo{author}{Mahdian, A.}, \bibinfo{author}{Krükow, A.},
  \bibinfo{author}{Denschlag, J.H.}, \bibinfo{year}{2021}.
\newblock \bibinfo{title}{Direct observation of swap cooling in
  atom{\textendash}ion collisions}.
\newblock \bibinfo{journal}{New Journal of Physics} \bibinfo{volume}{23},
  \bibinfo{pages}{065008}.
\newblock \URLprefix \url{https://doi.org/10.1088/1367-2630/ac0575},
  \DOIprefix\doi{10.1088/1367-2630/ac0575}.
%Type = Article
\bibitem[{Major and Dehmelt(1968)}]{Major:1968}
\bibinfo{author}{Major, F.G.}, \bibinfo{author}{Dehmelt, H.G.},
  \bibinfo{year}{1968}.
\newblock \bibinfo{title}{Exchange-collision technique for the rf spectroscopy
  of stored ions}.
\newblock \bibinfo{journal}{Phys. Rev.} \bibinfo{volume}{170},
  \bibinfo{pages}{91}.
\newblock \DOIprefix\doi{10.1103/PhysRev.170.91}.
%Type = Article
\bibitem[{Meekhof et~al.(1996)Meekhof, Monroe, King, Itano and
  Wineland}]{Meekhof:1996}
\bibinfo{author}{Meekhof, D.M.}, \bibinfo{author}{Monroe, C.},
  \bibinfo{author}{King, B.E.}, \bibinfo{author}{Itano, W.M.},
  \bibinfo{author}{Wineland, D.J.}, \bibinfo{year}{1996}.
\newblock \bibinfo{title}{Generation of nonclassical motional states of a
  trapped atom.}
\newblock \bibinfo{journal}{Phys.~Rev.~Lett.} \bibinfo{volume}{76},
  \bibinfo{pages}{1796--1799}.
\newblock \URLprefix
  \url{https://journals.aps.org/prl/abstract/10.1103/PhysRevLett.76.1796},
  \DOIprefix\doi{10.1103/PhysRevLett.76.1796}.
%Type = Phdthesis
\bibitem[{Meir(2016)}]{Meir:2016PhD}
\bibinfo{author}{Meir, Z.}, \bibinfo{year}{2016}.
\newblock \bibinfo{title}{Dynamics of a single ground-state cooled and trapped
  ion colliding with ultracold atoms: {A} micromotion tale}.
\newblock Ph.D. thesis. {Weizmann Institute}.
\newblock \URLprefix
  \url{https://www.weizmann.ac.il/complex/ozeri/sites/complex.ozeri/files/uploads/ziv_thesis.pdf}.
%Type = Article
\bibitem[{Meir et~al.(2017a)Meir, Sikorsky, Akerman, Ben-shlomi, Pinkas and
  Ozeri}]{Meir:2017}
\bibinfo{author}{Meir, Z.}, \bibinfo{author}{Sikorsky, T.},
  \bibinfo{author}{Akerman, N.}, \bibinfo{author}{Ben-shlomi, R.},
  \bibinfo{author}{Pinkas, M.}, \bibinfo{author}{Ozeri, R.},
  \bibinfo{year}{2017}a.
\newblock \bibinfo{title}{Single-shot energy measurement of a single atom and
  the direct reconstruction of its energy distribution}.
\newblock \bibinfo{journal}{Phys. Rev. A} \bibinfo{volume}{96},
  \bibinfo{pages}{020701(R)}.
\newblock \URLprefix
  \url{https://journals.aps.org/pra/abstract/10.1103/PhysRevA.96.020701},
  \DOIprefix\doi{10.1103/PhysRevA.96.020701}.
%Type = Article
\bibitem[{Meir et~al.(2016)Meir, Sikorsky, Ben-shlomi, Akerman, Dallal and
  Ozeri}]{Meir2016doa}
\bibinfo{author}{Meir, Z.}, \bibinfo{author}{Sikorsky, T.},
  \bibinfo{author}{Ben-shlomi, R.}, \bibinfo{author}{Akerman, N.},
  \bibinfo{author}{Dallal, Y.}, \bibinfo{author}{Ozeri, R.},
  \bibinfo{year}{2016}.
\newblock \bibinfo{title}{Dynamics of a ground-state cooled ion colliding with
  ultra-cold atoms}.
\newblock \bibinfo{journal}{Phys.~Rev.~Lett.} \bibinfo{volume}{117},
  \bibinfo{pages}{243401}.
\newblock \DOIprefix\doi{10.1103/PhysRevLett.117.243401}.
%Type = Article
\bibitem[{Meir et~al.(2017b)Meir, Sikorsky, Ben-shlomi, Akerman, Pinkas, Dallal
  and Ozeri}]{Meir2017eaf}
\bibinfo{author}{Meir, Z.}, \bibinfo{author}{Sikorsky, T.},
  \bibinfo{author}{Ben-shlomi, R.}, \bibinfo{author}{Akerman, N.},
  \bibinfo{author}{Pinkas, M.}, \bibinfo{author}{Dallal, Y.},
  \bibinfo{author}{Ozeri, R.}, \bibinfo{year}{2017}b.
\newblock \bibinfo{title}{Experimental apparatus for overlapping a ground-state
  cooled ion with ultracold atoms}.
\newblock \bibinfo{journal}{Journal of Modern Optics} \bibinfo{volume}{65},
  \bibinfo{pages}{501--519}.
\newblock \DOIprefix\doi{10.1080/09500340.2017.1397217}.
%Type = Book
\bibitem[{Metcalf and van~der Straten(1999)}]{Metcalf1999book}
\bibinfo{author}{Metcalf, H.J.}, \bibinfo{author}{van~der Straten, P.},
  \bibinfo{year}{1999}.
\newblock
  \bibinfo{title}{\href{http://www.springer.com/gp/book/9780387987286}{Laser
  Cooling and Trapping}}.
\newblock \bibinfo{publisher}{Springer, New York}.
\newblock \DOIprefix\doi{10.1007/978-1-4612-1470-0}.
%Type = Article
\bibitem[{Meyer and Wester(2017)}]{Meyer2017imr}
\bibinfo{author}{Meyer, J.}, \bibinfo{author}{Wester, R.},
  \bibinfo{year}{2017}.
\newblock \bibinfo{title}{Ion–molecule reaction dynamics}.
\newblock \bibinfo{journal}{Annual Review of Physical Chemistry}
  \bibinfo{volume}{68}, \bibinfo{pages}{333--353}.
\newblock \URLprefix
  \url{https://doi.org/10.1146/annurev-physchem-052516-044918},
  \DOIprefix\doi{10.1146/annurev-physchem-052516-044918},
  \href{http://arxiv.org/abs/https://doi.org/10.1146/annurev-physchem-052516-044918}{{\tt
  arXiv:https://doi.org/10.1146/annurev-physchem-052516-044918}}.
  \bibinfo{note}{pMID: 28463654}.
%Type = Article
\bibitem[{Midya et~al.(2016)Midya, Tomza, Schmidt and Lemeshko}]{Midya2016roc}
\bibinfo{author}{Midya, B.}, \bibinfo{author}{Tomza, M.},
  \bibinfo{author}{Schmidt, R.}, \bibinfo{author}{Lemeshko, M.},
  \bibinfo{year}{2016}.
\newblock \bibinfo{title}{Rotation of cold molecular ions inside a
  {B}ose-{E}instein condensate}.
\newblock \bibinfo{journal}{Phys. Rev. A} \bibinfo{volume}{94},
  \bibinfo{pages}{041601}.
\newblock \URLprefix \url{https://link.aps.org/doi/10.1103/PhysRevA.94.041601},
  \DOIprefix\doi{10.1103/PhysRevA.94.041601}.
%Type = Article
\bibitem[{Mills et~al.(2019)Mills, Puri, Li, Schowalter, Dunning, Schneider,
  Kotochigova and Hudson}]{Mills2019ees}
\bibinfo{author}{Mills, M.}, \bibinfo{author}{Puri, P.}, \bibinfo{author}{Li,
  M.}, \bibinfo{author}{Schowalter, S.J.}, \bibinfo{author}{Dunning, A.},
  \bibinfo{author}{Schneider, C.}, \bibinfo{author}{Kotochigova, S.},
  \bibinfo{author}{Hudson, E.R.}, \bibinfo{year}{2019}.
\newblock \bibinfo{title}{Engineering excited-state interactions at ultracold
  temperatures}.
\newblock \bibinfo{journal}{Phys. Rev. Lett.} \bibinfo{volume}{122},
  \bibinfo{pages}{233401}.
\newblock \URLprefix
  \url{https://link.aps.org/doi/10.1103/PhysRevLett.122.233401},
  \DOIprefix\doi{10.1103/PhysRevLett.122.233401}.
%Type = Article
\bibitem[{Mohammadi et~al.(2021)Mohammadi, Krükow, Mahdian, Dei{\ss},
  P{\'{e}}rez-R{\'{\i}}os, da~Silva, Raoult, Dulieu and
  Denschlag}]{Mohammadi:2021}
\bibinfo{author}{Mohammadi, A.}, \bibinfo{author}{Krükow, A.},
  \bibinfo{author}{Mahdian, A.}, \bibinfo{author}{Dei{\ss}, M.},
  \bibinfo{author}{P{\'{e}}rez-R{\'{\i}}os, J.}, \bibinfo{author}{da~Silva,
  H.}, \bibinfo{author}{Raoult, M.}, \bibinfo{author}{Dulieu, O.},
  \bibinfo{author}{Denschlag, J.H.}, \bibinfo{year}{2021}.
\newblock \bibinfo{title}{Life and death of a cold {BaRb}$^+$ molecule inside
  an ultracold cloud of {R}b atoms}.
\newblock \bibinfo{journal}{Phys. Rev. Research} \bibinfo{volume}{3},
  \bibinfo{pages}{013196}.
\newblock \URLprefix
  \url{https://journals.aps.org/prresearch/abstract/10.1103/PhysRevResearch.3.013196}.
%Type = Article
\bibitem[{Mohammadi et~al.(2019)Mohammadi, Wolf, Krükow, Deiß and
  Denschlag}]{Mohammadi:2019}
\bibinfo{author}{Mohammadi, A.}, \bibinfo{author}{Wolf, J.},
  \bibinfo{author}{Krükow, A.}, \bibinfo{author}{Deiß, M.},
  \bibinfo{author}{Denschlag, J.H.}, \bibinfo{year}{2019}.
\newblock \bibinfo{title}{Minimizing rf-induced excess micromotion of a trapped
  ion with the help of ultracold atoms}.
\newblock \bibinfo{journal}{Appl.~Phys.~B} \bibinfo{volume}{125},
  \bibinfo{pages}{122}.
\newblock \URLprefix \url{https://doi.org/10.1007/s00340-019-7223-y},
  \DOIprefix\doi{10.1007/s00340-019-7223-y}.
%Type = Incollection
\bibitem[{Mokhberi et~al.(2020)Mokhberi, Hennrich and
  Schmidt-Kaler}]{Mokhberi2020tri}
\bibinfo{author}{Mokhberi, A.}, \bibinfo{author}{Hennrich, M.},
  \bibinfo{author}{Schmidt-Kaler, F.}, \bibinfo{year}{2020}.
\newblock \bibinfo{title}{Chapter four - trapped rydberg ions: A new platform
  for quantum information processing}, \bibinfo{publisher}{Academic Press}.
  volume~\bibinfo{volume}{69} of \textit{\bibinfo{series}{Advances In Atomic,
  Molecular, and Optical Physics}}, pp. \bibinfo{pages}{233--306}.
\newblock \URLprefix
  \url{https://www.sciencedirect.com/science/article/pii/S1049250X20300045},
  \DOIprefix\doi{https://doi.org/10.1016/bs.aamop.2020.04.004}.
%Type = Article
\bibitem[{Monroe et~al.(2021)Monroe, Campbell, Duan, Gong, Gorshkov, Hess,
  Islam, Kim, Linke, Pagano, Richerme, Senko and Yao}]{Monroe2021pqs}
\bibinfo{author}{Monroe, C.}, \bibinfo{author}{Campbell, W.C.},
  \bibinfo{author}{Duan, L.M.}, \bibinfo{author}{Gong, Z.X.},
  \bibinfo{author}{Gorshkov, A.V.}, \bibinfo{author}{Hess, P.W.},
  \bibinfo{author}{Islam, R.}, \bibinfo{author}{Kim, K.},
  \bibinfo{author}{Linke, N.M.}, \bibinfo{author}{Pagano, G.},
  \bibinfo{author}{Richerme, P.}, \bibinfo{author}{Senko, C.},
  \bibinfo{author}{Yao, N.Y.}, \bibinfo{year}{2021}.
\newblock \bibinfo{title}{Programmable quantum simulations of spin systems with
  trapped ions}.
\newblock \bibinfo{journal}{Rev. Mod. Phys.} \bibinfo{volume}{93},
  \bibinfo{pages}{025001}.
\newblock \URLprefix
  \url{https://link.aps.org/doi/10.1103/RevModPhys.93.025001},
  \DOIprefix\doi{10.1103/RevModPhys.93.025001}.
%Type = Article
\bibitem[{Morigi and Walther(2001)}]{Morigi:2001}
\bibinfo{author}{Morigi, G.}, \bibinfo{author}{Walther, H.},
  \bibinfo{year}{2001}.
\newblock \bibinfo{title}{Two-species coulomb chains for quantum information}.
\newblock \bibinfo{journal}{Eur.~Phys.~J.~D} \bibinfo{volume}{13},
  \bibinfo{pages}{261–269}.
\newblock \URLprefix \url{https://doi.org/10.1007/s100530170275},
  \DOIprefix\doi{10.1007/s100530170275}.
%Type = Article
\bibitem[{Mur-Petit et~al.(2012)Mur-Petit, Garc{\'{\i}}a-Ripoll,
  P{\'{e}}rez-R{\'{\i}}os, Campos-Mart{\'{\i}}nez, Hern{\'{a}}ndez and
  Willitsch}]{MurPetit:2012}
\bibinfo{author}{Mur-Petit, J.}, \bibinfo{author}{Garc{\'{\i}}a-Ripoll, J.J.},
  \bibinfo{author}{P{\'{e}}rez-R{\'{\i}}os, J.},
  \bibinfo{author}{Campos-Mart{\'{\i}}nez, J.},
  \bibinfo{author}{Hern{\'{a}}ndez, M.I.}, \bibinfo{author}{Willitsch, S.},
  \bibinfo{year}{2012}.
\newblock \bibinfo{title}{Temperature-independent quantum logic for molecular
  spectroscopy}.
\newblock \bibinfo{journal}{Phys.~Rev.~A} \bibinfo{volume}{85},
  \bibinfo{pages}{022308}.
\newblock \DOIprefix\doi{10.1103/physreva.85.022308}.
%Type = Article
\bibitem[{Myerson et~al.(2008)Myerson, Szwer, Webster, Allcock, Curtis, Imreh,
  Sherman, Stacey, Steane and Lucas}]{Myerson:2008}
\bibinfo{author}{Myerson, A.H.}, \bibinfo{author}{Szwer, D.J.},
  \bibinfo{author}{Webster, S.C.}, \bibinfo{author}{Allcock, D.T.C.},
  \bibinfo{author}{Curtis, M.J.}, \bibinfo{author}{Imreh, G.},
  \bibinfo{author}{Sherman, J.A.}, \bibinfo{author}{Stacey, D.N.},
  \bibinfo{author}{Steane, A.M.}, \bibinfo{author}{Lucas, D.M.},
  \bibinfo{year}{2008}.
\newblock \bibinfo{title}{High-fidelity readout of trapped-ion qubits}.
\newblock \bibinfo{journal}{Phys.~Rev.~Lett.} \bibinfo{volume}{100},
  \bibinfo{pages}{200502}.
\newblock \URLprefix
  \url{https://journals.aps.org/prl/abstract/10.1103/PhysRevLett.100.200502},
  \DOIprefix\doi{10.1103/PhysRevLett.100.200502}.
%Type = Article
\bibitem[{Nadlinger et~al.(2021)Nadlinger, Drmota, Main, Nichol, Araneda,
  Srinivas, Stephenson, Ballance and Lucas}]{Nadlinger:2021}
\bibinfo{author}{Nadlinger, D.P.}, \bibinfo{author}{Drmota, P.},
  \bibinfo{author}{Main, D.}, \bibinfo{author}{Nichol, B.C.},
  \bibinfo{author}{Araneda, G.}, \bibinfo{author}{Srinivas, R.},
  \bibinfo{author}{Stephenson, L.J.}, \bibinfo{author}{Ballance, C.J.},
  \bibinfo{author}{Lucas, D.M.}, \bibinfo{year}{2021}.
\newblock \bibinfo{title}{Micromotion minimisation by synchronous detection of
  parametrically excited motion}.
\newblock \bibinfo{journal}{arXiv:2107.00056} \URLprefix
  \url{https://arxiv.org/abs/2107.00056}.
%Type = Article
\bibitem[{Najafian et~al.(2020)Najafian, Meir, Sinhal and
  Willitsch}]{Najafian2020iom}
\bibinfo{author}{Najafian, K.}, \bibinfo{author}{Meir, Z.},
  \bibinfo{author}{Sinhal, M.}, \bibinfo{author}{Willitsch, S.},
  \bibinfo{year}{2020}.
\newblock \bibinfo{title}{Identification of molecular quantum states using
  phase-sensitive forces}.
\newblock \bibinfo{journal}{Nature Communications} \bibinfo{volume}{11}.
\newblock \DOIprefix\doi{10.1038/s41467-020-18170-9}.
%Type = Article
\bibitem[{Narayanan et~al.(2011)Narayanan, Daniilidis, Möller, Clark, Ziesel,
  Singer, Schmidt-Kaler and Häffner}]{Narayanan:2011}
\bibinfo{author}{Narayanan, S.}, \bibinfo{author}{Daniilidis, N.},
  \bibinfo{author}{Möller, S.A.}, \bibinfo{author}{Clark, R.},
  \bibinfo{author}{Ziesel, F.}, \bibinfo{author}{Singer, K.},
  \bibinfo{author}{Schmidt-Kaler, F.}, \bibinfo{author}{Häffner, H.},
  \bibinfo{year}{2011}.
\newblock \bibinfo{title}{Electric field compensation and sensing with a single
  ion in a planar trap}.
\newblock \bibinfo{journal}{{Journal of Applied Physics}}
  \bibinfo{volume}{110}, \bibinfo{pages}{114909}.
\newblock \URLprefix \url{https://doi.org/10.1063/1.3665647},
  \DOIprefix\doi{10.1063/1.3665647}.
%Type = Article
\bibitem[{Negretti et~al.(2014)Negretti, Gerritsma, Idziaszek, Schmidt-Kaler
  and Calarco}]{Negretti:2014}
\bibinfo{author}{Negretti, A.}, \bibinfo{author}{Gerritsma, R.},
  \bibinfo{author}{Idziaszek, Z.}, \bibinfo{author}{Schmidt-Kaler, F.},
  \bibinfo{author}{Calarco, T.}, \bibinfo{year}{2014}.
\newblock \bibinfo{title}{Generalized {K}ronig-{P}enney model for ultracold
  atomic quantum systems}.
\newblock \bibinfo{journal}{Phys. Rev. B} \bibinfo{volume}{90},
  \bibinfo{pages}{155426}.
\newblock \DOIprefix\doi{10.1103/PhysRevB.90.155426}.
%Type = Article
\bibitem[{Niranjan et~al.(2021)Niranjan, Prakash and
  Rangwala}]{Niranjan2021aom}
\bibinfo{author}{Niranjan, M.}, \bibinfo{author}{Prakash, A.},
  \bibinfo{author}{Rangwala, S.A.}, \bibinfo{year}{2021}.
\newblock \bibinfo{title}{Analysis of multipolar linear {P}aul traps for
  ion–atom ultracold collision experiments}.
\newblock \bibinfo{journal}{Atoms} \bibinfo{volume}{9}.
\newblock \URLprefix \url{https://www.mdpi.com/2218-2004/9/3/38},
  \DOIprefix\doi{10.3390/atoms9030038}.
%Type = Article
\bibitem[{Novan et~al.(2012)Novan, Smith and Hadzibabic}]{Navon2021qgi}
\bibinfo{author}{Novan, N.}, \bibinfo{author}{Smith, R.P.},
  \bibinfo{author}{Hadzibabic, Z.}, \bibinfo{year}{2012}.
\newblock \bibinfo{title}{Quantum gases in optical boxes}.
\newblock \bibinfo{journal}{Nat. Phys.} \bibinfo{volume}{17},
  \bibinfo{pages}{1334--1341}.
\newblock \DOIprefix\doi{10.1038/s41567-021-01403-z}.
%Type = Article
\bibitem[{Nötzold et~al.(2020)Nötzold, Hassan, Tauch, Endres, Wester and
  Weidemüller}]{Notzold2020TMI}
\bibinfo{author}{Nötzold, M.}, \bibinfo{author}{Hassan, S.Z.},
  \bibinfo{author}{Tauch, J.}, \bibinfo{author}{Endres, E.},
  \bibinfo{author}{Wester, R.}, \bibinfo{author}{Weidemüller, M.},
  \bibinfo{year}{2020}.
\newblock \bibinfo{journal}{Appl. Sci.} \bibinfo{volume}{10}.
\newblock \URLprefix \url{https://doi.org/10.3390/app10155264}.
%Type = Article
\bibitem[{Oghittu et~al.(2021)Oghittu, Johannsen, Negretti and
  Gerritsma}]{Oghittu2021doa}
\bibinfo{author}{Oghittu, L.}, \bibinfo{author}{Johannsen, M.},
  \bibinfo{author}{Negretti, A.}, \bibinfo{author}{Gerritsma, R.},
  \bibinfo{year}{2021}.
\newblock \bibinfo{title}{Dynamics of a trapped ion in a quantum gas: Effects
  of particle statistics}.
\newblock \bibinfo{journal}{Phys. Rev. A} \bibinfo{volume}{104},
  \bibinfo{pages}{053314}.
\newblock \URLprefix
  \url{https://link.aps.org/doi/10.1103/PhysRevA.104.053314},
  \DOIprefix\doi{10.1103/PhysRevA.104.053314}.
%Type = Article
\bibitem[{{Paul}(1990)}]{Paul:1990}
\bibinfo{author}{{Paul}, W.}, \bibinfo{year}{1990}.
\newblock \bibinfo{title}{{Electromagnetic traps for charged and neutral
  particles}}.
\newblock \bibinfo{journal}{Rev.~Mod.~Phys.} \bibinfo{volume}{62},
  \bibinfo{pages}{531--540}.
\newblock \URLprefix \url{https://doi.org/10.1103/RevModPhys.62.531},
  \DOIprefix\doi{10.1103/RevModPhys.62.531}.
%Type = Article
\bibitem[{Pedregosa et~al.(2010)Pedregosa, Champenois, Houssin and
  Knoop}]{Pedregosa2010ACR}
\bibinfo{author}{Pedregosa, J.}, \bibinfo{author}{Champenois, C.},
  \bibinfo{author}{Houssin, M.}, \bibinfo{author}{Knoop, M.},
  \bibinfo{year}{2010}.
\newblock \bibinfo{journal}{Int. J. Mass Spectrom.} \bibinfo{volume}{290}.
\newblock \URLprefix \url{https://doi.org/10.1016/j.ijms.2009.12.009}.
%Type = Article
\bibitem[{Perego et~al.(2020)Perego, Duca and Sias}]{Perego2020eoi}
\bibinfo{author}{Perego, E.}, \bibinfo{author}{Duca, L.},
  \bibinfo{author}{Sias, C.}, \bibinfo{year}{2020}.
\newblock \bibinfo{title}{Electro-optical ion trap for experiments with
  atom-ion quantum hybrid systems}.
\newblock \bibinfo{journal}{Applied Sciences} \bibinfo{volume}{10}.
\newblock \URLprefix \url{https://www.mdpi.com/2076-3417/10/7/2222},
  \DOIprefix\doi{10.3390/app10072222}.
%Type = Article
\bibitem[{Pinkas et~al.(2020)Pinkas, Meir, Sikorsky, Ben-Shlomi, Akerman and
  Ozeri}]{Pinkas2020eoi}
\bibinfo{author}{Pinkas, M.}, \bibinfo{author}{Meir, Z.},
  \bibinfo{author}{Sikorsky, T.}, \bibinfo{author}{Ben-Shlomi, R.},
  \bibinfo{author}{Akerman, N.}, \bibinfo{author}{Ozeri, R.},
  \bibinfo{year}{2020}.
\newblock \bibinfo{title}{Effect of ion-trap parameters on energy distributions
  of ultra-cold atom{\textendash}ion mixtures}.
\newblock \bibinfo{journal}{New J. Phys.} \bibinfo{volume}{22},
  \bibinfo{pages}{013047}.
\newblock \DOIprefix\doi{10.1088/1367-2630/ab6792}.
%Type = Article
\bibitem[{Puri et~al.(2019)Puri, Mills, Simbotin, Jr., C\^ot\'e, Schneider,
  Suits and Hudson}]{Puri2019rbi}
\bibinfo{author}{Puri, P.}, \bibinfo{author}{Mills, M.},
  \bibinfo{author}{Simbotin, I.}, \bibinfo{author}{Jr., J.A.M.},
  \bibinfo{author}{C\^ot\'e, R.}, \bibinfo{author}{Schneider, C.},
  \bibinfo{author}{Suits, A.G.}, \bibinfo{author}{Hudson, E.R.},
  \bibinfo{year}{2019}.
\newblock \bibinfo{title}{Reaction blockading in a reaction between an excited
  atom and a charged molecule at low collision energy}.
\newblock \bibinfo{journal}{Nature Chemistry} \bibinfo{volume}{11}.
\newblock \DOIprefix\doi{10.1038/s41557-019-0264-3}.
%Type = Article
\bibitem[{Pérez-Ríos(2021a)}]{PerezRios2021cca}
\bibinfo{author}{Pérez-Ríos, J.}, \bibinfo{year}{2021}a.
\newblock \bibinfo{title}{Cold chemistry: a few-body perspective on impurity
  physics of a single ion in an ultracold bath}.
\newblock \bibinfo{journal}{Molecular Physics} \bibinfo{volume}{119},
  \bibinfo{pages}{e1881637}.
\newblock \URLprefix \url{https://doi.org/10.1080/00268976.2021.1881637},
  \DOIprefix\doi{10.1080/00268976.2021.1881637},
  \href{http://arxiv.org/abs/https://doi.org/10.1080/00268976.2021.1881637}{{\tt
  arXiv:https://doi.org/10.1080/00268976.2021.1881637}}.
%Type = Article
\bibitem[{Pérez-Ríos(2021b)}]{PerezRios2021efd}
\bibinfo{author}{Pérez-Ríos, J.}, \bibinfo{year}{2021}b.
\newblock \bibinfo{title}{Electric-field dissociation of weakly bound molecular
  ions}.
\newblock \bibinfo{journal}{Phys. Rev. A} \bibinfo{volume}{104},
  \bibinfo{pages}{L031302}.
\newblock \URLprefix
  \url{https://link.aps.org/doi/10.1103/PhysRevA.104.L031302},
  \DOIprefix\doi{10.1103/PhysRevA.104.L031302}.
%Type = Article
\bibitem[{Pérez-Ríos and Sanz(2013)}]{perezrios2013hda}
\bibinfo{author}{Pérez-Ríos, J.}, \bibinfo{author}{Sanz, A.S.},
  \bibinfo{year}{2013}.
\newblock \bibinfo{title}{How does a magnetic trap work?}
\newblock \bibinfo{journal}{American Journal of Physics} \bibinfo{volume}{81},
  \bibinfo{pages}{836--843}.
\newblock \URLprefix \url{https://doi.org/10.1119/1.4819167},
  \DOIprefix\doi{10.1119/1.4819167},
  \href{http://arxiv.org/abs/https://doi.org/10.1119/1.4819167}{{\tt
  arXiv:https://doi.org/10.1119/1.4819167}}.
%Type = Article
\bibitem[{Ratschbacher et~al.(2013)Ratschbacher, Sias, Carcagni, Silver, Zipkes
  and K\"ohl}]{Ratschbacher2013doa}
\bibinfo{author}{Ratschbacher, L.}, \bibinfo{author}{Sias, C.},
  \bibinfo{author}{Carcagni, L.}, \bibinfo{author}{Silver, J.M.},
  \bibinfo{author}{Zipkes, C.}, \bibinfo{author}{K\"ohl, M.},
  \bibinfo{year}{2013}.
\newblock \bibinfo{title}{Decoherence of a single-ion qubit immersed in a
  spin-polarized atomic bath}.
\newblock \bibinfo{journal}{Phys. Rev. Lett.} \bibinfo{volume}{110},
  \bibinfo{pages}{160402}.
\newblock \URLprefix
  \url{https://link.aps.org/doi/10.1103/PhysRevLett.110.160402},
  \DOIprefix\doi{10.1103/PhysRevLett.110.160402}.
%Type = Article
\bibitem[{Ratschbacher et~al.(2012)Ratschbacher, Zipkes, Sias and
  K\"ohl}]{Ratschbacher:2012}
\bibinfo{author}{Ratschbacher, L.}, \bibinfo{author}{Zipkes, C.},
  \bibinfo{author}{Sias, C.}, \bibinfo{author}{K\"ohl, M.},
  \bibinfo{year}{2012}.
\newblock \bibinfo{title}{Controlling chemical reactions of a single particle}.
\newblock \bibinfo{journal}{Nat. Phys.} \bibinfo{volume}{8},
  \bibinfo{pages}{649--652}.
\newblock \DOIprefix\doi{10.1038/nphys2373}.
%Type = Article
\bibitem[{Ravi et~al.(2012)Ravi, Lee, Sharma, Werth and Rangwala}]{Ravi2012cas}
\bibinfo{author}{Ravi, K.}, \bibinfo{author}{Lee, S.}, \bibinfo{author}{Sharma,
  A.}, \bibinfo{author}{Werth, G.}, \bibinfo{author}{Rangwala, S.},
  \bibinfo{year}{2012}.
\newblock \bibinfo{title}{Cooling and stabilization by collisions in a mixed
  ion–atom system}.
\newblock \bibinfo{journal}{Nat. Commun.} \bibinfo{volume}{3},
  \bibinfo{pages}{1126}.
\newblock \DOIprefix\doi{10.1038/ncomms2131}.
%Type = Article
\bibitem[{Roos et~al.(2000)Roos, Leibfried, Mundt, Schmidt-Kaler, Eschner and
  Blatt}]{Roos:2000a}
\bibinfo{author}{Roos, C.F.}, \bibinfo{author}{Leibfried, D.},
  \bibinfo{author}{Mundt, A.}, \bibinfo{author}{Schmidt-Kaler, F.},
  \bibinfo{author}{Eschner, J.}, \bibinfo{author}{Blatt, R.},
  \bibinfo{year}{2000}.
\newblock \bibinfo{title}{Experimental demonstration of ground state laser
  cooling with electromagnetically induced transparency.}
\newblock \bibinfo{journal}{Phys.~Rev.~Lett.} \bibinfo{volume}{85},
  \bibinfo{pages}{5547--5550}.
\newblock \URLprefix
  \url{https://journals.aps.org/prl/abstract/10.1103/PhysRevLett.85.5547},
  \DOIprefix\doi{10.1103/PhysRevLett.85.5547}.
%Type = Article
\bibitem[{Rouse and Willitsch(2017)}]{Rouse2017sed}
\bibinfo{author}{Rouse, I.}, \bibinfo{author}{Willitsch, S.},
  \bibinfo{year}{2017}.
\newblock \bibinfo{title}{Superstatistical energy distributions of an ion in an
  ultracold buffer gas}.
\newblock \bibinfo{journal}{Phys. Rev. Lett.} \bibinfo{volume}{118},
  \bibinfo{pages}{143401}.
\newblock \DOIprefix\doi{10.1103/PhysRevLett.118.143401}.
%Type = Article
\bibitem[{Rouse and Willitsch(2019)}]{Rouse2019ted}
\bibinfo{author}{Rouse, I.}, \bibinfo{author}{Willitsch, S.},
  \bibinfo{year}{2019}.
\newblock \bibinfo{title}{The energy distribution of an ion in a radiofrequency
  trap interacting with a nonuniform neutral buffer gas}.
\newblock \bibinfo{journal}{Molecular Physics} \bibinfo{volume}{117},
  \bibinfo{pages}{3120--3131}.
\newblock \URLprefix \url{https://doi.org/10.1080/00268976.2019.1581952},
  \DOIprefix\doi{10.1080/00268976.2019.1581952},
  \href{http://arxiv.org/abs/https://doi.org/10.1080/00268976.2019.1581952}{{\tt
  arXiv:https://doi.org/10.1080/00268976.2019.1581952}}.
%Type = Article
\bibitem[{Safronova et~al.(2018)Safronova, Budker, DeMille, Kimball, Derevianko
  and Clark}]{Safronova2018sfn}
\bibinfo{author}{Safronova, M.S.}, \bibinfo{author}{Budker, D.},
  \bibinfo{author}{DeMille, D.}, \bibinfo{author}{Kimball, D.F.J.},
  \bibinfo{author}{Derevianko, A.}, \bibinfo{author}{Clark, C.W.},
  \bibinfo{year}{2018}.
\newblock \bibinfo{title}{Search for new physics with atoms and molecules}.
\newblock \bibinfo{journal}{Rev. Mod. Phys.} \bibinfo{volume}{90},
  \bibinfo{pages}{025008}.
\newblock \URLprefix
  \url{https://link.aps.org/doi/10.1103/RevModPhys.90.025008},
  \DOIprefix\doi{10.1103/RevModPhys.90.025008}.
%Type = Article
\bibitem[{Saito et~al.(2017)Saito, Haze, Sasakawa, Nakai, Raoult, Silva, Dulieu
  and Mukaiyama}]{Saito:2017}
\bibinfo{author}{Saito, R.}, \bibinfo{author}{Haze, S.},
  \bibinfo{author}{Sasakawa, M.}, \bibinfo{author}{Nakai, R.},
  \bibinfo{author}{Raoult, M.}, \bibinfo{author}{Silva, H.D.},
  \bibinfo{author}{Dulieu, O.}, \bibinfo{author}{Mukaiyama, T.},
  \bibinfo{year}{2017}.
\newblock \bibinfo{title}{Characterization of charge-exchange collisions
  between ultracold $^6${L}i atoms and $^{40}${C}a$^+$ ions}.
\newblock \bibinfo{journal}{Phys.~Rev.~A} \bibinfo{volume}{95},
  \bibinfo{pages}{032709}.
\newblock \DOIprefix\doi{10.1103/PhysRevA.95.032709}.
%Type = Article
\bibitem[{Santos et~al.(2007)Santos, Giedke and Solano}]{Santos:2007}
\bibinfo{author}{Santos, M.F.}, \bibinfo{author}{Giedke, G.},
  \bibinfo{author}{Solano, E.}, \bibinfo{year}{2007}.
\newblock \bibinfo{title}{Noise-free measurement of harmonic oscillators with
  instantaneous interactions}.
\newblock \bibinfo{journal}{Phys.~Rev.~Lett.} \bibinfo{volume}{98},
  \bibinfo{pages}{020401}.
\newblock \URLprefix
  \url{https://journals.aps.org/prl/abstract/10.1103/PhysRevLett.98.020401},
  \DOIprefix\doi{10.1103/PhysRevLett.98.020401}.
%Type = Article
\bibitem[{Schaetz(2017)}]{Schaetz2017tia}
\bibinfo{author}{Schaetz, T.}, \bibinfo{year}{2017}.
\newblock \bibinfo{title}{Trapping ions and atoms optically}.
\newblock \bibinfo{journal}{Journal of Physics B: Atomic, Molecular and Optical
  Physics} \bibinfo{volume}{50}, \bibinfo{pages}{102001}.
\newblock \URLprefix \url{https://doi.org/10.1088/1361-6455/aa69b2},
  \DOIprefix\doi{10.1088/1361-6455/aa69b2}.
%Type = Article
\bibitem[{Schmid et~al.(2010)Schmid, H\"arter and Denschlag}]{Schmid2010doa}
\bibinfo{author}{Schmid, S.}, \bibinfo{author}{H\"arter, A.},
  \bibinfo{author}{Denschlag, J.H.}, \bibinfo{year}{2010}.
\newblock \bibinfo{title}{Dynamics of a cold trapped ion in a {B}ose-{E}instein
  condensate}.
\newblock \bibinfo{journal}{Phys. Rev. Lett.} \bibinfo{volume}{105},
  \bibinfo{pages}{133202}.
\newblock \URLprefix
  \url{https://link.aps.org/doi/10.1103/PhysRevLett.105.133202},
  \DOIprefix\doi{10.1103/PhysRevLett.105.133202}.
%Type = Article
\bibitem[{Schmid et~al.(2012)Schmid, Härter, Frisch, Hoinka and
  Denschlag}]{Schmid2012aaf}
\bibinfo{author}{Schmid, S.}, \bibinfo{author}{Härter, A.},
  \bibinfo{author}{Frisch, A.}, \bibinfo{author}{Hoinka, S.},
  \bibinfo{author}{Denschlag, J.H.}, \bibinfo{year}{2012}.
\newblock \bibinfo{title}{An apparatus for immersing trapped ions into an
  ultracold gas of neutral atoms}.
\newblock \bibinfo{journal}{Review of Scientific Instruments}
  \bibinfo{volume}{83}, \bibinfo{pages}{053108}.
\newblock \URLprefix \url{https://doi.org/10.1063/1.4718356},
  \DOIprefix\doi{10.1063/1.4718356},
  \href{http://arxiv.org/abs/https://doi.org/10.1063/1.4718356}{{\tt
  arXiv:https://doi.org/10.1063/1.4718356}}.
%Type = Article
\bibitem[{Schmid et~al.(2018)Schmid, Veit, Zuber, L\"ow, Pfau, Tarana and
  Tomza}]{Schmid2018rmf}
\bibinfo{author}{Schmid, T.}, \bibinfo{author}{Veit, C.},
  \bibinfo{author}{Zuber, N.}, \bibinfo{author}{L\"ow, R.},
  \bibinfo{author}{Pfau, T.}, \bibinfo{author}{Tarana, M.},
  \bibinfo{author}{Tomza, M.}, \bibinfo{year}{2018}.
\newblock \bibinfo{title}{Rydberg molecules for ion-atom scattering in the
  ultracold regime}.
\newblock \bibinfo{journal}{Phys.~Rev.~Lett.} \bibinfo{volume}{120},
  \bibinfo{pages}{153401}.
\newblock \DOIprefix\doi{10.1103/PhysRevLett.120.153401}.
%Type = Article
\bibitem[{Schmidt et~al.(2020a)Schmidt, H\"onig, Weckesser, Thielemann, Schaetz
  and Karpa}]{Schmidt2020msr}
\bibinfo{author}{Schmidt, J.}, \bibinfo{author}{H\"onig, D.},
  \bibinfo{author}{Weckesser, P.}, \bibinfo{author}{Thielemann, F.},
  \bibinfo{author}{Schaetz, T.}, \bibinfo{author}{Karpa, L.},
  \bibinfo{year}{2020}a.
\newblock \bibinfo{title}{Mass-selective removal of ions from {P}aul traps
  using parametric excitation}.
\newblock \bibinfo{journal}{Appl. Phys. B} \bibinfo{volume}{11},
  \bibinfo{pages}{176--126}.
\newblock \DOIprefix\doi{10.1007/s00340-020-07491-8}.
%Type = Article
\bibitem[{Schmidt et~al.(2020b)Schmidt, Weckesser, Thielemann, Sch\"atz and
  Karpa}]{Schmidt2020otf}
\bibinfo{author}{Schmidt, J.}, \bibinfo{author}{Weckesser, P.},
  \bibinfo{author}{Thielemann, F.}, \bibinfo{author}{Sch\"atz, T.},
  \bibinfo{author}{Karpa, L.}, \bibinfo{year}{2020}b.
\newblock \bibinfo{title}{Optical traps for sympathetic cooling of ions with
  ultracold neutral atoms}.
\newblock \bibinfo{journal}{Phys. Rev. Lett.} \bibinfo{volume}{124},
  \bibinfo{pages}{053402}.
\newblock \URLprefix
  \url{https://link.aps.org/doi/10.1103/PhysRevLett.124.053402},
  \DOIprefix\doi{10.1103/PhysRevLett.124.053402}.
%Type = Article
\bibitem[{Schmidt et~al.(2005)Schmidt, Rosenband, Langer, Itano, Bergquist and
  Wineland}]{Schmidt:2005}
\bibinfo{author}{Schmidt, P.O.}, \bibinfo{author}{Rosenband, T.},
  \bibinfo{author}{Langer, C.}, \bibinfo{author}{Itano, W.M.},
  \bibinfo{author}{Bergquist, J.C.}, \bibinfo{author}{Wineland, D.J.},
  \bibinfo{year}{2005}.
\newblock \bibinfo{title}{{S}pectroscopy using quantum logic.}
\newblock \bibinfo{journal}{Science} \bibinfo{volume}{309},
  \bibinfo{pages}{749--752}.
\newblock \DOIprefix\doi{https://doi.org/10.1126/science.1114375}.
%Type = Article
\bibitem[{Schmitz et~al.(2009)Schmitz, Matjeschk, Schneider, Glueckert, Huber
  and Schaetz}]{Schmitz:2009}
\bibinfo{author}{Schmitz, H.}, \bibinfo{author}{Matjeschk, R.},
  \bibinfo{author}{Schneider, C.}, \bibinfo{author}{Glueckert, J.},
  \bibinfo{author}{Huber, T.}, \bibinfo{author}{Schaetz, T.},
  \bibinfo{year}{2009}.
\newblock \bibinfo{title}{Quantum walk of a trapped ion in phase space}.
\newblock \bibinfo{journal}{Phys.~Rev.~Lett.} \bibinfo{volume}{103},
  \bibinfo{pages}{090504}.
\newblock \URLprefix
  \url{https://journals.aps.org/prl/abstract/10.1103/PhysRevLett.103.090504},
  \DOIprefix\doi{10.1103/PhysRevLett.103.090504}.
%Type = Article
\bibitem[{Schneider et~al.(2010)Schneider, Enderlein, Huber and
  Schaetz}]{Schneider2010oto}
\bibinfo{author}{Schneider, C.}, \bibinfo{author}{Enderlein, M.},
  \bibinfo{author}{Huber, T.}, \bibinfo{author}{Schaetz, T.},
  \bibinfo{year}{2010}.
\newblock \bibinfo{title}{Optical trapping of an ion}.
\newblock \bibinfo{journal}{Nature Photonics} \bibinfo{volume}{4},
  \bibinfo{pages}{772}.
\newblock \URLprefix \url{https://www.nature.com/articles/nphoton.2010.236},
  \DOIprefix\doi{10.1038/nphoton.2010.236}.
%Type = Article
\bibitem[{Schowalter et~al.(2016)Schowalter, Dunning, Chen, Puri, Schneider and
  Hudson}]{Schowalter2016bsb}
\bibinfo{author}{Schowalter, S.J.}, \bibinfo{author}{Dunning, A.J.},
  \bibinfo{author}{Chen, K.}, \bibinfo{author}{Puri, P.},
  \bibinfo{author}{Schneider, C.}, \bibinfo{author}{Hudson, E.R.},
  \bibinfo{year}{2016}.
\newblock \bibinfo{title}{Blue-sky bifurcation of ion energies and the limits
  of neutral-gas sympathetic cooling of trapped ions}.
\newblock \bibinfo{journal}{Nat. Commun.} \bibinfo{volume}{7},
  \bibinfo{pages}{12448}.
\newblock \DOIprefix\doi{10.1038/ncomms12448}.
%Type = Article
\bibitem[{Schreck and Druten(2021)}]{Schreck2021lcf}
\bibinfo{author}{Schreck, F.}, \bibinfo{author}{Druten, K.v.},
  \bibinfo{year}{2021}.
\newblock \bibinfo{title}{Laser cooling for quantum gases}.
\newblock \bibinfo{journal}{Nat. Phys.} \bibinfo{volume}{17},
  \bibinfo{pages}{1296--1304}.
\newblock \DOIprefix\doi{10.1038/s41567-021-01379-w}.
%Type = Article
\bibitem[{Schurer et~al.(2014)Schurer, Schmelcher and Negretti}]{Schurer2014}
\bibinfo{author}{Schurer, J.M.}, \bibinfo{author}{Schmelcher, P.},
  \bibinfo{author}{Negretti, A.}, \bibinfo{year}{2014}.
\newblock \bibinfo{title}{Ground-state properties of ultracold trapped bosons
  with an immersed ionic impurity}.
\newblock \bibinfo{journal}{Phys. Rev. A} \bibinfo{volume}{90},
  \bibinfo{pages}{033601}.
\newblock \DOIprefix\doi{10.1103/PhysRevA.90.033601}.
%Type = Article
\bibitem[{Secker et~al.(2017)Secker, Ewald, Joger, F\"urst, Feldker and
  Gerritsma}]{Secker:2017}
\bibinfo{author}{Secker, T.}, \bibinfo{author}{Ewald, N.},
  \bibinfo{author}{Joger, J.}, \bibinfo{author}{F\"urst, H.},
  \bibinfo{author}{Feldker, T.}, \bibinfo{author}{Gerritsma, R.},
  \bibinfo{year}{2017}.
\newblock \bibinfo{title}{Trapped ions in {R}ydberg-dressed atomic gases}.
\newblock \bibinfo{journal}{Phys.~Rev.~Lett.} \bibinfo{volume}{118},
  \bibinfo{pages}{263201}.
\newblock \DOIprefix\doi{10.1103/PhysRevLett.118.263201}.
%Type = Article
\bibitem[{Secker et~al.(2016)Secker, Gerritsma, Glaetzle and
  Negretti}]{Secker:2016}
\bibinfo{author}{Secker, T.}, \bibinfo{author}{Gerritsma, R.},
  \bibinfo{author}{Glaetzle, A.W.}, \bibinfo{author}{Negretti, A.},
  \bibinfo{year}{2016}.
\newblock \bibinfo{title}{Controlled long-range interactions between {R}ydberg
  atoms and ions}.
\newblock \bibinfo{journal}{Phys. Rev. A} \bibinfo{volume}{94},
  \bibinfo{pages}{013420}.
\newblock \URLprefix \url{https://link.aps.org/doi/10.1103/PhysRevA.94.013420},
  \DOIprefix\doi{10.1103/PhysRevA.94.013420}.
%Type = Article
\bibitem[{\ifmmode~\acute{S}\else \'{S}\fi{}mia\l{}kowski and
  Tomza(2020)}]{Smialkowski2020iac}
\bibinfo{author}{\ifmmode~\acute{S}\else \'{S}\fi{}mia\l{}kowski, M.},
  \bibinfo{author}{Tomza, M.}, \bibinfo{year}{2020}.
\newblock \bibinfo{title}{Interactions and chemical reactions in ionic
  alkali-metal and alkaline-earth-metal diatomic $a{B}^{+}$ and triatomic
  ${A}_{2}{B}^{+}$ systems}.
\newblock \bibinfo{journal}{Phys. Rev. A} \bibinfo{volume}{101},
  \bibinfo{pages}{012501}.
\newblock \URLprefix
  \url{https://link.aps.org/doi/10.1103/PhysRevA.101.012501},
  \DOIprefix\doi{10.1103/PhysRevA.101.012501}.
%Type = Inbook
\bibitem[{Sias and Köhl()}]{Sias2014hqs}
\bibinfo{author}{Sias, C.}, \bibinfo{author}{Köhl, M.}, .
\newblock \bibinfo{title}{Hybrid Quantum Systems of Atoms and Ions}. chapter
  \bibinfo{chapter}{Chapter 12}.
\newblock pp. \bibinfo{pages}{267--291}.
\newblock \DOIprefix\doi{10.1142/9781783264766_0012}.
%Type = Article
\bibitem[{Sikorsky et~al.(2018)Sikorsky, Meir, Ben-shlomi, Akerman and
  Ozeri}]{Sikorsky2018qca}
\bibinfo{author}{Sikorsky, T.}, \bibinfo{author}{Meir, Z.},
  \bibinfo{author}{Ben-shlomi, R.}, \bibinfo{author}{Akerman, N.},
  \bibinfo{author}{Ozeri, R.}, \bibinfo{year}{2018}.
\newblock \bibinfo{title}{Spin-controlled atom--ion chemistry}.
\newblock \bibinfo{journal}{Nat. Comm.} \bibinfo{volume}{9},
  \bibinfo{pages}{920}.
\newblock \DOIprefix\doi{10.1038/s41467-018-03373-y}.
%Type = Article
\bibitem[{{Sikorsky} et~al.(2018){Sikorsky}, {Morita}, {Meir}, {Buchachenko},
  {Ben-shlomi}, {Akerman}, {Narevicius}, {Tscherbul} and
  {Ozeri}}]{Sikorsky2018plb}
\bibinfo{author}{{Sikorsky}, T.}, \bibinfo{author}{{Morita}, M.},
  \bibinfo{author}{{Meir}, Z.}, \bibinfo{author}{{Buchachenko}, A.A.},
  \bibinfo{author}{{Ben-shlomi}, R.}, \bibinfo{author}{{Akerman}, N.},
  \bibinfo{author}{{Narevicius}, E.}, \bibinfo{author}{{Tscherbul}, T.V.},
  \bibinfo{author}{{Ozeri}, R.}, \bibinfo{year}{2018}.
\newblock \bibinfo{title}{{Phase-locking between different partial-waves in
  atom-ion spin-exchange collisions}}.
\newblock \bibinfo{journal}{Phys.~Rev.~Lett.} \bibinfo{volume}{121},
  \bibinfo{pages}{173402}.
\newblock \DOIprefix\doi{10.1103/PhysRevLett.121.173402}.
%Type = Article
\bibitem[{da~Silva~Jr et~al.(2017)da~Silva~Jr, Raoult, Mohammadi, Denchlag and
  Dulieu}]{Silva:2017}
\bibinfo{author}{da~Silva~Jr, H.}, \bibinfo{author}{Raoult, M.},
  \bibinfo{author}{Mohammadi, A.}, \bibinfo{author}{Denchlag, J.H.},
  \bibinfo{author}{Dulieu, O.}, \bibinfo{year}{2017}.
\newblock \bibinfo{title}{Photodissociation of cold {RbBa}$^+$ ions produced in
  a hybrid cold atom-ion trap}.
\newblock \bibinfo{journal}{J. Phys.: Conf. Ser.} \bibinfo{volume}{875},
  \bibinfo{pages}{082013}.
\newblock \URLprefix \url{https://doi.org/10.1088/1742-6596/875/9/082013},
  \DOIprefix\doi{10.1088/1742-6596/875/9/082013}.
%Type = Article
\bibitem[{Sinhal et~al.(2020)Sinhal, Meir, Najafian, Hegi and
  Willitsch}]{Sinhal:2020}
\bibinfo{author}{Sinhal, M.}, \bibinfo{author}{Meir, Z.},
  \bibinfo{author}{Najafian, K.}, \bibinfo{author}{Hegi, G.},
  \bibinfo{author}{Willitsch, S.}, \bibinfo{year}{2020}.
\newblock \bibinfo{title}{Quantum-nondemolition state detection and
  spectroscopy of single trapped molecules}.
\newblock \bibinfo{journal}{Science} \bibinfo{volume}{367},
  \bibinfo{pages}{1213--1218}.
\newblock \DOIprefix\doi{10.1126/science.aaz9837}.
%Type = Article
\bibitem[{Smith et~al.(2005)Smith, Makarov and Lin}]{Smith2005cin}
\bibinfo{author}{Smith, W.W.}, \bibinfo{author}{Makarov, O.P.},
  \bibinfo{author}{Lin, J.}, \bibinfo{year}{2005}.
\newblock \bibinfo{title}{Cold ion–neutral collisions in a hybrid trap}.
\newblock \bibinfo{journal}{Journal of Modern Optics} \bibinfo{volume}{52},
  \bibinfo{pages}{2253--2260}.
\newblock \URLprefix \url{https://doi.org/10.1080/09500340500275850},
  \DOIprefix\doi{10.1080/09500340500275850},
  \href{http://arxiv.org/abs/https://doi.org/10.1080/09500340500275850}{{\tt
  arXiv:https://doi.org/10.1080/09500340500275850}}.
%Type = Incollection
\bibitem[{Stamper-Kurn and Thywissen(2012)}]{StamperKurn2012emo}
\bibinfo{author}{Stamper-Kurn, D.M.}, \bibinfo{author}{Thywissen, J.},
  \bibinfo{year}{2012}.
\newblock
  \bibinfo{title}{\href{https://doi.org/10.1016/B978-0-444-53857-4.00001-5}{Chapter
  1 - Experimental Methods of Ultracold Atomic Physics}}, in:
  \bibinfo{editor}{Kathryn~Levin, A.L.F.}, \bibinfo{editor}{Stamper-Kurn, D.M.}
  (Eds.), \bibinfo{booktitle}{Ultracold Bosonic and Fermionic Gases}.
  \bibinfo{publisher}{Elsevier}. volume~\bibinfo{volume}{5} of
  \textit{\bibinfo{series}{Contemporary Concepts of Condensed Matter Science}},
  p.~\bibinfo{pages}{1}.
\newblock \URLprefix
  \url{https://www.sciencedirect.com/science/article/pii/B9780444538574000015},
  \DOIprefix\doi{https://doi.org/10.1016/B978-0-444-53857-4.00001-5}.
%Type = Article
\bibitem[{Stamper-Kurn and Ueda(2013)}]{StamperKurn2013sbg}
\bibinfo{author}{Stamper-Kurn, D.M.}, \bibinfo{author}{Ueda, M.},
  \bibinfo{year}{2013}.
\newblock \bibinfo{title}{Spinor {B}ose gases: Symmetries, magnetism, and
  quantum dynamics}.
\newblock \bibinfo{journal}{Rev. Mod. Phys.} \bibinfo{volume}{85},
  \bibinfo{pages}{1191--1244}.
\newblock \URLprefix \url{https://link.aps.org/doi/10.1103/RevModPhys.85.1191},
  \DOIprefix\doi{10.1103/RevModPhys.85.1191}.
%Type = Article
\bibitem[{Tanaka et~al.(2012)Tanaka, Masuda, Akimoto, Koda, Ibaraki and
  Urabe}]{Tanaka:2012}
\bibinfo{author}{Tanaka, U.}, \bibinfo{author}{Masuda, K.},
  \bibinfo{author}{Akimoto, Y.}, \bibinfo{author}{Koda, K.},
  \bibinfo{author}{Ibaraki, Y.}, \bibinfo{author}{Urabe, S.},
  \bibinfo{year}{2012}.
\newblock \bibinfo{title}{Micromotion compensation in a surface electrode trap
  by parametric excitation of trapped ions}.
\newblock \bibinfo{journal}{Applied Physics B} \bibinfo{volume}{107},
  \bibinfo{pages}{907--912}.
\newblock \URLprefix
  \url{https://link.springer.com/article/10.1007/s00340-011-4762-2},
  \DOIprefix\doi{10.1007/S00340-011-4762-2}.
%Type = Article
\bibitem[{Ticknor et~al.(2004)Ticknor, Regal, Jin and Bohn}]{Ticknor2004mso}
\bibinfo{author}{Ticknor, C.}, \bibinfo{author}{Regal, C.A.},
  \bibinfo{author}{Jin, D.S.}, \bibinfo{author}{Bohn, J.L.},
  \bibinfo{year}{2004}.
\newblock \bibinfo{title}{Multiplet structure of {F}eshbach resonances in
  nonzero partial waves}.
\newblock \bibinfo{journal}{Phys. Rev. A} \bibinfo{volume}{69},
  \bibinfo{pages}{042712}.
\newblock \URLprefix \url{https://link.aps.org/doi/10.1103/PhysRevA.69.042712},
  \DOIprefix\doi{10.1103/PhysRevA.69.042712}.
%Type = Article
\bibitem[{{Tomza} et~al.(2019){Tomza}, {Jachymski}, {Gerritsma}, {Negretti},
  {Calarco}, {Idziaszek} and {Julienne}}]{Tomza2019chi}
\bibinfo{author}{{Tomza}, M.}, \bibinfo{author}{{Jachymski}, K.},
  \bibinfo{author}{{Gerritsma}, R.}, \bibinfo{author}{{Negretti}, A.},
  \bibinfo{author}{{Calarco}, T.}, \bibinfo{author}{{Idziaszek}, Z.},
  \bibinfo{author}{{Julienne}, P.S.}, \bibinfo{year}{2019}.
\newblock \bibinfo{title}{{Cold hybrid ion-atom systems}}.
\newblock \bibinfo{journal}{Rev. Mod. Phys.} \bibinfo{volume}{91},
  \bibinfo{pages}{035001}.
\newblock \DOIprefix\doi{10.1103/RevModPhys.91.035001}.
%Type = Article
\bibitem[{Tomza et~al.(2015)Tomza, Koch and Moszynski}]{Tomza2015cib}
\bibinfo{author}{Tomza, M.}, \bibinfo{author}{Koch, C.P.},
  \bibinfo{author}{Moszynski, R.}, \bibinfo{year}{2015}.
\newblock \bibinfo{title}{Cold interactions between an {Y}b$^+$ ion and a {L}i
  atom: {P}rospects for sympathetic cooling, radiative association, and
  {F}eshbach resonances}.
\newblock \bibinfo{journal}{Phys.~Rev.~A} \bibinfo{volume}{91},
  \bibinfo{pages}{042706}.
\newblock \DOIprefix\doi{10.1103/PhysRevA.91.042706}.
%Type = Article
\bibitem[{Tomza and Lisaj(2020)}]{Tomza2020iac}
\bibinfo{author}{Tomza, M.}, \bibinfo{author}{Lisaj, M.}, \bibinfo{year}{2020}.
\newblock \bibinfo{title}{Interactions and charge-transfer dynamics of an
  ${\mathrm{al}}^{+}$ ion immersed in ultracold {R}b and {S}r atoms}.
\newblock \bibinfo{journal}{Phys. Rev. A} \bibinfo{volume}{101},
  \bibinfo{pages}{012705}.
\newblock \URLprefix
  \url{https://link.aps.org/doi/10.1103/PhysRevA.101.012705},
  \DOIprefix\doi{10.1103/PhysRevA.101.012705}.
%Type = Article
\bibitem[{T{\"o}rm{\"a}(2016)}]{Torma2016pou}
\bibinfo{author}{T{\"o}rm{\"a}, P.}, \bibinfo{year}{2016}.
\newblock \bibinfo{title}{Physics of ultracold {F}ermi gases revealed by
  spectroscopies}.
\newblock \bibinfo{journal}{Phys. Scr.} \bibinfo{volume}{91},
  \bibinfo{pages}{043006}.
\newblock \URLprefix \url{http://stacks.iop.org/1402-4896/91/i=4/a=043006},
  \DOIprefix\doi{10.1088/0031-8949/91/4/043006}.
%Type = Book
\bibitem[{T{\"{o}}rm{\"{a}} and Sengstock(2014)}]{Torma2014qge}
\bibinfo{author}{T{\"{o}}rm{\"{a}}, P.}, \bibinfo{author}{Sengstock, K.},
  \bibinfo{year}{2014}.
\newblock
  \bibinfo{title}{\href{https://www.worldscientific.com/doi/abs/10.1142/9781783264766}{Quantum
  Gas Experiments: Exploring Many-Body States}}.
\newblock \bibinfo{publisher}{Imperial College Press}.
\newblock \URLprefix
  \url{https://www.worldscientific.com/doi/abs/10.1142/9781783264766},
  \DOIprefix\doi{10.1142/9781783264766_fmatter}.
%Type = Article
\bibitem[{Trimby et~al.(2022)Trimby, Hirzler, Fuerst, Safavi-Naini, Gerritsma
  and Lous}]{Trimby2022bgc}
\bibinfo{author}{Trimby, E.}, \bibinfo{author}{Hirzler, H.},
  \bibinfo{author}{Fuerst, H.A.}, \bibinfo{author}{Safavi-Naini, A.},
  \bibinfo{author}{Gerritsma, R.}, \bibinfo{author}{Lous, R.S.},
  \bibinfo{year}{2022}.
\newblock \bibinfo{title}{Buffer gas cooling of ions in radio frequency traps
  using ultracold atoms}.
\newblock \bibinfo{journal}{New Journal of Physics} \URLprefix
  \url{http://iopscience.iop.org/article/10.1088/1367-2630/ac5759}.
%Type = Article
\bibitem[{Tscherbul et~al.(2016)Tscherbul, Brumer and
  Buchachenko}]{Tscherbul2016soi}
\bibinfo{author}{Tscherbul, T.V.}, \bibinfo{author}{Brumer, P.},
  \bibinfo{author}{Buchachenko, A.A.}, \bibinfo{year}{2016}.
\newblock \bibinfo{title}{Spin-orbit interactions and quantum spin dynamics in
  cold ion-atom collisions}.
\newblock \bibinfo{journal}{Phys. Rev. Lett.} \bibinfo{volume}{117},
  \bibinfo{pages}{143201}.
\newblock \DOIprefix\doi{10.1103/PhysRevLett.117.143201}.
%Type = Article
\bibitem[{Turchette et~al.(2000)Turchette, Kielpinski, King, Leibfried,
  Meekhof, Myatt, Rowe, Sackett, Wood, Itano, Monroe and
  Wineland}]{Turchette:2000}
\bibinfo{author}{Turchette, Q.A.}, \bibinfo{author}{Kielpinski},
  \bibinfo{author}{King, B.E.}, \bibinfo{author}{Leibfried, D.},
  \bibinfo{author}{Meekhof, D.M.}, \bibinfo{author}{Myatt, C.J.},
  \bibinfo{author}{Rowe, M.A.}, \bibinfo{author}{Sackett, C.A.},
  \bibinfo{author}{Wood, C.S.}, \bibinfo{author}{Itano, W.M.},
  \bibinfo{author}{Monroe, C.}, \bibinfo{author}{Wineland, D.J.},
  \bibinfo{year}{2000}.
\newblock \bibinfo{title}{Heating of trapped ions from the quantum ground
  state}.
\newblock \bibinfo{journal}{Phys.~Rev.~A} \bibinfo{volume}{61},
  \bibinfo{pages}{063418}.
\newblock \URLprefix
  \url{https://journals.aps.org/pra/abstract/10.1103/PhysRevA.61.063418},
  \DOIprefix\doi{10.1103/PhysRevA.61.063418}.
%Type = Article
\bibitem[{Turlapov(2012)}]{Turlapov2012fgo}
\bibinfo{author}{Turlapov, A.V.}, \bibinfo{year}{2012}.
\newblock \bibinfo{title}{Fermi gas of atoms}.
\newblock \bibinfo{journal}{JETP Lett.} \bibinfo{volume}{95},
  \bibinfo{pages}{96}.
\newblock \URLprefix \url{https://doi.org/10.1134/S0021364012020105},
  \DOIprefix\doi{10.1134/S0021364012020105}.
%Type = Article
\bibitem[{Vale and Zwierlein(2021)}]{Vale2021spo}
\bibinfo{author}{Vale, C.J.}, \bibinfo{author}{Zwierlein, M.},
  \bibinfo{year}{2021}.
\newblock \bibinfo{title}{Spectroscopic probes of quantum gases}.
\newblock \bibinfo{journal}{Nat. Phys.} \bibinfo{volume}{17},
  \bibinfo{pages}{1305--1315}.
\newblock \DOIprefix\doi{10.1038/s41567-021-01434-6}.
%Type = Article
\bibitem[{Veit et~al.(2021)Veit, Zuber, Herrera-Sancho, Anasuri, Schmid,
  Meinert, L\"ow and Pfau}]{Veit2021pim}
\bibinfo{author}{Veit, C.}, \bibinfo{author}{Zuber, N.},
  \bibinfo{author}{Herrera-Sancho, O.A.}, \bibinfo{author}{Anasuri, V.S.V.},
  \bibinfo{author}{Schmid, T.}, \bibinfo{author}{Meinert, F.},
  \bibinfo{author}{L\"ow, R.}, \bibinfo{author}{Pfau, T.},
  \bibinfo{year}{2021}.
\newblock \bibinfo{title}{Pulsed ion microscope to probe quantum gases}.
\newblock \bibinfo{journal}{Phys. Rev. X} \bibinfo{volume}{11},
  \bibinfo{pages}{011036}.
\newblock \URLprefix \url{https://link.aps.org/doi/10.1103/PhysRevX.11.011036},
  \DOIprefix\doi{10.1103/PhysRevX.11.011036}.
%Type = Article
\bibitem[{Wallentowitz and Vogel(1995)}]{Wallentowitz:1995}
\bibinfo{author}{Wallentowitz, S.}, \bibinfo{author}{Vogel, W.},
  \bibinfo{year}{1995}.
\newblock \bibinfo{title}{Reconstruction of the quantum mechanical state of a
  trapped ion}.
\newblock \bibinfo{journal}{Phys.~Rev.~Lett.} \bibinfo{volume}{75},
  \bibinfo{pages}{2932--2935}.
\newblock \URLprefix
  \url{https://journals.aps.org/prl/abstract/10.1103/PhysRevLett.75.2932},
  \DOIprefix\doi{10.1103/PhysRevLett.75.2932}.
%Type = Article
\bibitem[{Wang et~al.(2020)Wang, Dei{\ss}, Raithel and Denschlag}]{Wang2020oco}
\bibinfo{author}{Wang, L.}, \bibinfo{author}{Dei{\ss}, M.},
  \bibinfo{author}{Raithel, G.}, \bibinfo{author}{Denschlag, J.H.},
  \bibinfo{year}{2020}.
\newblock \bibinfo{title}{Optical control of atom-ion collisions using a
  {R}ydberg state}.
\newblock \bibinfo{journal}{Journal of Physics B: Atomic, Molecular and Optical
  Physics} \bibinfo{volume}{53}, \bibinfo{pages}{134005}.
\newblock \URLprefix \url{https://doi.org/10.1088/1361-6455/ab7d24},
  \DOIprefix\doi{10.1088/1361-6455/ab7d24}.
%Type = Incollection
\bibitem[{Wang et~al.(2013)Wang, D’Incao and Esry}]{Wang2013ufb}
\bibinfo{author}{Wang, Y.}, \bibinfo{author}{D’Incao, J.P.},
  \bibinfo{author}{Esry, B.D.}, \bibinfo{year}{2013}.
\newblock \bibinfo{title}{Chapter 1 - ultracold few-body systems}, in:
  \bibinfo{editor}{Arimondo, E.}, \bibinfo{editor}{Berman, P.R.},
  \bibinfo{editor}{Lin, C.C.} (Eds.), \bibinfo{booktitle}{Advances in Atomic,
  Molecular, and Optical Physics}. \bibinfo{publisher}{Academic Press}.
  volume~\bibinfo{volume}{62} of \textit{\bibinfo{series}{Advances In Atomic,
  Molecular, and Optical Physics}}, pp. \bibinfo{pages}{1--115}.
\newblock \URLprefix
  \url{https://www.sciencedirect.com/science/article/pii/B9780124080904000013},
  \DOIprefix\doi{https://doi.org/10.1016/B978-0-12-408090-4.00001-3}.
%Type = Article
\bibitem[{Weckesser et~al.(2021a)Weckesser, Thielemann, Hoenig, Lambrecht,
  Karpa and Schaetz}]{Weckesser2021tsa}
\bibinfo{author}{Weckesser, P.}, \bibinfo{author}{Thielemann, F.},
  \bibinfo{author}{Hoenig, D.}, \bibinfo{author}{Lambrecht, A.},
  \bibinfo{author}{Karpa, L.}, \bibinfo{author}{Schaetz, T.},
  \bibinfo{year}{2021}a.
\newblock \bibinfo{title}{Trapping, shaping, and isolating of an ion {C}oulomb
  crystal via state-selective optical potentials}.
\newblock \bibinfo{journal}{Phys. Rev. A} \bibinfo{volume}{103},
  \bibinfo{pages}{013112}.
\newblock \URLprefix
  \url{https://link.aps.org/doi/10.1103/PhysRevA.103.013112},
  \DOIprefix\doi{10.1103/PhysRevA.103.013112}.
%Type = Article
\bibitem[{Weckesser et~al.(2021b)Weckesser, Thielemann, Wiater, Wojciechowska,
  Karpa, Jachymski, Tomza, Walker and Schaetz}]{Weckesser2021oof}
\bibinfo{author}{Weckesser, P.}, \bibinfo{author}{Thielemann, F.},
  \bibinfo{author}{Wiater, D.}, \bibinfo{author}{Wojciechowska, A.},
  \bibinfo{author}{Karpa, L.}, \bibinfo{author}{Jachymski, K.},
  \bibinfo{author}{Tomza, M.}, \bibinfo{author}{Walker, T.},
  \bibinfo{author}{Schaetz, T.}, \bibinfo{year}{2021}b.
\newblock \bibinfo{title}{Observation of {F}eshbach resonances between a single
  ion and ultracold atoms}.
\newblock \bibinfo{journal}{Nature} \bibinfo{volume}{600},
  \bibinfo{pages}{429}.
\newblock \DOIprefix\doi{10.1038/s41586-021-04112-y}.
%Type = Article
\bibitem[{Weinstock(1976)}]{Weinstock:1976}
\bibinfo{author}{Weinstock, R.}, \bibinfo{year}{1976}.
\newblock \bibinfo{title}{{On a fallacious proof of Earnshaw's theorem}}.
\newblock \bibinfo{journal}{Am.~J.~ of Phys.} \bibinfo{volume}{44},
  \bibinfo{pages}{392–393}.
\newblock \URLprefix \url{https://aapt.scitation.org/doi/10.1119/1.10449},
  \DOIprefix\doi{10.1119/1.10449}.
%Type = Article
\bibitem[{Wesenberg et~al.(2007)Wesenberg, Epstein, Leibfried, Blakestad,
  Britton, Home, Itano, Jost, Knill, Langer, Ozeri, Seidelin and
  Wineland}]{Wesenberg:2007}
\bibinfo{author}{Wesenberg, J.H.}, \bibinfo{author}{Epstein, R.J.},
  \bibinfo{author}{Leibfried, D.}, \bibinfo{author}{Blakestad, R.B.},
  \bibinfo{author}{Britton, J.}, \bibinfo{author}{Home, J.P.},
  \bibinfo{author}{Itano, W.M.}, \bibinfo{author}{Jost, J.D.},
  \bibinfo{author}{Knill, E.}, \bibinfo{author}{Langer, C.},
  \bibinfo{author}{Ozeri, R.}, \bibinfo{author}{Seidelin, S.},
  \bibinfo{author}{Wineland, D.J.}, \bibinfo{year}{2007}.
\newblock \bibinfo{title}{Fluorescence during {D}oppler cooling of a single
  trapped atom}.
\newblock \bibinfo{journal}{Phys.~Rev.~A} \bibinfo{volume}{76},
  \bibinfo{pages}{053416}.
\newblock \URLprefix
  \url{https://journals.aps.org/pra/abstract/10.1103/PhysRevA.76.053416},
  \DOIprefix\doi{10.1103/PhysRevA.76.053416}.
%Type = Article
\bibitem[{Wester(2009)}]{Wester2009RMT}
\bibinfo{author}{Wester, R.}, \bibinfo{year}{2009}.
\newblock \bibinfo{journal}{J. Phys. B: At. Mol. Opt. Phys.}
  \bibinfo{volume}{42}.
\newblock \URLprefix \url{https://doi.org/10.1088/0953-4075/42/15/154001}.
%Type = Article
\bibitem[{Willitsch(2015)}]{Willitsch:2015}
\bibinfo{author}{Willitsch, S.}, \bibinfo{year}{2015}.
\newblock \bibinfo{title}{Ion-atom hybrid systems}.
\newblock \bibinfo{journal}{Proc. Int. Sch. Phys. Enrico Fermi}
  \bibinfo{volume}{189}, \bibinfo{pages}{255 -- 268}.
\newblock \DOIprefix\doi{10.3254/978-1-61499-526-5-255}.
%Type = Article
\bibitem[{Wolf et~al.(2016)Wolf, Wan, Heip, Gebert, Shi and
  Schmidt}]{Wolf:2016}
\bibinfo{author}{Wolf, F.}, \bibinfo{author}{Wan, Y.}, \bibinfo{author}{Heip,
  J.C.}, \bibinfo{author}{Gebert, F.}, \bibinfo{author}{Shi, C.},
  \bibinfo{author}{Schmidt, P.O.}, \bibinfo{year}{2016}.
\newblock \bibinfo{title}{Non-destructive state detection for quantum logic
  spectroscopy of molecular ions}.
\newblock \bibinfo{journal}{Nature} \bibinfo{volume}{530},
  \bibinfo{pages}{457--460}.
\newblock \DOIprefix\doi{10.1038/nature16513}.
%Type = Article
\bibitem[{Z{\"a}hringer et~al.(2010)Z{\"a}hringer, Kirchmair, Gerritsma,
  Solano, Blatt and Roos}]{Zaehringer:2010}
\bibinfo{author}{Z{\"a}hringer, F.}, \bibinfo{author}{Kirchmair, G.},
  \bibinfo{author}{Gerritsma, R.}, \bibinfo{author}{Solano, E.},
  \bibinfo{author}{Blatt, R.}, \bibinfo{author}{Roos, C.F.},
  \bibinfo{year}{2010}.
\newblock \bibinfo{title}{Realization of a quantum walk with one and two
  trapped ions}.
\newblock \bibinfo{journal}{Phys.~Rev.~Lett.} \bibinfo{volume}{104},
  \bibinfo{pages}{100503}.
\newblock \URLprefix
  \url{https://journals.aps.org/prl/abstract/10.1103/PhysRevLett.104.100503},
  \DOIprefix\doi{10.1103/PhysRevLett.104.100503}.
%Type = Incollection
\bibitem[{Zhang and Willitsch(2017)}]{Zhang:2017}
\bibinfo{author}{Zhang, D.}, \bibinfo{author}{Willitsch, S.},
  \bibinfo{year}{2017}.
\newblock \bibinfo{title}{Cold ion chemistry, in low energy and low temperature
  molecular scattering}, \bibinfo{publisher}{RSC Publishing, London}.
\newblock \URLprefix \url{https://doi.org/10.1039/9781782626800-00496},
  \DOIprefix\doi{10.1039/9781782626800-00496}.
%Type = Article
\bibitem[{Zipkes et~al.(2010)Zipkes, Paltzer, Sias and
  K{\"o}hl}]{Zipkes2010ats}
\bibinfo{author}{Zipkes, C.}, \bibinfo{author}{Paltzer, S.},
  \bibinfo{author}{Sias, C.}, \bibinfo{author}{K{\"o}hl, M.},
  \bibinfo{year}{2010}.
\newblock \bibinfo{title}{A trapped single ion inside a {B}ose-{E}instein
  condensate}.
\newblock \bibinfo{journal}{Nature} \bibinfo{volume}{464},
  \bibinfo{pages}{388--391}.
\newblock \DOIprefix\doi{doi:10.1038/nature08865}.
%Type = Article
\bibitem[{Zipkes et~al.(2011)Zipkes, Ratschbacher, Sias and
  K{\"o}hl}]{Zipkes2011koa}
\bibinfo{author}{Zipkes, C.}, \bibinfo{author}{Ratschbacher, L.},
  \bibinfo{author}{Sias, C.}, \bibinfo{author}{K{\"o}hl, M.},
  \bibinfo{year}{2011}.
\newblock \bibinfo{title}{Kinetics of a single trapped ion in an ultracold
  buffer gas}.
\newblock \bibinfo{journal}{New J.~Phys.} \bibinfo{volume}{13},
  \bibinfo{pages}{053020}.
\newblock \DOIprefix\doi{10.1088/1367-2630/13/5/053020}.
%Type = Article
\bibitem[{Zuber et~al.(2021)Zuber, Anasuri, Berngruber, Zou, Meinert, L{\"o}w
  and Pfau}]{Zuber2021sio}
\bibinfo{author}{Zuber, N.}, \bibinfo{author}{Anasuri, V.S.V.},
  \bibinfo{author}{Berngruber, M.}, \bibinfo{author}{Zou, Y.Q.},
  \bibinfo{author}{Meinert, F.}, \bibinfo{author}{L{\"o}w, R.},
  \bibinfo{author}{Pfau, T.}, \bibinfo{year}{2021}.
\newblock \bibinfo{title}{Spatial imaging of a novel type of molecular ions}.
\newblock \bibinfo{journal}{arXiv:2111.02680}
  \DOIprefix\doi{https://doi.org/10.48550/arXiv.2111.02680}.
%Type = Book
\bibitem[{Zwerger(2012)}]{Zwerger2012tbb}
\bibinfo{editor}{Zwerger, W.} (Ed.), \bibinfo{year}{2012}.
\newblock
  \bibinfo{title}{\href{http://www.springer.com/de/book/9783642219771}{The
  BCS-BEC Crossover and the Unitary Fermi Gas}}. volume \bibinfo{volume}{836}
  of \textit{\bibinfo{series}{Lecture Notes in Physics}}.
\newblock \bibinfo{publisher}{Springer, Berlin Heidelberg}.

\end{thebibliography}

\end{document}